\documentclass[%
 reprint,
 amsmath,amssymb,
 aps,
]{revtex4-2}

\usepackage{graphicx}% Include figure files
\usepackage{dcolumn}% Align table columns on decimal point
\usepackage{bm}% bold math
\usepackage{xcolor}
\usepackage{hyperref}% add hypertext capabilities
\usepackage{booktabs}

\begin{document}

\preprint{APS/123-QED}

\title{
%Quantum-Correlated Polariton States in the Visible and Mid-Infrared Spectral Ranges
%\\
Entangled Polariton States in the Visible and Mid-Infrared Spectral Ranges
}

\author{Vladislav~Yu.~Shishkov\textsuperscript{1}
}
\email{vladislav.shishkov@aalto.fi}

\author{Oleg Kotov\textsuperscript{2,3}}

\author{Emily Haughton\textsuperscript{1}}

\author{Darius Urbonas\textsuperscript{4}}

\author{\\Lee A. Rozema\textsuperscript{5}}

\author{Francisco J. Garcia-Vidal\textsuperscript{2,3}}

\author{Johannes Feist\textsuperscript{2,3}}

\author{Anton~V.~Zasedatelev\textsuperscript{1}}
\email{anton.zasedatelev@aalto.fi}

\affiliation{
\textsuperscript{1}Department of Applied Physics, Aalto University School of Science, FI-00076 Espoo, Finland
}
\affiliation{
\textsuperscript{2}Departamento de Física Teórica de la Materia Condensada, Universidad Autónoma de Madrid, E-28049 Madrid, Spain
}
\affiliation{
\textsuperscript{3}Condensed Matter Physics Center (IFIMAC), Universidad Autónoma de Madrid, E-28049 Madrid, Spain
}
\affiliation{
\textsuperscript{4}IBM Research Europe – Zurich, Säumerstrasse 4, 8803 Rüschlikon, Switzerland
}
\affiliation{
\textsuperscript{5}University of Vienna, Faculty of Physics, Vienna Center for Quantum Science and Technology (VCQ) \& Research Platform TURIS, Boltzmanngasse 5, 1090 Vienna, Austria
}

\date{\today}

\begin{abstract}

Entanglement generation in polariton systems is fundamentally constrained by high losses and decoherence, which typically outweigh polariton nonlinearities. 
Here, we propose a conceptually different approach that uses optomechanical interactions, rather than polariton–polariton interactions, to generate entangled polaritons. 
Our double-resonant scheme relies on strong exciton-phonon coupling, found in both inorganic and molecular semiconductors, enabling room-temperature generation of spectrally disparate photon pairs. 
The quantum coherent and delocalized nature of polariton states inside optical cavities ensures efficient single-mode outcoupling and allows for \textit{unconditional} quantum state preparation -- not relying on any post-selection or projective measurements.
When conditioned on exciton-polariton emission, single phonon-polariton states can be prepared that subsequently yield bright, heralded single-photon emission in the mid-IR/THz. 
We introduce a double-resonant optomechanical platform that enables scalable, room-temperature quantum polaritonics without relying on conventional excitonic nonlinearities.

%\textcolor{blue}{in the abstract we need to say that we propose double resonant polariton system with optomechanical interaction that is capable of}

%1. generation entangled polaritons at room temperature

%2. doesn't require any polariton-polariton nonlinearity

%3. near-deterministic photon pair generation

%4. disparate mid-IR - visible photons, with possibility to extend to any spectral range via non-resonant

%5. almost ideal outcoupling into a single mode due to delocalized nature of polariton states, meaning collective entangled states are produced 

%6. as consequence of p.5 no \textit{"which-path"} information can by acquired in the detection, therefore our setup offers measurement-free quantum state preparation (\textit{in the pulsed reqime}).

\end{abstract}

\maketitle

{\it Introduction.} 
Quantum technologies are currently transforming communication, sensing, and computation. Among the various platforms, photonic approaches are especially promising. 
With respect to quantum communication, photons remain the leading candidate for interfacing different quantum systems and transmitting quantum information between remote parties~\cite{wang2020integrated}.
Photonic quantum computing is also advancing~\cite{madsen2022quantum, psiquantum2025manufacturable}, but it is hindered by the absence of strong, controllable photon–photon interactions required for deterministic two‑qubit gates and for generating entanglement. 
Matter-based systems, by contrast, readily provide interactions, but are hampered by decoherence. Hybrid quasiparticles known as polaritons, coherent superpositions of photonic and electronic degrees of freedom, inherit the best of both worlds: a light effective mass and high group velocity from their photonic component, together with strong nonlinearities from their matter component~\cite{carusotto2013quantum}. 
This seems to make polaritons uniquely poised to serve both as a nonlinear medium and as an interface between quantum communication channels and quantum computing devices.

Early theoretical work predicted that polariton–polariton interactions in solid‑state microcavities could naturally produce entangled states ~\cite{PhysRevB.63.041303,PhysRevB.69.245304,PhysRevLett.100.170505}. 
In principle, parametric scattering of polaritons is sufficiently strong because of their matter component ~\cite{PhysRevLett.84.1547,PhysRevLett.87.127403,saba2001high,sun2017direct}. 
In practice, however, the situation proved more complex: the desired quantum correlations are entirely washed out by competing processes such as thermalization via acoustic phonons, incoherent scattering by disordered states, and still-large cavity losses that exceed the pair‑particle scattering rate ~\cite{PhysRevA.69.063807, suarez2016entanglement}. Consequently, to date entanglement has not been demonstrated in solid‑state polariton systems.

Meanwhile, the quantum-coherent nature of polaritons at the single-particle level has been explored in various settings~\cite{cuevas2018first,munoz2019emergence,delteil2019towards}. 
An important experimental test involved injecting externally entangled photons into a polariton system~\cite{cuevas2018first}, demonstrating that polaritons can inherit and transmit quantum information. The measured concurrence remained high (0.806), indicating strong preservation of quantum correlations within exciton-polariton systems.

\begin{figure}
\includegraphics[width=1\linewidth]{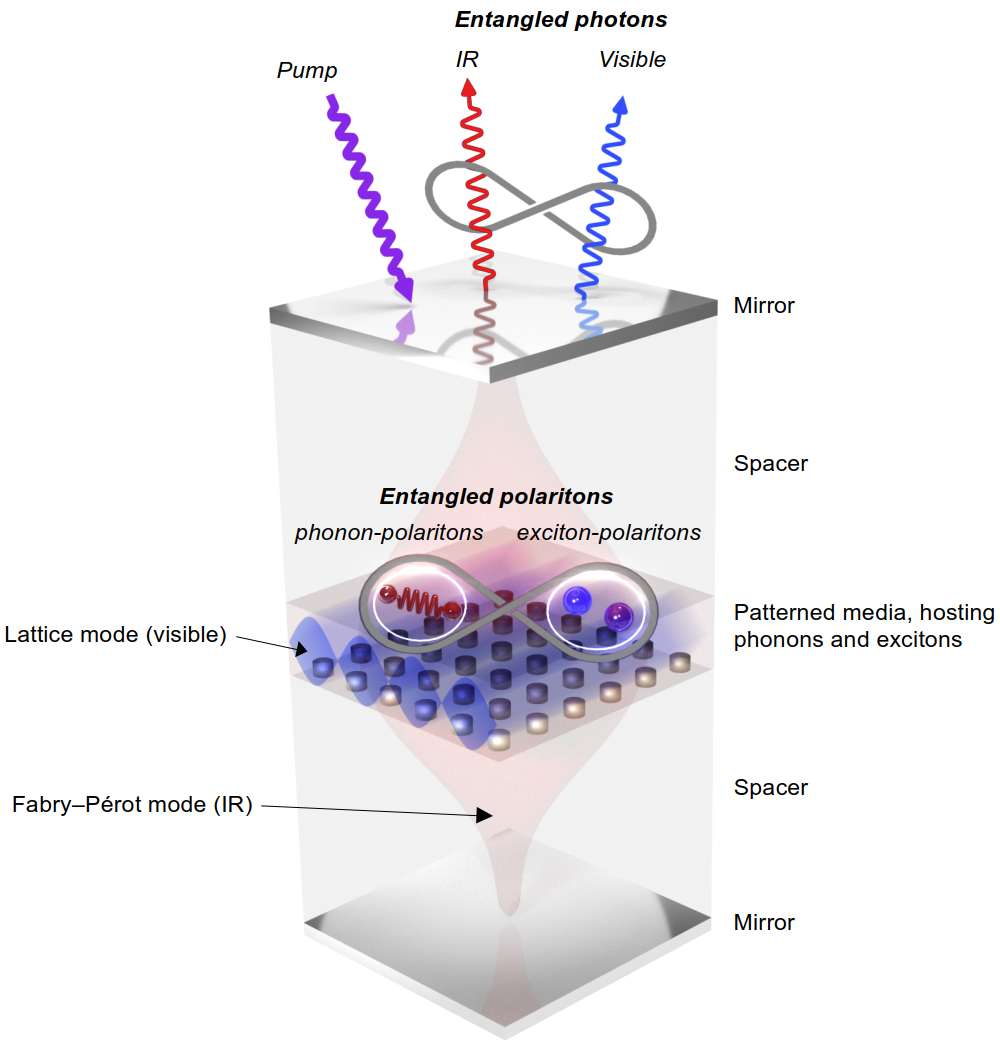}
\caption{
{\bf Double resonant polariton optomechanical system.}
Double-resonant polariton system hosting optomechanically interacting exciton-polaritons and phonon-polaritons that generates entangled polariton states and emits entangled photons under coherent pumping.
}
    \label{fig:Fig1}
\end{figure}

Therefore, the ability to generate entanglement intrinsically within a polariton system remains a central challenge in solid-state quantum optics, resolution of which would provide a novel route towards scalable quantum technologies. 
Polariton blockade could potentially isolate two-particle states, leading to quantum-correlated photon pairs~\cite{munoz2019emergence,delteil2019towards}. 
The interaction strength required to achieve this, quantified by the interaction over loss ratio $g/\gamma$, must exceed unity. 
With current estimates of $g/\gamma\sim0.1$, the blockade effect was evidenced by antibunching at the level $g^{(2)}(0)=0.9$~\cite{munoz2019emergence,delteil2019towards}, which so far keeps these proposals at bay. 
Enhancement via Feshbach resonances involving biexciton or triexciton bound states~\cite{takemura2014polaritonic} could push the system toward the strong single-particle interaction regime $g/\gamma>1$~\cite{carusotto2010feshbach}, to be achieved. 
In such a regime, cascaded emission from anharmonic ladders of interacting polaritons could serve as a source of entangled photon pairs~\cite{scarpelli2024probing}, similar to the biexciton cascades in quantum dots~\cite{akopian2006entangled}. 
Strong light-matter coupling with Rydberg excitons in $\rm Cu_2O$ offers an alternative, promising route toward the quantum regime at the single-polariton level, albeit under deep cryogenic conditions~\cite{kazimierczuk2014giant, heckotter2018rydberg, heckotter2021asymmetric,orfanakis2022rydberg}. 
Similarly, dipolar polaritons~\cite{rosenberg2018strongly} based on indirect excitons in coupled quantum wells or van der Waals heterostructures offer tunable long-range interactions via permanent dipole moments, with single particle nonlinearities approaching unity at cryogenic temperatures~\cite{high2012spontaneous, unuchek2018room}.

\begin{figure}
\includegraphics[width=0.98\linewidth]{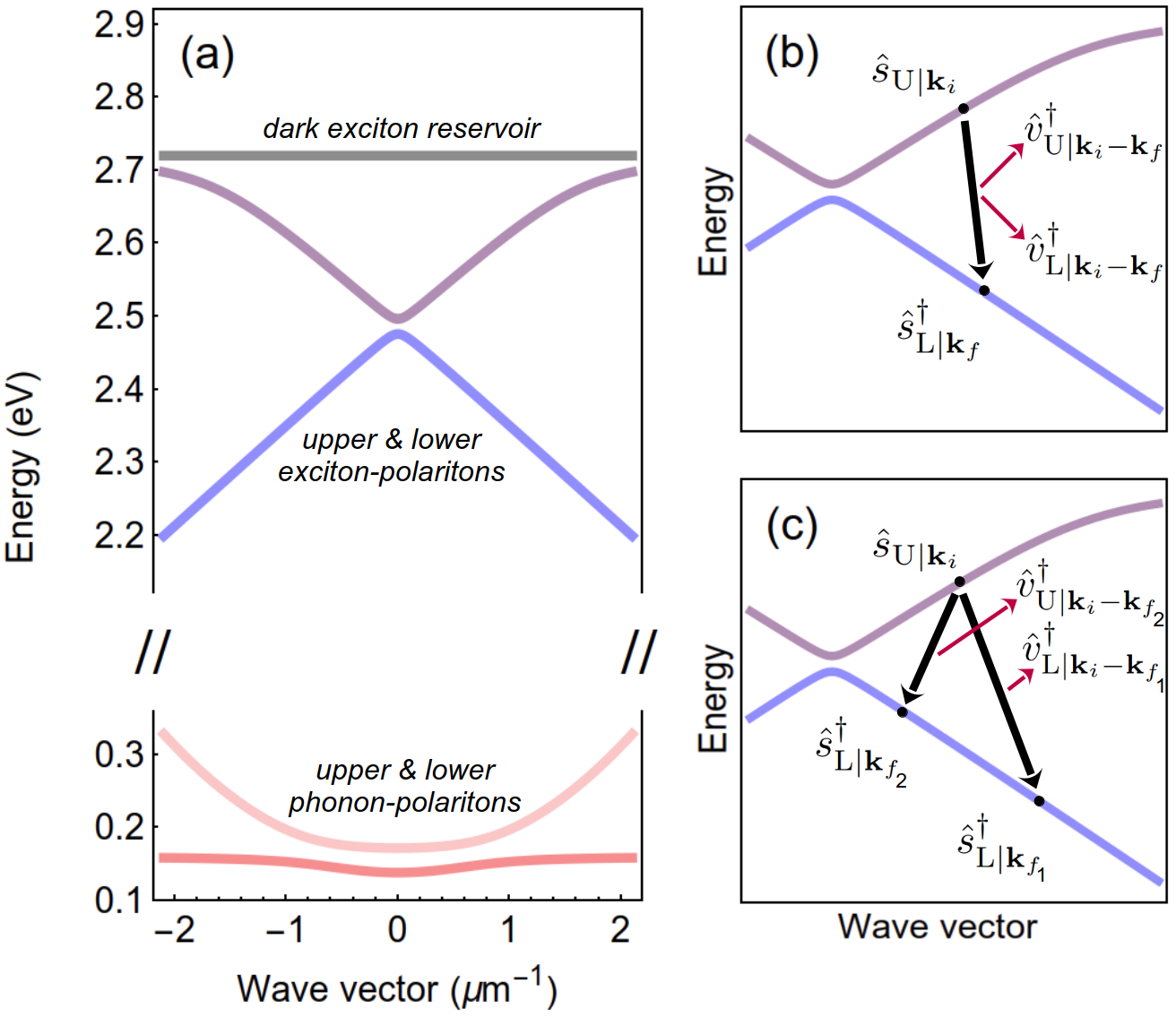}
\caption{
{\bf Dispersion relations.}
(a) Dispersions of upper (purple line) and lower (blue line) exciton-polarotons and upper (pink line) and lower (red line) phonon-polaritons.
Gray line in (a) shows the resonant frequency of excitons.  
(b) Generation of exciton- and phonon-polariton pairs.
(c) Generation of entangled exciton- and phonon-polariton Bell pairs.
Operators in (b) and (c) are introduced in Eq.~(\ref{H_optomech_f}).
Parameters are typical for room-temperature polariton systems; for concreteness, we use those of molecular polariton cavities with strong sideband-resolved optomechanical interaction~\cite{zasedatelev2019room, zasedatelev2021single}.
}
    \label{fig: dispersions}
\end{figure}

Beyond the single-particle level and discrete variables, the macroscopic quantum coherence of polaritons has been demonstrated through collective phenomena such as Bose–Einstein condensation~\cite{kasprzak2006bose}, superfluidity~\cite{amo2009superfluidity, lerario2017room}, quantized vortices~\cite{lagoudakis2008quantized} among others~\cite{carusotto2013quantum}. These observations confirm that polaritons can form collective quantum states, motivating proposals to harness such states for entanglement generation. In particular, a recent theoretical scheme shows that spin squeezing can be realized in a resonantly pumped spinor polariton condensate~\cite{feng2021sensitive} leveraging the system's intrinsic nonlinearities to drive one-axis twisting dynamics analogous to those used in atomic BECs~\cite{PhysRevA.104.043323}. Two-mode squeezing below the standard quantum limit can be used as experimental witness for bi-partite entanglement in continuous variables~\cite{feng2021sensitive,PhysRevA.108.053301}. Achieving strong squeezing in polariton BEC represents a significant challenge, though first evidences for quadrature squeezing in polariton systems has already been demonstrated~\cite{boulier2014polariton,adiyatullin2017periodic}.

In this work, we propose a conceptually new approach to generate entangled polaritons, fundamentally distinct from conventional schemes relying on polariton-polariton nonlinearities. Our platform leverages a double-resonant polariton system with intrinsic optomechanical interaction, where strong exciton-phonon coupling -- naturally present in both inorganic and molecular semiconductors -- enables robust entanglement generation at room temperature. 
In conventional optomechanics, photons interact with mechanical motion via radiation pressure or dispersive coupling~\cite{aspelmeyer2014cavity}. In our system, both the optical and mechanical modes are no longer pure -- they are hybrid light-matter excitations. 
The optical mode is an exciton-polariton, a coherent superposition of cavity photons and electronic excitations in matter. 
The mechanical mode arises from collective vibrational type excitations within the matter, delocalized {\it phonons}. 
The optomechanical interaction between them allows for the bright emission of spectrally disparate entangled photon pairs (e.g., mid-IR and visible), with spectral flexibility enabled by non-resonant excitation. Crucially, the delocalized nature of polariton states ensures efficient outcoupling into a single optical mode, inherently erasing which-path information from individual emitters coupled to the cavity. This enables quantum state preparation at room temperature without the need for mode post-selection.

We present a new paradigm for highly efficient generation of entangled light under ambient conditions.

\begin{figure}
\includegraphics[width=1\linewidth]{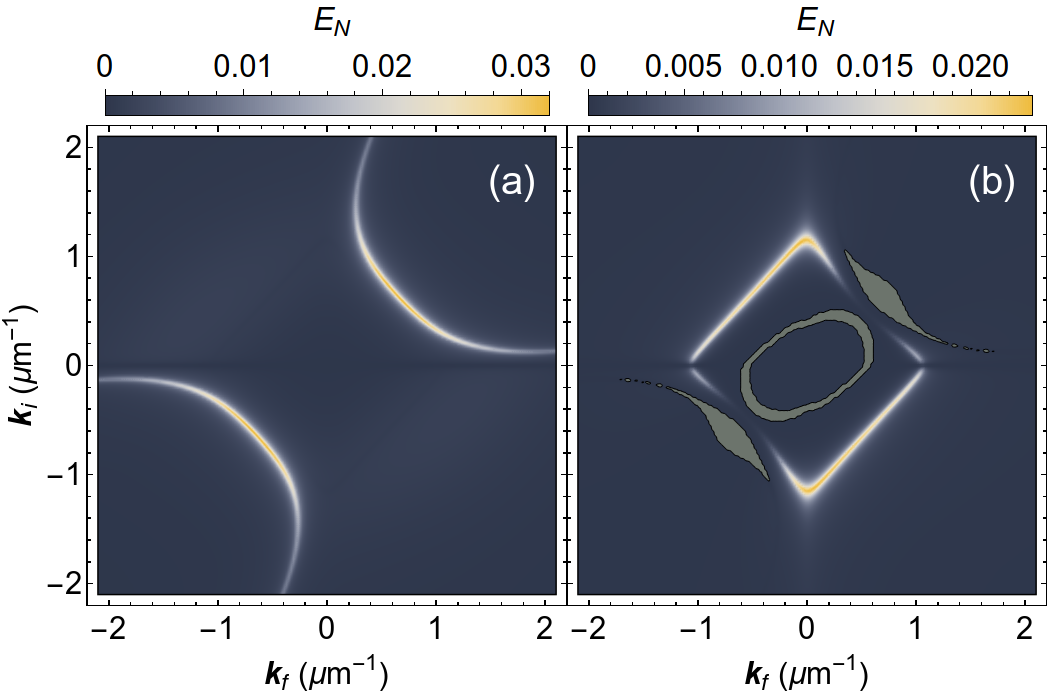}
\caption{
{\bf Entangled polaritons.}
Logarithmic negativity, $E_N$, for lower exciton-polaritons with the wave vector ${\bf k}_f$ and (a) upper, (b) lower phonon-polaritons with the wave vector ${\bf k}_i-{\bf k}_f$ under CW excitation, $n_{s_{\rm U}|{\bf k}_i} = 1630$.  
Green regions indicate $E_N=0$. 
Parameters are specified in Appendix~\ref{appendix: Hamiltonian}.
}
    \label{fig: LogNegMat}
\end{figure}

\begin{figure*}
\includegraphics[width=0.75\linewidth]{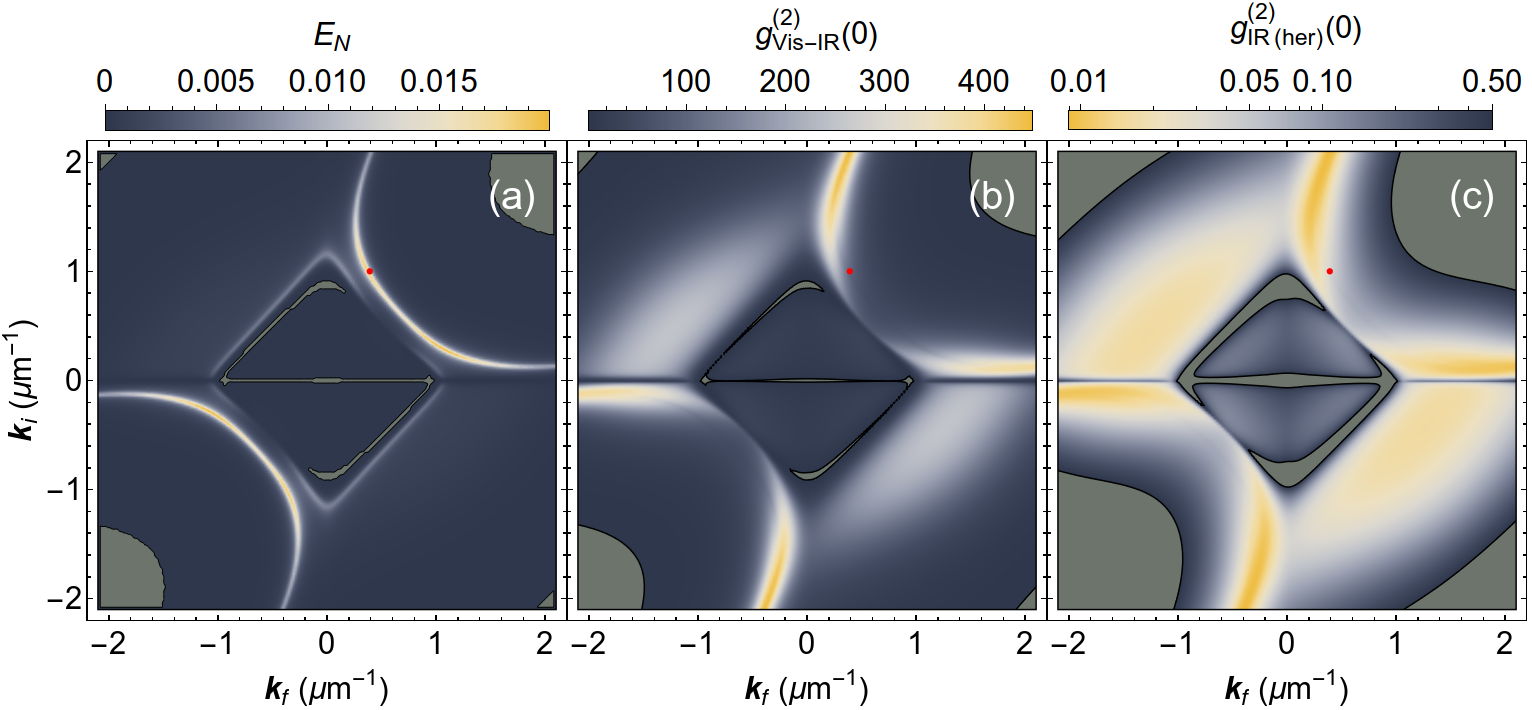}
\caption{
{\bf Entangled light emitted by the double resonant system.}
(a) Logarithmic negativity, $E_N$, for visible light with in-plane wave vector ${\bf k}_f$ and IR light in-plane wave vector ${\bf k}_i-{\bf k}_f$.
Green regions indicate areas where $E_N=0$.
(b) $g^{(2)}_{\rm Vis-IR}(0)$ for visible and mid-IR light.
Green regions indicate areas $g^{(2)}_{\rm Vis-IR}(0)\leq 2$. 
(c) $g^{(2)}_\text{IR(her)}(0)$ for visible and mid-IR light.
Green regions indicate areas $g^{(2)}_\text{IR(her)}(0) \geq 0.5$. 
The parameters are the same as in Fig.~\ref{fig: LogNegMat}, with ${\rm SNR_{Vis}} = \langle \hat n_{s_{\rm L}|{\bf k}_f} \rangle/N^{(\rm bg)}_{\rm Vis}$, $N^{(\rm bg)}_{\rm Vis}=10^{-6}$ and ${\rm SNR_{IR}} = \langle \hat n_{{\rm IR}|{\bf k}_i-{\bf k}_f} \rangle/N^{(\rm bg)}_{\rm IR}$, $N^{(\rm bg)}_{\rm IR} = 10^{-3}$.
%The dissipation rates of the vibrations, $\gamma_{\rm Vib}$, and the dissipation rate of mid-IR cavity, $\gamma_{\rm IR}$, are shown in Fig.~\ref{fig:Fig1}(e).
}
    \label{fig:correlations_CW}
\end{figure*}

{\it Double resonant polariton system.}
The proposed system is a cavity coupled to matter, hosting excitons and phonons.
The hosted phonons are Raman and IR active. For the exciton-polariton states considered throughout this work, we use parameters from existing polariton platforms based on ladder-type conjugated polymer (methyl-substituted ladder-type poly(p-phenylene), MeLPPP)~\cite{scherf1991polyarylenes} that have successfully demonstrated strong sideband-resolved optomechanical interactions at room temperature~\cite{zasedatelev2019room, zasedatelev2021single}. 
Similar conditions can be found in various material platforms: in molecular crystals, such as anthracene/pentacene, strong coupling with $\Omega_R\!\sim\!100$–$500$\,meV is routine, while intramolecular modes at $40$–$200$\,meV ($300$–$1600\,\mathrm{cm^{-1}}$) exhibit $S\!\sim\!0.5$–$2$ \cite{lidzey1998strong,kenacohen2010room,kenacohen2008prl,clark2006huang,wu2016polarons}; TMD monolayers (e.g. MoSe$_2$) show $\Omega_R\!\sim\!20$–$100$\,meV and room-temperature exciton–phonon coupling around $S\!\approx\!1$ \cite{dufferwiel2015vdw,kim2023mose2}; wide-gap polar inorganics (GaN, ZnO) provide Fr\"ohlich-coupled LO phonons at $\sim\!60$–$100$\,meV with effective $S\!\sim\!0.1$–$0.5$ \cite{verdi2015frohlich,christopoulos2007gann}; and halide perovskites combine strong-coupling with rich IR-active/Raman modes and recently observed ultrastrong multimode phonon-polaritons \cite{soci2023nanoletters, wong2020acsmaterialslett,kim2025multimode, bujalance2024strong}.
Appendix~\ref{appendix: coupled mode theory} provides an extensive discussion of material-specific considerations. The double resonant system and microscopic mechanisms behind the polariton entanglement are not limited to a specific material platform.

Figure~\ref{fig:Fig1} shows the hybrid design of the cavity supporting lattice resonance (LR) modes with conical dispersion in visible range~\cite{sanvitto2022, berghuis2023room, maggiolini2023, hakala2018, vakevainen2020, trypogeorgos2025, lindel2025close} and Fabry--Perot modes with quadratic dispersion in IR range~\cite{simpkins2015spanning, long2015coherent, shalabney2015coherent}~(Appendix~\ref{appendix: coupled mode theory}).
These modes provide strong light-matter interaction with both excitons and phonons, resulting in the formation of exciton-polaritons and phonon-polaritons, characterized by in-plane momenta $\hbar {\bf k}_{\parallel}$ (hereinafter $\hbar{\bf k}$), as well as a reservoir of dark states~(Fig.~\ref{fig: dispersions}(a))~(Appendix~\ref{appendix: Hamiltonian}).
The parametric coupling between Raman-active phonons and excitons~\cite{wang2016influence, wang2018temperature, wang2019multiphonon, wang2020multiphonon, kirton2013nonequilibrium, kirton2015thermalization, shishkov2024optomech} leads to the interaction between exciton-polaritons and phonon-polaritons.
The resultant Hamiltonian is $\hat H = \hat H_{\rm OM} + \hat H_{\rm Dark}$, with $\hat H_{\rm Dark}$ governing the quantum dynamics of the dark states.
The optomechanical Hamiltonian is
\begin{equation} \label{H_strong}
\hat H_{\rm OM}
=
\hat H_\text{exc-pol}
+
\hat H_\text{phon-pol}
+
\hat H_{\rm int},
\end{equation}
where $\hat H_\text{exc-pol}$ is the Hamiltonian of the upper and lower exciton-polaritons, including the coherent drive of the upper exciton-polariton state with the wave vector ${\bf k}_i$, $\hat H_\text{phon-pol}$ is the Hamiltonian of the upper and lower phonon-polaritons, and $\hat H_{\rm int}$ is the Hamiltonian of the optomechanical interaction between exciton-polaritons and phonon-polaritons in blue detuned configuration~(Fig.~\ref{fig:Fig1}).
Below the threshold of optomechanical instability, the dynamics of directly driven upper exciton-polaritons becomes almost independent of the evolution of the lower exciton-polaritons and phonon-polaritons, allowing us to exclude the upper exciton-polaritons adiabatically and obtain the linearized optomechanical Hamiltonian, $\hat H_{\rm OM} - \hbar \omega_{s_{\rm U}|{\bf k}_i} \hat n_{s_{\rm L}|{\bf k}_f} \approx \sum_{{\bf k}_f} \hat H_{{\rm OM}|{\bf k}_i,{\bf k}_f}$~(Appendix~\ref{appendix: Hamiltonian})
\begin{multline} \label{H_optomech_f}
\hat H_{{\rm OM}|{\bf k}_i,{\bf k}_f}
=
\underbrace{
\hbar 
\left(
\omega_{s_{\rm L}|{\bf k}_f}
-
\omega_{s_{\rm U}|{\bf k}_i}
\right)
\hat n_{s_{\rm L}|{\bf k}_f}
}
_
\text{lower exciton-polaritons}
\\
+
\underbrace{
\hbar \omega_{v_{\rm U}|{\bf k}_i-{\bf k}_f}
\hat n_{v_{\rm U}|{\bf k}_i-{\bf k}_f} 
+
\hbar \omega_{v_{\rm L}|{\bf k}_i-{\bf k}_f}
\hat n_{v_{\rm L}|{\bf k}_i-{\bf k}_f} 
}
_
\text{upper \& lower phonon-polaritons}
\\
+
\underbrace{
\hbar 
\left(
G_{{\rm U}|{\bf k}_f{\bf k}_i}
\hat v_{{\rm U}|{\bf k}_i-{\bf k}_f}^\dag 
+
G_{{\rm L}|{\bf k}_f{\bf k}_i}
\hat v_{{\rm L}|{\bf k}_i-{\bf k}_f}^\dag 
\right)
\hat s_{{\rm L}|{\bf k}_f}^\dag
+
h.c.
}
_
\text{optomechanical interaction}
\, ,
\end{multline}  
where $\omega_{s_{\rm U}|{\bf k}}$, $\omega_{s_{\rm L}|{\bf k}}$, $\omega_{v_{\rm U}|{\bf q}}$, and $\omega_{v_{\rm L}|{\bf q}}$ are the frequencies of the upper and lower exciton-polariton with wave vector $\bf k$, upper and lower phonon-polariton with wave vector $\bf q$ respectively, $\hat s_{{\rm L}|{\bf k}}$, $\hat v_{{\rm U}|{\bf q}}$, and $\hat v_{{\rm L}|{\bf q}}$ are the corresponding annihilation bosonic operators and $\hat n_{s_{\rm L}|{\bf k}} = \hat s^\dag_{{\rm L}|{\bf k}} \hat s_{{\rm L}|{\bf k}}$, $\hat n_{v_{\rm U}|{\bf q}} = \hat v^\dag_{{\rm U}|{\bf q}} \hat v_{{\rm U}|{\bf q}}$, and $\hat n_{v_{\rm L}|{\bf q}} = \hat v^\dag_{{\rm L}|{\bf q}} \hat v_{{\rm L}|{\bf q}}$ are the number operators, $ G_{{\rm U}|{\bf k}_f{\bf k}_i}(t)= g_{{\rm U}|{\bf k}_f{\bf k}_i}\sqrt{n_{s_{\rm U}|{\bf k}_i}(t)}$ and $ G_{{\rm L}|{\bf k}_f{\bf k}_i}(t)= g_{{\rm L}|{\bf k}_f{\bf k}_i}\sqrt{n_{s_{\rm U}|{\bf k}_i}(t)}$ are collective optomechanical coupling constants for upper and lower phonon-polaritons respectively, with $n_{{\rm U}|{\bf k}_i}(t)$ being the number of upper exciton-polaritons directly excited by the coherent drive, $g_{{\rm U}|{\bf k}_f{\bf k}_i}$ and $g_{{\rm L}|{\bf k}_f{\bf k}_i}$ are single-polariton optomechanical coupling constants for upper and lower phonon-polaritons respectively.

Unlike conventional optomechanics, where photons couple to mechanical motion via radiation pressure, dispersive, or dissipative interactions~\cite{aspelmeyer2014cavity, primo2020quasinormal, roelli2024nanocavities}, our system features hybrid light-matter excitations in both modes. 
The optical mode is an exciton-polariton, a coherent mix of cavity photons and electronic excitations, while the mechanical mode consists of delocalized collective vibrations {\it phonons}. 
Polariton systems under the optomechanical interaction have demonstrated phonon lasing~\cite{chafatinos2020polariton, papuccio2025polariton, kuznetsov2023microcavity,santos2023polaromechanics}, time-crystals~\cite{carraro2024solid}, and have recently been proposed for sympathetic Bose--Einstein condensation of phonons at room temperature~\cite{shishkov2024sympathetic}, coherent vibrational control over polaritons~\cite{shishkov2024optomech}, and entanglement generation between phonons and exciton-polaritons~\cite{huang2025room}.
Besides the aforementioned optomechanical physics of exciton-polaritons and phonons, our system expresses strong interaction between phonons and the IR cavity, leading to the formation of phonon-polaritons~\cite{del2016exploiting}.
Thus, we enter a new regime in optomechanics where both optical and mechanical degrees of freedom are hybrid light-matter states, allowing the manipulation with IR light and direct optical readout of mechanical states. 

%\newpage

{\it Polariton Bell pairs.}
The interaction term in Hamiltonian~(\ref{H_optomech_f}) is a two-mode squeezing Hamiltonian, leading to the generation of \textit{time-energy entangled} exciton- and phonon-polaritons pairs: 
\begin{equation}
| \Psi_{{\bf k}_f} \rangle =
%\psi_{{\rm U}|{\bf k}_f{\bf k}_i}
\psi_{\rm U}
|1_{s_{\rm L}|{\bf k}_f}1_{v_{\rm U}|{\bf k}_i-{\bf k}_f}\rangle
+
%\psi_{{\rm L}|{\bf k}_f{\bf k}_i}
\psi_{\rm L}
|1_{s_{\rm L}|{\bf k}_f}1_{v_{\rm L}|{\bf k}_i-{\bf k}_f}\rangle,
\end{equation}
where $\psi_{\rm U} \propto G_{{\rm U}|{\bf k}_f{\bf k}_i}$ and $\psi_{\rm L} \propto G_{{\rm L}|{\bf k}_f{\bf k}_i}$ are the quantum mechanical amplitudes, $|1_{s_{\rm L}|{\bf k}_f}1_{v_{\rm U}|{\bf k}_i-{\bf k}_f}\rangle$ and $|1_{s_{\rm L}|{\bf k}_f}1_{v_{\rm L}|{\bf k}_i-{\bf k}_f}\rangle$ are the paired states of lower exciton-polaritons and upper phonon-polaritons or lower phonon-polaritons, respectively ~(Fig.~\ref{fig: dispersions}(b)).
Given the scattering process requires phase-matching conditions, the optomechanical interaction for the polariton states with different momenta ${\bf k}_f$ naturally results in Bell pair generation \textit{entangled over energy-momentum} degrees of freedom $
|1_{s_{\rm L}|{\bf k}_{f1}}1_{v_{\rm U}|{\bf k}_i-{\bf k}_{f1}}\rangle
+
|1_{s_{\rm L}|{\bf k}_{f2}}1_{v_{\rm L}|{\bf k}_i-{\bf k}_{f2}}\rangle$~(Fig.~\ref{fig: dispersions}(c)).

{\it Entanglement in continuous variables.}
The full quantum simulations, provided in Appendix~\ref{appendix: correlations} and Appendix~\ref{appendix: logneg}, show the presence of continuous-variable entanglement between the quadratures of exciton-polaritons with wave vector ${\bf k}_f $ and phonon-polaritons with wave vector ${\bf k}_i - {\bf k}_f$. This is evidenced by the logarithmic negativity, $E_N$,~\cite{barzanjeh2019stationary, maziero2016computing, vidal2002computable, weedbrook2012gaussian, plenio2005logarithmic} which emerges for a wide parameter range in the $\{ {\bf k}_i, {\bf k}_f \}$ plane (Fig.~\ref{fig: LogNegMat}).
This entanglement between polaritons is a unique feature of the proposed double strongly coupled system.

{\it Entangled photons.}
The light emitted by polaritons inherits this entanglement, providing a way to measure the entanglement of polaritons.
The statistical properties of both lower exciton-polaritons with wave vector ${\bf k}$ and their emission are fully characterized by the operator $\hat s_{{\rm L}|{\bf k}}$, the light coming from upper and lower phonon-polariton states with in-plane wave vector ${\bf q}$ is fully characterized by the phonon-polariton operators $\hat v_{{\rm U}|{\bf q}}$ and $\hat v_{{\rm L}|{\bf q}}$.
We also explore the statistical properties of IR light emitted from {\it both} upper and lower phonon-polaritons.
In this case, it is the operator of the IR cavity, $\hat a_{{\rm IR}|{\bf q}} $, that fully characterizes the statistical properties of emitted IR light
\begin{equation} \label{aIR}
\hat a_{{\rm IR}|{\bf q}} 
=
\hat v_{{\rm L}|{\bf q}} \cos \phi_{{\bf q}} 
+
\hat v_{{\rm U}|{\bf q}} \sin \phi_{{\bf q}}, 
\end{equation}
where $\phi_{\bf q}$ determines the contribution of upper and lower phonon-polaritons to IR emission~(Appendix~\ref{appendix: Hamiltonian}).

\begin{figure*}
\includegraphics[width=0.75\linewidth]{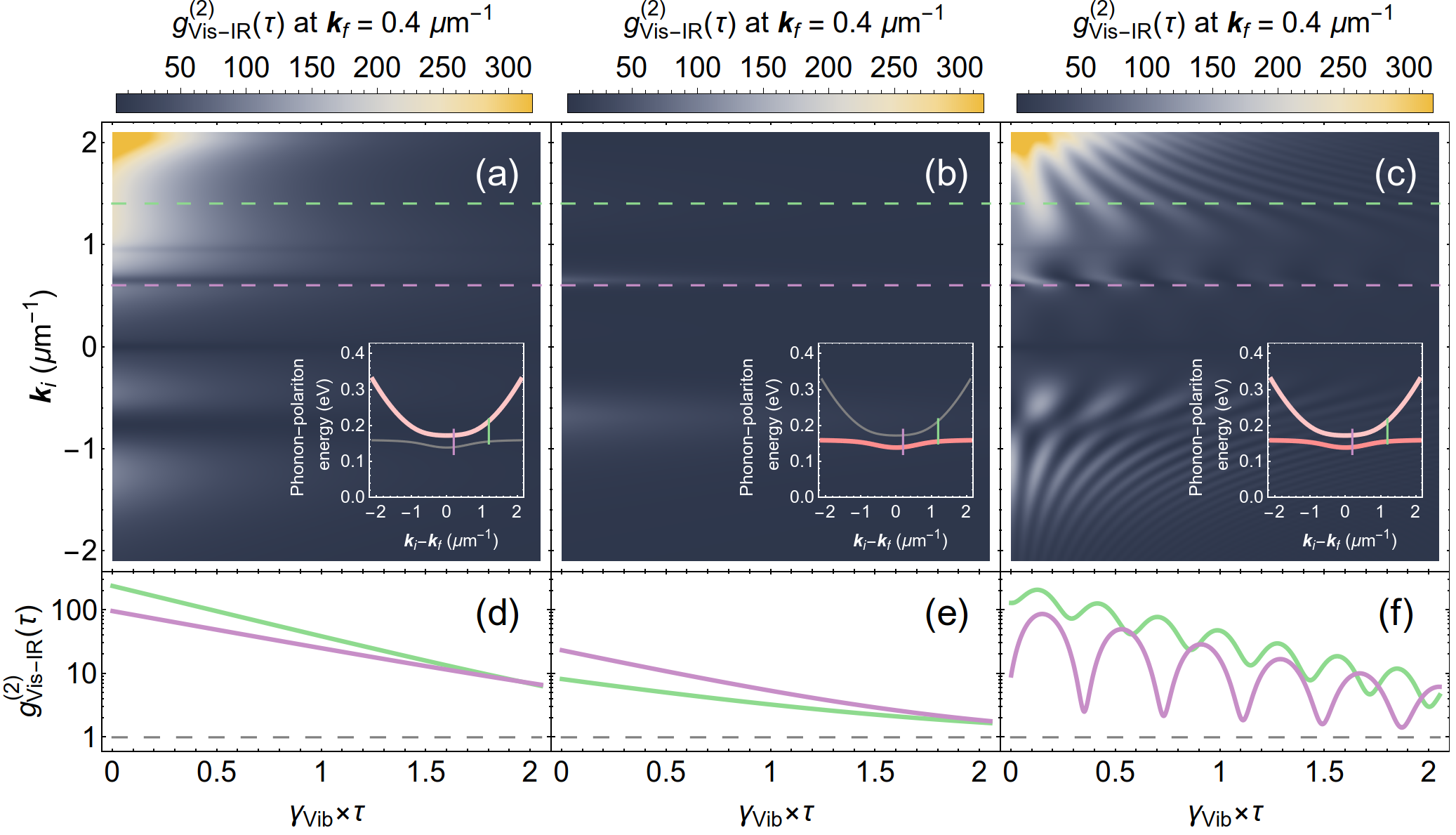}
\caption{
{\bf Quantum interference}.
$g^{(2)}_{\rm Vis-IR}(\tau)$as a function of time delay and wave vector ${\bf k}_i$ with IR light coming from (a) upper, (b) lower, and (c) both phonon-polariton states.
Insets in (a-c): The dispersion of phonon-polaritons, with the vertical purple and green lines mark the phonon-polariton states assisting the transition of energy, corresponding to the green and purple dashed lines in (a-c).
(d-f) $g^{(2)}_{\rm Vis-IR}(\tau)$ for fixed ${\bf k}_i$ corresponding to the green and purple dashed lines in (a-c).
The parameters are the same as in Fig.~\ref{fig:correlations_CW}.
%The coherent drive excites lower exciton-polaritons with in-plane wave vector ${\bf k}_{i}$ which undergo inelastic scattering to the lower exciton-polariton state with in-plane wave vector ${\bf k}_{f}$ via optomechanical interaction phonon-polariton states.
}
    \label{fig:correlations_CW_t}
\end{figure*}

The entanglement of the emitted light from the double cavity is sufficient evidence of the polaritons' entanglement, even when the light is the subject to uncorrelated background noise~(Appendix~\ref{appendix: logneg}).
In practice, background photons originating, for example, from additional light generated via the reservoir of dark states \cite{herrera2017dark, shishkov2024sympathetic, shishkov2024room} mix with entangled photons emitted by the polariton states and could potentially hide their quantum signatures.
To quantify this, we define the signal-to-noise ratios $\rm SNR_{Vis}$ and $\rm SNR_{IR}$ in the visible and IR ranges \cite{panyukov2022heralded} and explicitly incorporate them into all photon-correlation functions and quadrature covariance matrices, adding background-noise terms where appropriate (Appendix \ref{appendix: correlations}). Here, we emphasize that the presence of additional uncorrelated noise means that observing entanglement in the light leaking from the double cavity directly verifies entanglement between the polariton states and, consequently, their constituent electronic and vibrational matter components.

Figure~\ref{fig:correlations_CW}(a) shows the background adjusted logarithmic negativity, $E_N$, for the emitted visible light with wave vector ${\bf k}_f$ and IR light with the wave vector ${\bf k}_i-{\bf k}_f$. 
$E_N$ is positive in a wide range in $\{ {\bf k}_i, {\bf k}_f \}$ plane, evidencing the entanglement of the emitted light.
Moreover, the pattern of $E_N$ in $\{ {\bf k}_i, {\bf k}_f \}$ plane for the emitted light (Fig.~\ref{fig:correlations_CW}(a)) combines the patterns of $E_N$ for polaritons shown in Fig.~\ref{fig: LogNegMat}.

{\it Entanglement in discrete variables.}
While the non-zero logarithmic negativity is a continuous variable form of a sufficient and necessary condition for two-partite entanglement, it has been shown that the positivity of $E_N$ also implies the presence of discrete-variable entanglement, i.e. $E_N>0$ is equivalent to a violation of Cauchy--Schwartz inequality, $g^{(2)}_\text{Vis}(0)g^{(2)}_\text{IR}(0) \ge \left[ g^{(2)}_\text{Vis-IR}(0) \right]^2$~\cite{wasak2014cauchy, kheruntsyan2012violation}.
We denote the second-order autocorrelation functions of the visible and IR light as $g^{(2)}_\text{Vis}(0)$ and $g^{(2)}_\text{IR}(0)$, and the second-order cross-correlation function as~(Appendix~\ref{appendix: g2})
\begin{equation} \label{Vis-IR correlations}
g^{(2)}_\text{Vis-IR}(\tau) = 1 +
\frac
{
|\left\langle 
\hat a_{{\rm IR}|{\bf k}_{i}-{\bf k}_{f}}(t+\tau)
\hat s_{{\rm L}|{\bf k}_{f}}(t)
\right\rangle|^2
}
{
\langle 
\hat n_{s_{\rm L}|{\bf k}_f}(t)
\rangle 
\langle  
\hat n_{{\rm IR}|{\bf k}_{i}-{\bf k}_{f}}(t+\tau)
\rangle
},
\end{equation}
where $\hat n_{{\rm IR}|{\bf q}}=\hat a_{{\rm IR}|{\bf q}}^\dag \hat a_{{\rm IR}|{\bf q}}$ and $\tau>0$, meaning that the visible light is detected first (heralded measurement). 
Violating the Cauchy–Schwartz inequality means that the particle number between the IR and visible modes is entangled. The Bell state example discussed above is a particular example of this type of entanglement.
Since we consider the operation of the system below the threshold of optomechanical instability, we have $g^{(2)}_\text{Vis}(0) = g^{(2)}_\text{IR}(0) = 2$ (Appendix~\ref{appendix: g2}).
Therefore, the criteria for photon-number entanglement between emitted visible and IR light becomes $g^{(2)}_\text{Vis-IR}(0) > 2$.

Figure~\ref{fig:correlations_CW}(b) shows $g^{(2)}_\text{Vis-IR}(0)$, fully capturing the genuine quantum nature of our double-resonant polariton system with multi-mode optomechanical interaction.
The region in $\{{\bf k}_{i},{\bf k}_{f}\}$ plane, where $g^{(2)}_\text{Vis-IR}(0)>2$ is wide and coincides to the region where visible and IR photons are entangled~(Fig.~\ref{fig:correlations_CW}(a,b)).
Moreover, $g^{(2)}_\text{Vis-IR}(0)$ approaches its fundamental limit at the finite temperature $1/n_{v_{\rm U}|{\bf k}_i-{\bf k}_f}^{\rm th}\approx 500$, $1/n_{v_{\rm L}|{\bf k}_i-{\bf k}_f}^{\rm th}\approx 500$~\cite{shishkov2021enhancement1, shishkov2021enhancement2}.
These high values of the second-order cross-correlation function can be used as a resource to construct a heralded single-photon mid-IR light source.
Indeed, heralding the IR light with the wave vector ${\bf k}_i-{\bf k}_f$ by detecting the visible light with the wave vector ${\bf k}_f$ leads to the second-order autocorrelation function of heralded IR light~(Appendix~\ref{appendix: g2})
\begin{equation}
g^{(2)}_\text{IR(her)}(0) = 
\frac{4}{g^{(2)}_\text{Vis-IR}(0)}
-
\frac{2}{[g^{(2)}_\text{Vis-IR}(0)]^2}.
\end{equation}
Figure~\ref{fig:correlations_CW}(c) shows that for some wave vectors ${\bf k}_i$ and ${\bf k}_f$ $g^{(2)}_\text{IR(her)}(0)$ can become as low as $10^{-2}$.

The involvement of both upper and lower phonon-polaritons in the inelastic light scattering leads to the oscillations of $g^{(2)}_\text{Vis-IR}(\tau)$ with the frequency difference between upper and lower phonon-polariton states, $\omega_{v_{\rm U}|{\bf k}_i-{\bf k}_f} - \omega_{v_{\rm L}|{\bf k}_i-{\bf k}_f}$,~(Fig.~\ref{fig:correlations_CW_t}(c)). 
%This phenomenon is a distinct result of quantum interference between two scattering pathways: one via the upper phonon-polariton state and the other via the lower phonon-polariton state.
These quantum beats in $g^{(2)}_\text{Vis-IR}(\tau)$ are the quantum version of Young's interference experiment, where upper and lower phonon-polariton states play the role of two slits, providing different pathways for the inelastic light scattering.
If we block the IR light from the lower or upper photon-polariton state, the quantum beats in $g^{(2)}_\text{Vis-IR}(\tau)$ completely disappear~(Fig.~\ref{fig:correlations_CW_t}(a,b)).
Recent experiments~\cite{vento2023measurement} showed similar quantum beats of the second-order cross-correlation function between Stokes--anti-Stokes light at room temperature in a liquid.

\begin{figure}
\includegraphics[width=0.57\linewidth]{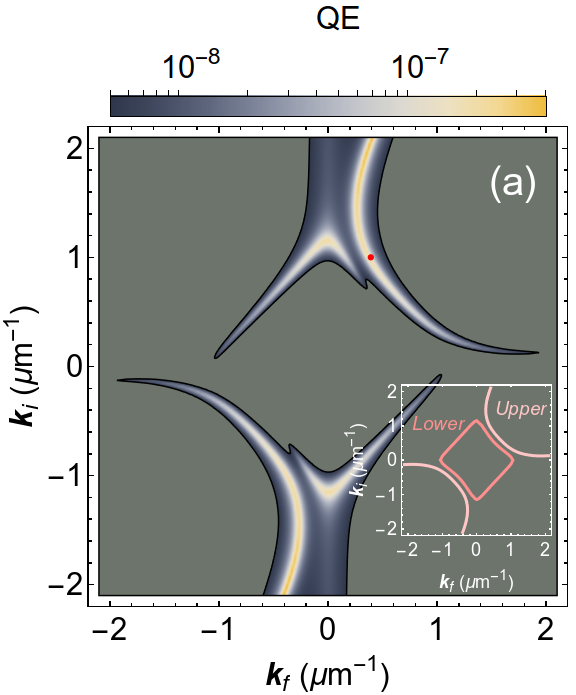}
\caption{
{\bf Quantum efficiency of entangled polariton state generation.}
Quantum efficiency of transition from the upper exciton-polariton state ${\bf k}_i$ to the lower exciton-polariton state ${\bf k}_f$.
Green color marks the regions ${\rm QE} \leq 5\cdot10^{-9}$.
Inset: Energy-momentum matching condition for this transition in $\{ {\bf k}_i, {\bf k}_f \}$ plane.
The parameters are the same as in Fig.~\ref{fig: LogNegMat}.
}
    \label{fig: QE_CW}
\end{figure}

\begin{figure}
\includegraphics[width=0.9\linewidth]{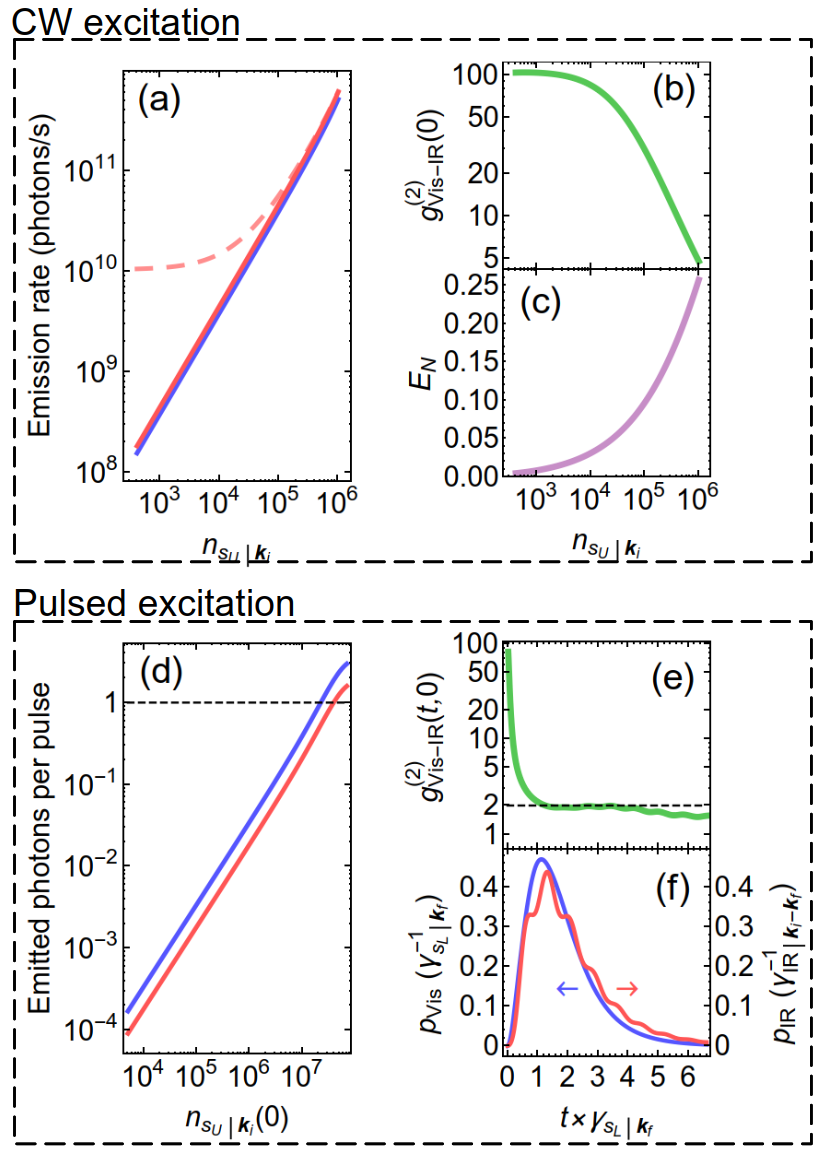}
\caption{
{\bf Entangled photon pairs rate.}
Pump power dependence under CW excitation of (a) emission rates of visible light (blue line), IR light (red dashed line), excess IR light (red line), (b) $g^{(2)}_\text{Vis-IR}(0)$, (c) $E_N$.
(d) Pump power dependence under pulsed excitation of the number of emitted photons per pulse in the visible spectral region (blue line) and mid-IR spectral region (red line). 
Black dashed line marks emission of one photon per pulse.
(e) Time dependence of $g^{(2)}_\text{Vis-IR}(t, 0)$, at pumping $n_{s_{\rm U}|{\bf k}_i}(0) = 4 \cdot 10^7$.
Black dashed line marks $g^{(2)}_\text{Vis-IR}(t, 0) = 2$.
(f) Normalized temporal intensity profiles of emitting photons in visible (blue line) and in IR (red line) spectral ranges. The normalization is $\int_0^{+\infty}p_\alpha(t)dt = 1$.
The parameters are the same as in Fig.~\ref{fig:correlations_CW} with ${\bf k}_i=1~{\rm \mu m}^{-1}$ and ${\bf k}_f=0.4~{\rm \mu m}^{-1}$, marked by red dots in Fig.~\ref{fig:correlations_CW}.
}
    \label{fig:Count rate and g2}
\end{figure}

{\it Quantum efficiency of entangled polaritons generation}.
The brightness of the proposed source of entangled states requires a high transition rate from the excited upper exciton-polariton state to a lower exciton-polariton state (Fig.~\ref{fig: dispersions}(b)).
To quantify this transition, we introduce the quantum efficiency as the steady-state ratio between the flows of polaritons from a lower exciton-polariton state and an upper exciton-polariton state into the dissipative reservoir
%, ${\rm QE}$, as the ratio between the loss flow of upper exciton-polaritons from the state ${\bf k}_i$ and the energy flow of lower exciton-polaritons from the state ${\bf k}_f$
\begin{equation} \label{QE}
{\rm QE} = 
\frac
{
\gamma_{s_{\rm L}|{\bf k}_f} n_{s_{\rm L}|{\bf k}_f}
}{
\gamma_{s_{\rm U}|{\bf k}_i} n_{s_{\rm U}|{\bf k}_i}
},
\end{equation}
where $\gamma_{s_{\rm L}|{\bf k}_f}$ and $\gamma_{s_{\rm U}|{\bf k}_i}$ are the dissipation rates of lower and upper exciton-polariton states and $n_{s_{\rm L}|{\bf k}_f} = \langle \hat n_{s_{\rm L}|{\bf k}_f} \rangle$ in steady-state.
%Both lower and upper phonon-polaritons exhibit significant matter component, characterized by Hopfield coefficients~(Appendix~\ref{appendix: Hamiltonian}), leading to comparable contributions to the quantum efficiency. 
The region of maximum quantum efficiency~(Fig.~\ref{fig: QE_CW}) is determined by the energy-momentum conservation between excited upper exciton-polaritons and resulting lower exciton- and phonon-polaritons~(inset in Fig.~\ref{fig: QE_CW}).
As a result, the quantum efficiency peaks (Fig.~\ref{fig: QE_CW}) in the same regions of $\{{\bf k}_{i},{\bf k}_{f}\}$ plane where the polariton states are maximally entangled~(Fig.~\ref{fig: LogNegMat}), the emitted light is maximally entangled~(Fig.~\ref{fig:correlations_CW}(a)), and $g^{(2)}_\text{Vis-IR}(0)$ is high~(Fig.~\ref{fig:correlations_CW}(b)).

\begin{figure*}
\includegraphics[width=0.8\linewidth]{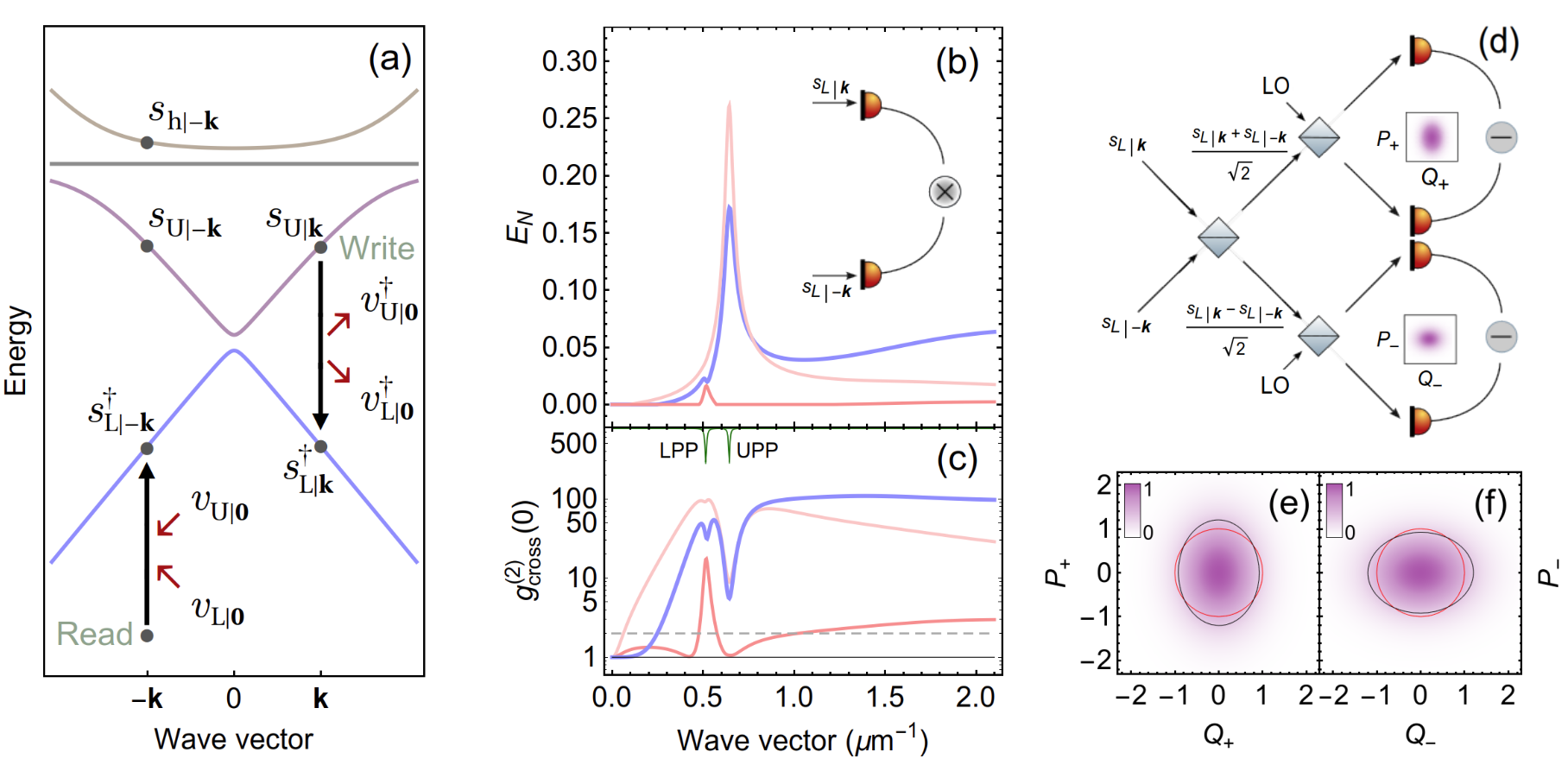}
\caption{
{\bf Entanglement witness of polaritons in the visible spectral range.}
(a)~Generation of the entanglement between exciton-polaritons with the wave vectors ${\bf k}$ and $-{\bf k}$.
Brown line is higher exciton-polaritons, other lines are the same as in Fig.~\ref{fig: dispersions}.
(b-c)~The logarithmic negativity (b) and second-order cross-correlation function (c) between light emitted by exciton-polarions with wave vectors $\bf k$ and $-{\bf k}$, given by Eq.~(\ref{Vis-Vis correlations}), (blue line), between light emitted by exciton-polarions with wave vectors $\bf k$ and upper (pink line) or lower (red line) phonon-polaritons with the wave vector ${\bf q}={\bf 0}$.
Inset in (b): schemes for measurement of the exciton-polaritons entanglement in discrete and in continuous variables.
Green line in (c) shows the resonant conditions for the lower and upper phonon-polaritons.
(e-f) Wigner function of symmetric (e) or antisymmetric (f) exciton-polariton modes that can be reconstructed with the measurement scheme (d).
Black lines in (e) and (f) shows the quadratures, at which the Wigner function drops by $1/e$ from its maximum, red lines are the same for an ideal vacuum state.
The parameters are the same as in Fig.~\ref{fig: LogNegMat}, with uncorrelated background light in the visible and mid-IR spectral ranges $N^{(\rm bg)}_{\rm Vis}=10^{-3}$, $N^{(\rm bg)}_{\rm IR} = 10^{-3}$, and the number of polaritons directly excited by the coherent drive $n_{s_{\rm U}|{\bf k}} = 10^6$, $n_{s_{\rm L}|-{\bf k}}^{\rm drive} = 10^5$.
}
    \label{fig:WitnessJoined}
\end{figure*}

Meeting the energy-momentum matching condition for ${\bf k}_i$ and ${\bf k}_f$, even for a moderate incident photon rate of coherent drive $\gamma_{s_{\rm U}|{\bf k}_i} n_{s_{\rm U}|{\bf k}_i} \approx 10^{17}~{\rm photons/s} \approx 10~{\rm mW}$, leads to the emission rate of paired photons around $10^{10}~{\rm photons/s} \approx 1~{\rm nW}$~(Fig.~\ref{fig:Count rate and g2}(a)). 
This corresponds to a visible–mid-IR pair production rate of $\sim 1~{\rm GHz/mW}$, at least two orders of magnitude higher than that of bulk crystals currently used for quantum sensing and imaging with undetected photons~\cite{kviatkovsky2020microscopy, lemos2014quantum, vanselow2020frequency}. In addition, our system enables the generation of spectrally distant photon pairs, with one photon in the so-called {\it fingerprint} spectral region~\cite{paterova2022broadband, kalashnikov2016infrared}.
%currently available cutting-edge devices for photon pair generation in mid-IR~\cite{sua2017direct, mancinelli2017mid, kumar2021mid, arahata2021wavelength, hildenstein2025efficient}.
%Moreover, our system allows entering the undetected wavelength into the so-called fingerprint region

The high intensities of the coherent drive, though, may change the population of vibrational states, promoting stimulated processes and decreasing the second-order cross-correlation function between visible and mid-IR light~(Fig.~\ref{fig:Count rate and g2}(b)).
Nevertheless, the logarithmic negativity grows with the intensity of the coherent drive~(Fig.~\ref{fig:Count rate and g2}(c)).

{\it Highly-efficient entangled photon pairs generation}.
To explore the possibility of the nearly deterministic visible and IR paired photon generation in our system, we consider a short light pulse, exciting the upper exciton-polaritons, such that $n_{s_{\rm U}|{\bf k}_i}(t) = n_{s_{\rm U}|{\bf k}_i}(0)e^{-\gamma_{s_{\rm U}|{\bf k}_i}t}$ in Hamiltonian~(\ref{H_optomech_f}).
We numerically solve the master equation for the Hamiltonian~(\ref{H_optomech_f})~(Appendix~\ref{appendix: correlations}).
Figure~\ref{fig:Count rate and g2}(d) demonstrates bright visible--IR photon generation with at least one photon pair emitted per excitation pulse.
The generated paired photons are nonclassically correlated~(Fig.~\ref{fig:Count rate and g2}(e)), while their temporal intensity profiles almost match~(Fig.~\ref{fig:Count rate and g2}(f)).

{\it Alternative entanglement witnesses.}
Probing of the quantum entanglement between exciton-polaritons and phonon-polaritons via their emitted light requires the ability to detect single photons in the mid-IR spectral range, which is limited at the current level of technology~\cite{dello2022advances, xie2024temporal, taylor2023low, chang2022efficient, nowakowski2025single, liu2024highly}.
An alternative method of probing these correlations is to use an additional Raman probe and analyze Stokes--anti-Stokes correlations~\cite{riedinger2016non, tarrago2020bell}.
Nonclassical Stokes--anti-Stokes correlations have been observed in different physical systems, including organic molecules~\cite{saraiva2017photonic, de2024properties}, showing their robustness against room-temperature fluctuations and disorder~\cite{kasperczyk2016temporal, saraiva2017photonic, vento2023measurement}.

We propose the protocol for the transfer of entanglement between exciton-polaritons and phonon-polaritons to the entanglement between exciton-polaritons with different wave vectors.
In this protocol, we apply a {\it write} coherent drive beam with the frequency $\omega_{\rm write} = \omega_{s_{\rm U}|{\bf k}}$ to excite lower exciton-polaritons with the wave vector $\bf k$ and phonon-polaritons with the wave vector ${\bf q} = {\bf 0}$~(Fig.~\ref{fig:WitnessJoined}(a))  and create the entanglement between these two states~(Fig.~\ref{fig:Fig1}).
Then, we apply a {\it read} coherent drive with the frequency $\omega_{\rm read} = 2\omega_{s_{\rm L}|{\bf -k}} - \omega_{s_{\rm U}|{\bf -k}}$ and the wave vector $-{\bf k}$~(Fig.~\ref{fig:WitnessJoined}(a)) to map this entanglement onto the entanglement between lower exciton-polariton states with wave vectors ${\bf k}$ and $-{\bf k}$~(Appendix~\ref{appendix: witness}).
The detection schemes plotted in the inset of Fig.~\ref{fig:WitnessJoined}(b) allows probing this entanglement in continuous variables~(Fig.~\ref{fig:WitnessJoined}(b)) through logarithmic negativity, $E_N$, and discrete variables~(Fig.~\ref{fig:WitnessJoined}(c)) through the second-order cross correlation function,
\begin{equation} \label{Vis-Vis correlations}
g^{(2)}_\text{Vis-Vis}(0) = 1 +
\frac
{
|\left\langle 
\hat s_{{\rm L}|{\bf -k}}(t)
\hat s_{{\rm L}|{\bf k}}(t)
\right\rangle|^2
}
{
\langle 
\hat n_{s_{\rm L}|{\bf -k}}(t)
\rangle 
\langle 
\hat n_{s_{\rm L}|{\bf k}}(t)
\rangle 
}.
\end{equation}
This protocol quantitatively preserves the entanglement, meaning that both $g^{(2)}_\text{Vis-Vis}(0)$ and $E_N$ follow closely their exciton- and phonon-polaritons counterparts~(Fig.~\ref{fig:WitnessJoined}(b,c)).

Non-zero logarithmic negativity between exciton-polaritons with the wave vectors ${\bf k}$ and $-{\bf k}$~(Fig.~\ref{fig:WitnessJoined}(b)) emerges due to an effective two-mode squeezing process between these exciton-polaritons.
The scheme plotted in Fig.~\ref{fig:WitnessJoined}(d) allows us to transfer the two-mode squeezing between exciton-polaritons into one-mode squeezing of symmetric or antisymmetric exciton-polariton modes with the quadratures $\{ Q_+, P_+ \}$ and $\{ Q_-, P_- \}$ corresponding to the annihilation operators $(\hat s_{{\rm L}|{\bf k}} + \hat s_{{\rm L}|{\bf -k}})/\sqrt2$ and $(\hat s_{{\rm L}|{\bf k}} - \hat s_{{\rm L}|{\bf -k}})/\sqrt2$, respectively.
The Wigner functions plotted in~(Fig.~\ref{fig:WitnessJoined}(e,f)) for the quadratures $\{ Q_+, P_+ \}$ and $\{ Q_-, P_- \}$ clearly show that these one-mode squeezed states go beyond the standard quantum limit (SQL)~\cite{scully1997quantum}.
The variance of $Q_+$~(black curve in Fig.~\ref{fig:WitnessJoined}(e)) is below the SQL~(red curve in Fig.~\ref{fig:WitnessJoined}(e)) and the variance of $P_-$~(black curve in Fig.~\ref{fig:WitnessJoined}(f)) is below the SQL (red in Fig.~\ref{fig:WitnessJoined}(f)). 
Another feature of this entanglement is that the second-order cross-correlation function dips at resonantly assisted transitions~(Fig.~\ref{fig:WitnessJoined}(d)), which agrees with the Stokes--anti-Stokes correlations measured in cavityless spontaneous Raman scattering~\cite{saraiva2017photonic}.

{\it Conclusion.} In summary, we have proposed and analyzed a double-resonant polariton system that enables entanglement generation between exciton-polaritons and phonon-polaritons under coherent optical pumping. The scheme exploits intrinsic optomechanical interactions arising from strong exciton–phonon coupling, enabling robust entanglement generation at room temperature.
Indeed, the system proves to be resilient against thermal fluctuations due to the high phonon energies, and, owing to the strong light-matter interaction, quantum correlations remain robust across a wide range of parameters.
Our theoretical model demonstrates that this system emits spectrally disparate entangled photon pairs, spanning the visible and mid-IR regions, with distinct correlations arising from quantum interference between upper and lower phonon-polariton pathways. We demonstrate entanglement by crossing the separability boundary in logarithmic negativity and through strong violation of the Cauchy–Schwarz inequality in second-order photon correlation function.

Importantly, the double-resonant optomechanical system enables highly-efficient generation of entangled photon pairs spanning the visible and mid-IR spectral regions. This is particularly relevant for quantum sensing, imaging, and spectroscopy in the infrared, where direct detection is often challenging~\cite{defienne2024advances, defienne2019quantum, lemos2014quantum, kviatkovsky2020microscopy, kviatkovsky2022mid, gilaberte2023experimental}. A key limitation for mid-IR quantum technologies is the lack of materials with suitable bandgaps to efficiently produce high-brightness single photons in this range~\cite{lu2019infrared}. Recently, several proposals addressed this problem suggesting quantum emitters coupled to nanocavities as quantum light sources in mid-IR/THz~\cite{groiseau2024single}, including on-demand single-photon photon sources~\cite{iles2025demand,groiseau2025deterministic}.
In this context, heralding visible photons from strongly correlated visible–mid-IR pairs~\cite{goldschmidt2008spectrally, kaneda2016heralded, clark2013heralded, mosley2008heralded, meany2014hybrid} offers a route to create bright and chip-scale single-photon sources in the mid-IR and THz ranges.

\begin{acknowledgments}
We acknowledge financial support by the Spanish Ministerio de Ciencia y Universidades—Agencia Estatal de Investigación through grants PID2021-125894NB-I00, PID2024-161142NB-I00, EUR2023-143478, and CEX2023-001316-M through the María de Maeztu program for Units of Excellence in R\&D, as well as from the European Union’s Horizon Europe research and innovation programme under Grant Agreement No. 101070700 (MIRAQLS).
Sh.V.Yu. thanks the Magnus Ehrnrooth foundation. E.H. and A.V.Z. acknowledge support by Finnish Research Impact Foundation within the Tandem Industry Academia (TIA) Seed Project no.~667. We thank Stanislav~Yu.~Shishkov for assisting the preparation of Fig.~\ref{fig:Fig1}.
\end{acknowledgments}

\clearpage

\appendix
\begin{widetext}

\section{Dispersion engineering for the double resonant polariton system} \label{appendix: coupled mode theory}

The feasibility of double-resonant polariton optomechanics requires (i) large visible-range excitonic dipole moments to enable strong light--matter coupling with vacuum Rabi splittings $\Omega_R \sim 100$\,meV, and (ii) IR/THz-active lattice or intramolecular vibrations with a dipole moments $\mu_{\mathrm{IR}} \sim 0.1\,\mathrm{D}$ and Huang--Rhys factors $S \sim 0.5$ at phonon frequencies $\omega_{\mathrm{ph}}\sim 40\text{--}200$\,meV ($300$--$1600\,\mathrm{cm^{-1}}$). These conditions are met across several material platforms. In molecular systems, such as crystalline anthracene/pentacene, $\pi$-conjugated polymers, and densely packed organic films, Frenkel excitons yield very large oscillator strengths at room temperature, achieving strong coupling up to $\Omega_R\sim 0.1\text{--}0.5$\,eV \cite{lidzey1998strong,kenacohen2010room,kenacohen2008prl}. Intramolecular vibrations are often strongly coupled to electronic transitions with $S\!\sim\!0.5\text{--}2$ and typically exhibit a rich set of IR-allowed modes within $600$--$2000\,\mathrm{cm^{-1}}$, owing to symmetry breaking that is generally present to some extent in molecular solids \cite{clark2006huang,wu2016polarons}. Transition-metal dichalcogenides (e.g., monolayer MoSe$_2$) combine stable Wannier excitons at $1.7$--$2.0$\,eV with routinely observed strong coupling (Rabi splittings $\sim 20\text{--}100$\,meV) and room-temperature exciton--phonon coupling of order $S\!\approx\!1$ for selected optical phonons, matching our target parameters \cite{dufferwiel2015vdw,kim2023mose2}. Wide-gap polar semiconductors such as GaN and ZnO are another class of materials that offer strong Fr\"ohlich coupling to LO phonons at $500$--$750\,\mathrm{cm^{-1}}$ and well-studied room-temperature polariton physics; effective $S$ values of $\sim 0.1\text{--}0.5$ can be enhanced by confinement or doping \cite{verdi2015frohlich,christopoulos2007gann}.
Hybrid and inorganic lead-halide perovskites supply large oscillator strengths together with soft, anharmonic lattices that naturally host strong carrier–lattice (Fr\"ohlich) coupling and rich IR-active phonon spectra. In 3D MAPbI$_3$, strong-coupling microcavities and even electrically injected polaritons have been realized, while quasi-2D PEA$_2$PbI$_4$ raises exciton binding ($\sim 0.2$–$0.3$\,eV) and oscillator strength; organic-cation bending modes around $\sim 1500\,\mathrm{cm^{-1}}$ then act as the mid-IR mode for our scheme \cite{soci2023nanoletters}. Importantly, multimode phonon-polaritons in the ultrastrong-coupling regime were reported recently in lead-halide films, independently confirming that perovskite phonons can reach the coupling strengths and IR dipoles our mid-IR channel requires \cite{kim2025multimode}. The combined FTIR and Raman studies showcase IR-active vibrational modes and Raman modes in MAPbI$_3$ across $\sim 500$–$1600\,\mathrm{cm^{-1}}$, underpinning the availability of suitable $\omega_{\mathrm{ph}}$ and $S$ \cite{wong2020acsmaterialslett}.

Altogether, across the organic, TMD, and polar–inorganic systems discussed above, the required set of parameters can be condensed to the following: $\{\mu_{\mathrm{vis}}\!\to\!\Omega_R,\ \mu_{\mathrm{IR}},\ S\ \text{at}\ \omega_{\mathrm{ph}}\}$, which has already demonstrated experimentally over a broad scope of materials systems of different nature, as we discuss above. For concreteness in the simulations, we adopt parameters from molecular platforms that have shown sideband-resolved optomechanical dynamics coupling vibrational modes to polariton states \cite{zasedatelev2019room,zasedatelev2021single,shishkov2024optomech}; given the breadth of viable systems, this choice does not limit the generality of the concept.

Regardless of the spectral range, Fabry--Perot cavity has the photon dispersion $\hbar \omega_{\bf k} = \hbar \omega_{{\bf k}={\bf 0}} + \alpha {\bf k}^2$, where ${\bf k}$ is the wave vector in the mirrors' plane.
Nevertheless, the dispersion coefficient $\alpha$ strongly depends on a particular spectral range: the dispersion becomes steeper as the fundamental frequency of the cavity, $\omega_{{\bf k}={\bf 0}}$, shifts from visible to mid-IR spectral range.
For a typical experimental setup with strong coupling $\alpha$ for a Fabry--Perot cavity in the mid-IR spectral range is around $10$ times smaller than $\alpha$ for a Fabry--Perot cavity in the visible spectral range~\cite{simpkins2015spanning, long2015coherent, shalabney2015coherent, kena2008strong, keeling2020bose, zasedatelev2019room, zasedatelev2021single}.

The emission of strongly correlated mid-IR photons and visible photons as a result of coherent resonant excitation of some visible mode of the cavity requires the simultaneous fulfillment of the phase matching condition, ${\bf k}_{{\rm Vis}|i} = {\bf k}_{{\rm Vis}|f}+{\bf k}_{\rm IR}$, and energy matching condition, $\omega_{{\rm Vis}|i} = \omega_{{\rm Vis}|f} + \omega_{\rm IR}$, where wave vector and frequency of coherent resonant excitation are ${\bf k}_{{\rm Vis}|i}$ and $\omega_{{\rm Vis}|i}$, wave vector and frequency of emitted visible photons are ${\bf k}_{{\rm Vis}|f}$ and $\omega_{{\rm Vis}|f}$, and wave vector and frequency of emitted mid-IR photons are ${\bf k}_{\rm IR}$ and $\omega_{\rm IR}$. 
Due to the difference in steepness of the visible and mid-IR dispersion curves in a double Fabry--Perot cavity, the simultaneous fulfillment of the phase and energy matching conditions is not compatible with other essential requirements, such as operating far from the resonant frequency of dark excitons.
Resolving this issue requires broadening the dispersion of the mid-IR cavity, narrowing the dispersion of the visible cavity, or engineering the dispersion of both the visible and mid-IR cavities.
Here, we use a patterned media, providing lattice resonance (LR) modes for the visible light, as an alternative to a Fabry--Perot resonator, while preserving a Fabry-Perot resonator for the mid-IR light (Fig.~\ref{fig:Fig1}(a) and Fig.~\ref{fig:All visible modes}(a,b)). 
A patterned medium serves as the periodic optical potential for visible light, leading to a Bloch structure of the LR modes and enabling conical intersections at small wave vectors and high frequencies~\cite{berghuis2023room}. 
Working at small wave vectors (in the first Brillouin zone) allows achieving maximum efficiency in translating the optical wave vector, while conical intersection shown in Fig.~\ref{fig:All visible modes}(c) allows obtaining sufficient differences in energy with moderate changes in the wave vector (for the transition between the upper and lower parts of the cone). 
A rigorous calculation of such a structure requires modeling the entire system; however, to show the fulfillment of the matching conditions, we will limit ourselves to a simplified model based on the coupled mode theory and the available experimental data. 
The structure shown in Fig.~\ref{fig:All visible modes}(a,b) hosts all Bloch LR visible modes with the dispersion $\hbar\omega_j=\pm \frac{\hbar c}{n_{eff}}(k\pm 2\pi j/a)$, where $c$ is the speed of light in vacuum, $a$ is the lattice period, $n_{eff}=1.42 $ is the effective refractive index of the LR modes, and $j=1,2,3..$.
We will use only first-order ($j=1$) counter-propagating modes, and we will be interested in the area near the point of the intersection ($k=0$,  $\hbar\omega= 2.55~{\rm eV}$) as shown in Fig.~\ref{fig:All visible modes}(c) by the red circle.
%We will use these visible modes and operate in the vicinity of $k = 0~{\rm \mu m^{-1}}$ to fulfill the phase matching and energy matching conditions (red circle in Fig.~\ref{fig:All visible modes}(c)).
Regarding the mid-IR cavity, we consider a typical for phonon-polaritons Fabry--Perot cavity~\cite{simpkins2015spanning}, consisting of metal mirrors separated by an IR-transparent spacer around the internal layer of patterned media (Fig.~\ref{fig:Fig1}(a) and Fig.~\ref{fig:All visible modes}(a)). 
The mirrors for IR light can be a metal mirrors with the thickness $\sim20~{\rm nm}$, that are thick enough to create a good Fabry--Perot cavity for mid-IR photons~\cite{long2015coherent}, but thin enough to allow for the optical pump in the visible range ~\cite{munkhbat2021}.

\begin{figure*}
\includegraphics[width=0.8\linewidth]{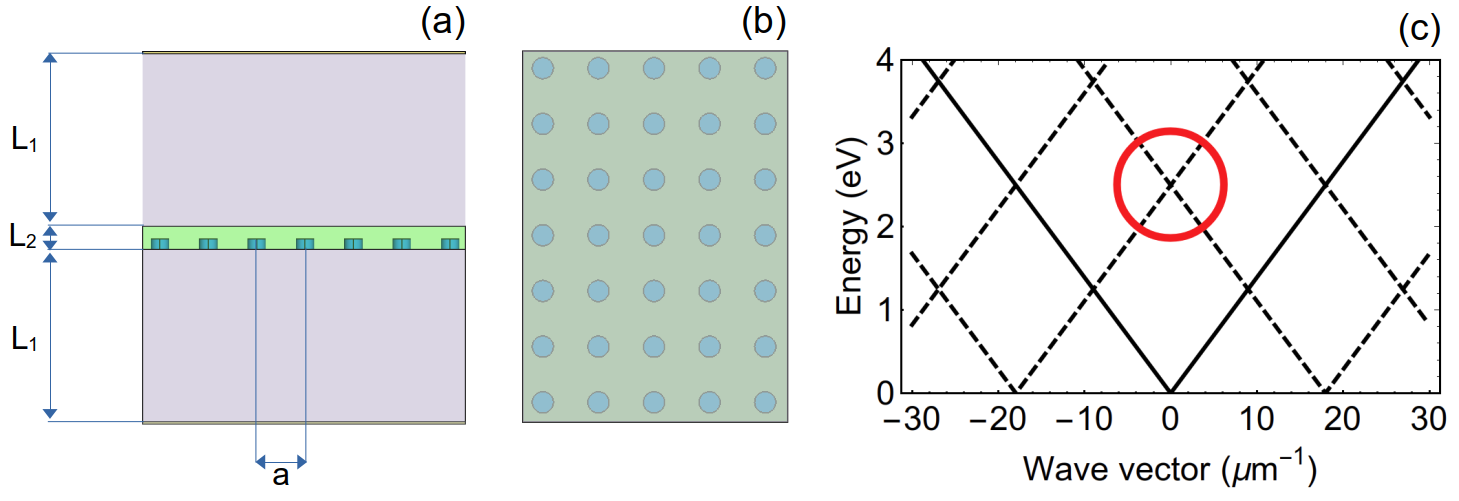}
\caption{
{\bf Cavity design and lattice resonance visible modes.}
(a) Side view of the structure, thin layers on top and bottom are $\sim20~{\rm nm}$ mirrors for mid-IR light, gray area is spacer, green area is media hosting excitons and phonons, and blue circles depict the lattice structure.
The parameters are $a=0.35~{\rm \mu m}$, $L_1=1.5~{\rm \mu m}$, $L_2=0.4~{\rm \mu m}$.
(b) Top view of the structure.
(c) Dispersion curves of the visible LR modes of the structure.
To fulfill the phase matching and energy matching conditions, we will use the region in the dispersion plane marked with red circle.
}
    \label{fig:All visible modes}
\end{figure*}

We assume that the photonic crystal structure does not change the quadratic dispersion of the mid-IR cavity near its fundamental mode. 
We also neglect the interaction of LR modes with the Mie modes of the dielectric nanodisks (which may form the bound state in the continuum in similar setups~\cite{sanvitto2022, berghuis2023room}), assuming that these modes are far from the frequency region we are interested in. 
Finally, we neglect the energy gap at ${\bf k}={\bf 0}$ for LR modes since it is typically negligeble in such structures~\cite{sanvitto2022,berghuis2023room,maggiolini2023,hakala2018, vakevainen2020, trypogeorgos2025}. 
The resultant dispersion curves for visible and mid-IR modes are
\begin{equation} \label{Vis left}
\hbar \omega_{{\rm VisL}|{\bf k}} 
= 
-\frac{\hbar c}{n_{\rm eff}}
\left(
k - \frac{2\pi}{a}
\right),
\end{equation}
\begin{equation} \label{Vis right}
\hbar \omega_{{\rm VisR}|{\bf k}} 
= 
\frac{\hbar c}{n_{\rm eff}}
\left(
k + \frac{2\pi}{a}
\right),
\end{equation}
\begin{equation} \label{Vis right}
\hbar \omega_{{\rm IR}|{\bf k}} 
= 
\hbar \omega_{{\rm IR}|{\bf k}={\bf 0}} 
+
\alpha_{\rm IR} {\bf k}^2,
\end{equation}
where $n_{\rm eff}=1.42$, $a = 0.35~{\rm \mu m}$, $\hbar \omega_{{\rm IR}|{\bf k}={\bf 0}} = 0.14~{\rm eV}$, and $\alpha_{\rm IR} = 0.04~{\rm eV \cdot \mu m^2}$.

The corresponding Hamiltonian of the electromagnetic modes is
\begin{equation} \label{H cav}
\hat H_{\rm cav}
=
\sum\limits_{\bf k} 
\left(
\hbar {\omega_{{\rm VisL}|{\bf k}}}
\hat a_{{\rm VisL}|{\bf k}}^\dag {{\hat a}_{{\rm VisL}|{\bf k}}}
+
\hbar {\omega_{{\rm VisR}|{\bf k}}}
\hat a_{{\rm VisR}|{\bf k}}^\dag {{\hat a}_{{\rm VisR}|{\bf k}}}
+
\hbar {\omega_{{\rm IR}|{\bf k}}}
\hat a_{{\rm IR}|{\bf k}}^\dag {{\hat a}_{{\rm IR}|{\bf k}}}
\right)
+
\hbar \Omega'
\left(
\hat a_{{\rm Vis}|{\bf k}_{i}}^\dag e^{-i\omega_\Omega t} 
+ 
\hat a_{{\rm Vis}|{\bf k}_{i}} e^{i\omega_\Omega t}
\right)
\end{equation}
where $\hat a_{{\rm{VisL}}|{\bf{k}}}^\dag$ and $\hat a_{{\rm{VisR}}|{\bf{k}}}^\dag$ (${\hat a_{{\rm{VisL}}|{\bf{k}}}}$ and ${\hat a_{{\rm{VisR}}|{\bf{k}}}}$) are the creation (annihilation) operators of a photon in the optical cavity with the wave vector ${\bf{k}}$, $\hat a_{{\rm{IR}}|{\bf{k}}}^\dag$ (${\hat a_{{\rm{IR}}|{\bf{k}}}}$) is the creation (annihilation) operator of a photon in the IR cavity with the wave vector ${\bf{k}}$.  
The commutation relations $\left[ \hat a_{{\rm VisL}|{\bf k}},\hat a_{{\rm VisL}|{\bf k'}}^\dag \right] = \delta _{{\bf k},{\bf k'}}$, $\left[ \hat a_{{\rm VisR}|{\bf k}},\hat a_{{\rm VisR}|{\bf k'}}^\dag \right] = \delta _{{\bf k},{\bf k'}}$, and $\left[ \hat a_{{\rm IR}|{\bf k}},\hat a_{{\rm IR}|{\bf k'}}^\dag \right] = \delta _{{\bf k},{\bf k'}}$ hold for the operators in the visible and mid-IR spectral ranges.
In addition, we consider the coherent excitation of the visible mode of the structure with wave vector ${\bf k}_i$ by monochromatic light with the frequency $\omega_\Omega$, which we consider in detail in Appendix~\ref{appendix: Hamiltonian}.
Without the loss of generality, we consider the excitation of the waveguide mode propagating to the right in Fig.~\ref{fig:All visible modes}. 
The parameter $\Omega'$ is proportional to the electric field of the incident light. 
We set the dissipation rates of the visible cavity modes $\gamma_{{\rm VisL}|{\bf k}}$, $\gamma_{{\rm VisR}|{\bf k}}$, and mid-IR cavity modes $\gamma_{{\rm IR}|{\bf k}}$, to $\gamma_{{\rm VisL}|{\bf k}} = \gamma_{{\rm VisR}|{\bf k}} = 3~{\rm meV}$ and $\gamma_{{\rm IR}|{\bf k}} = 4~{\rm meV}$, which is in agreement with the recent experiments~\cite{sanvitto2022,berghuis2023room, simpkins2015spanning, long2015coherent, shalabney2015coherent, zasedatelev2021single}.

\section{Microscopic theory of optomechanically interacting exciton-polaritons and phonon-polaritons} \label{appendix: Hamiltonian}

The full Hamiltonian of the system is
\begin{multline}\label{Full Hamiltonian simple form}
\hat H =  
\hat H_{\rm cav}
+
\sum\limits_{j=1}^{N_{\rm exc}} 
\left[
{\hbar {\omega_{\rm exc}}\hat \sigma_{{\rm Exc}{j}}^\dag {{\hat \sigma}_{{\rm Exc}{j}}}}
+
{\hbar {\omega_{\rm Vib}}\hat b_{{\rm Vib}j}^\dag {\hat b_{{\rm Vib}j}}}
+
{\hbar \Lambda \omega_{\rm Vib}\hat \sigma_{{\rm Exc}j}^\dag {{\hat \sigma }_{{\rm Exc} {j}}}\left( {{\hat b_{{\rm Vib}j}} + \hat b_{{\rm Vib}j}^\dag } \right)}
\right]
\\
+
\sum\limits_{j=1}^{N_{\rm exc}} 
\sum\limits_{{\bf k}} 
\left\{
\hbar \Omega^{(1)}_{{\rm Vis}j{\bf k}}
\left[ 
\hat \sigma_{{\rm Exc}{j}}^\dag 
\left( \hat a_{{\rm VisL}|{\bf k}} + \hat a_{{\rm VisR}|{\bf k}} \right)
e^{i{\bf k}{\bf r}_j}
+
h.c.
\right]
+
\hbar \Omega^{(1)}_{{\rm IR}j{\bf k}}
\left( 
\hat b_{{\rm Vib}{j}}^\dag \hat a_{{\rm IR}|{\bf k}} e^{i{\bf k}{\bf r}_j}
+
h.c.
\right)
\right\}
,
\end{multline}
where $\hat H_{\rm cav}$ is defined in Eq.~(\ref{H cav}), $N_{\rm exc}$ is the total number of excitonic states in the illuminated region.
We assume, that each excitonic state has the frequency ${\omega _{{\rm{exc}}}}$ independently on the other excitons~\cite{yamamoto2003semiconductor}.
For the $j$-th emitter $\hat \sigma_{{\rm{Exc}}{j}}^\dag$ (${\hat \sigma_{{\rm{Exc}}{j}}}$) is the creation (annihilation) operator of the exciton. 
Exciton operators obey the commutation relation $\left[{{\hat \sigma}_{{\rm{Exc}}{j}}}, {{\hat \sigma}_{{\rm{Exc}}{j'}}^\dag}\right] = {\delta _{{j},{j'}}}\left(1-2{{\hat \sigma}_{{\rm{Exc}}{j'}}^\dag}{{\hat \sigma}_{{\rm{Exc}}{j}}}\right)$. 
Below, we consider the case of small probability for the exciton being in an excited state, $\left\langle \hat \sigma^\dag_{{\rm{Exc}}j} \hat \sigma_{{\rm{Exc}}j} \right\rangle \ll 1 $. 
In this case, the approximate commutation relation $\left[ {{{\hat \sigma}_{{\rm{Exc}}{j}}},\hat \sigma_{{\rm{Exc}}{j'}}^\dag } \right] \approx {\delta _{{j},{j'}}}$ is valid~\cite{combescot2008microscopic}. 
We use the rotating wave approximation in Hamiltonian~(\ref{Full Hamiltonian simple form}) to describe light--matter interaction.
Also, we assume that each emitter hosts one vibrational mode with the frequency $\omega_{\rm Vib}$.
The operator $\hat b_{{\rm Vib}j}^\dag$ (${\hat b_{{\rm Vib}j}}$) is the creation (annihilation) vibrational state of $j$th emitter.
Here, we approximate each vibrational state by a harmonic oscillator.
The constant $\Lambda$ is the square root of the Huang--Rhys factor, $S$,~\cite{kirton2013nonequilibrium, cwik2016excitonic}.
Vector ${{\bf r}_j}$ points to the position of the $j$-th emitter, $\Omega_{{\rm Vis}j{\bf k}}^{(1)}$ is a single-exciton Rabi frequency of the interaction with the optical cavity~\cite{scully1997quantum} identical for both counter-propagating modes in the visible spectral region due to symmetry of the structure and $\Omega_{{\rm IR}j{\bf k}}^{(1)}$ is a single-vibration Rabi frequency of the interaction with the mid-IR cavity photons. 
We consider the parameters of the system, that are typical for the experimental setups: $\Lambda = 1$~\cite{guha2003temperature, tereshchenkov2024thermalization, kena2008strong, kena2010room, coropceanu2002hole}, $N_{\rm exc} = 10^8$~\cite{zasedatelev2021single}, $\hbar\omega_{\rm Vib} = 200~{\rm meV}$~\cite{zasedatelev2019room, zasedatelev2021single, guha2003temperature, coropceanu2002hole, xia2023ladder}, and $\hbar\omega_{\rm exc} - \Lambda^2\hbar\omega_{\rm Vib} = 2.72~{\rm eV}$~\cite{plumhof2014room, zasedatelev2019room, zasedatelev2021single, xia2023ladder, tereshchenkov2024thermalization}. 

Following~\cite{shishkov2024sympathetic, shishkov2024room, kirton2013nonequilibrium, wu2016polarons, tereshchenkov2024thermalization}, we consider the interaction between excitons, phonons, and the visible cavity and introduce the higher exciton-polaritons ($\hat s_{{\rm h}|{\bf k}}$), upper exciton-polaritons ($\hat s_{{\rm U}|{\bf k}}$), the lower exciton-polaritons ($\hat s_{{\rm L}|{\bf k}}$), and bright vibrational states ($\hat c_{{\rm Vib}|{\bf q}}$), which are phase-coherent, many-body delocalized states with well-defined in-plane momentum, corresponding to the eigenstates of the cavity
\begin{equation} \label{upper exciton-polaritons}
\hat s_{{\rm h}|{\bf k}} 
= 
X_{s_{\rm h}|{\bf k}, {\rm VisL}|{\bf k}} 
\hat a_{{\rm VisL}|{\bf k}} 
+
X_{s_{\rm h}|{\bf k}, {\rm VisR}|{\bf k}} 
\hat a_{{\rm VisR}|{\bf k}}  
+
X_{s_{\rm h}|{\bf k}, {\rm Exc}}
\hat c_{{\rm Exc}|{\bf k}}, 
\end{equation}
\begin{equation} \label{upper exciton-polaritons}
\hat s_{{\rm U}|{\bf k}} 
= 
X_{s_{\rm U}|{\bf k}, {\rm VisL}|{\bf k}} 
\hat a_{{\rm VisL}|{\bf k}} 
+
X_{s_{\rm U}|{\bf k}, {\rm VisR}|{\bf k}} 
\hat a_{{\rm VisR}|{\bf k}}  
+
X_{s_{\rm U}|{\bf k}, {\rm Exc}}
\hat c_{{\rm Exc}|{\bf k}}, 
\end{equation}
\begin{equation} \label{lower exciton-polaritons}
\hat s_{{\rm L}|{\bf k}} 
= 
X_{s_{\rm L}|{\bf k}, {\rm VisL}|{\bf k}} 
\hat a_{{\rm VisL}|{\bf k}} 
+
X_{s_{\rm L}|{\bf k}, {\rm VisR}|{\bf k}} 
\hat a_{{\rm VisR}|{\bf k}}  
+
X_{s_{\rm L}|{\bf k}, {\rm Exc}}
\hat c_{{\rm Exc}|{\bf k}}, 
\end{equation}
\begin{equation} 
\hat c_{{\rm Vib}|{\bf k}} = 
\frac{1} {\sqrt{N_{\rm exc}}}
\sum_{j=1}^{N_{\rm exc}} 
\left(
\hat b_{{\rm Vib}j} 
+
\Lambda \hat \sigma^\dag_{{\rm Exc}j} \hat \sigma_{{\rm Exc}j} 
\right)
e^{-i{\bf k}{\bf r}_j},
\end{equation}
where we introduced the collective Rabi frequency for the visible cavity $
\Omega_{\rm Vis} = \sqrt{\sum_{j=1}^{N_{\rm exc}}\left( \Omega_{{\rm Vis}j{\bf k}}^{(1)} \right)^2}$ assumed its $\bf k$-independence.
We set $\hbar\Omega_{\rm Vis} = 50~{\rm meV}$ which is consistent with the experimental data of molecular polaritonic systems~\cite{plumhof2014room, sanvitto2016road, zasedatelev2019room, zasedatelev2021single}.
We also denoted~\cite{shishkov2024room}
\begin{equation}
\hat c_{{\rm Exc}|{\bf k}} =
\sum_{j=1}^{N_{\rm exc}}  
\frac{\Omega_{{\rm Vis}j{\bf k}}^{(1)}}{\Omega_{\rm Vis}}
\hat \sigma_{{\rm Exc}j} e^{\Lambda (\hat b_{{\rm Vib}j}^\dag - \hat b_{{\rm Vib}j})} 
{e^{-i{\bf k}{\bf r}_j}}.
\end{equation}
The coefficients $X_{\alpha}$ are the elements of the unitary matrix $W_{{\rm Vis}|{\bf k}}$ 
\begin{equation} \label{unitary transformation}
W_{{\rm Vis}|{\bf k}} = 
\begin{pmatrix}
X_{s_{\rm L}|{\bf k}, {\rm Exc}} &  X_{s_{\rm U}|{\bf k}, {\rm Exc}} &  X_{s_{\rm h}|{\bf k}, {\rm Exc}} \\
X_{s_{\rm L}|{\bf k}, {\rm VisR}|{\bf k}} & X_{s_{\rm U}|{\bf k}, {\rm VisR}|{\bf k}} & X_{s_{\rm h}|{\bf k}, {\rm VisR}|{\bf k}}  \\
X_{s_{\rm L}|{\bf k}, {\rm VisL}|{\bf k}} & X_{s_{\rm U}|{\bf k}, {\rm VisL}|{\bf k}} & X_{s_{\rm h}|{\bf k}, {\rm VisL}|{\bf k}}.  
\end{pmatrix}
\end{equation}
The unitary matrix $W_{{\rm Vis}|{\bf k}}$ diagonalize the matrix 
\begin{equation} \label{Mvis}
M_{{\rm Vis}|{\bf k}} = 
\begin{pmatrix}
\omega_{\rm exc} - \Lambda^2\omega_{\rm Vib} & \Omega_{\rm Vis} & \Omega_{\rm Vis} \\
\Omega_{\rm Vis} & \omega_{{\rm VisR}|{\bf k}} & 0 \\
\Omega_{\rm Vis} & 0 & \omega_{{\rm VisL}|{\bf k}}  
\end{pmatrix},
\end{equation}
meaning
\begin{equation}
W^{-1}_{{\rm Vis}|{\bf k}} M_{{\rm Vis}|{\bf k}} W_{{\rm Vis}|{\bf k}} = 
\begin{pmatrix}
\omega_{s_{\rm L}|{\bf k}} & 0 & 0 \\
0 & \omega_{s_{\rm U}|{\bf k}} & 0 \\
0 & 0 & \omega_{s_{\rm h}|{\bf k}}  
\end{pmatrix},
\end{equation}
where $\omega_{s_{\rm h}|{\bf k}}$, $\omega_{s_{\rm U}|{\bf k}}$, and $\omega_{s_{\rm L}|{\bf k}} $ are the eigenfrequencies of higher, upper, and lower exciton-polaritons with wave vector $\bf k$.

In what follows, we assume that higher exciton-polaritons do not affect the quantum dynamics of the system in our setup and omit these states.
The reason for this omission is that the density of states of dark states significantly overcomes the density of states of higher excitons, while their resonant frequencies are almost overlapping.
Thus, the Hamiltonian~(\ref{Full Hamiltonian simple form}) takes the form
\begin{multline} \label{H_weak}
\hat H 
=
\hat H_{\rm Dark}
+
\sum\limits_{\bf k} 
\hbar \omega_{s_{\rm U}|{\bf k}}
\hat s_{{\rm U}|{\bf k}}^\dag \hat s_{{\rm U}|{\bf k}}
+
\sum\limits_{\bf k} 
\hbar \omega_{s_{\rm L}|{\bf k}}
\hat s_{{\rm L}|{\bf k}}^\dag \hat s_{{\rm L}|{\bf k}}
+
\hbar 
\Omega_{{\bf k}_{i}}
\left(
\hat s_{{\rm U}|{\bf k}_{i}}^\dag e^{-i\omega_{s_{\rm U}|{\bf k}_{i}} t} + 
\hat s_{{\rm U}|{\bf k}_{i}} e^{i\omega_{s_{\rm U}|{\bf k}_{i}} t}
\right) 
\\
+
\hbar 
\frac{\Lambda \Omega_{\rm Vis}}{\sqrt{N_{\rm exc}}}
\sum_{{\bf k}'-{\bf k}={\bf q}}
\left[
\left(
X_{s_{\rm U}|{\bf k'}, {\rm Exc}}^* 
\hat s_{{\rm U}|{\bf k'}}^\dag
+ 
X_{s_{\rm L}|{\bf k'}, {\rm Exc}}^*
\hat s_{{\rm L}|{\bf k'}}^\dag
\right)
\left(
\hat c^\dag_{{\rm Vib}|{\bf -q}}
-
\hat c_{{\rm Vib}|{\bf q}}
\right)
\left(
X_{s_{\rm U}|{\bf k}, {\rm Vis}|{\bf k}}
\hat s_{{\rm U}|{\bf k}}
+ 
X_{s_{\rm L}|{\bf k}, {\rm Vis}|{\bf k}}
\hat s_{{\rm L}|{\bf k}}
\right)
+
h.c.
\right]
\\
-
\hbar 
\frac{\Lambda \Omega_{\rm IR}}{\sqrt{N_{\rm exc}}}
\sum_{{\bf k}'-{\bf k}={\bf q}}
\left[
\left(
X_{s_{\rm U}|{\bf k'}, {\rm Exc}}^* 
\hat s_{{\rm U}|{\bf k'}}^\dag
+ 
X_{s_{\rm L}|{\bf k'}, {\rm Exc}}^*
\hat s_{{\rm L}|{\bf k'}}^\dag
\right)
\hat a^\dag_{{\rm IR}|{\bf q}}
\left(
X_{s_{\rm U}|{\bf k}, {\rm Exc}} 
\hat s_{{\rm U}|{\bf k}}
+ 
X_{s_{\rm L}|{\bf k}, {\rm Exc}}
\hat s_{{\rm L}|{\bf k}}
\right)
+
h.c.
\right]
\\
+
\sum\limits_{\bf q} 
\hbar {\omega_{\rm Vib}}
\hat c_{{\rm Vib}|{\bf q}}^\dag \hat c_{{\rm Vib}|{\bf q}} 
+
\sum\limits_{\bf q} 
\hbar {\omega_{{\rm IR}|{\bf q}}}
\hat a_{{\rm IR}|{\bf q}}^\dag {{\hat a}_{{\rm IR}|{\bf q}}}
+
\sum\limits_{{\bf q}} 
\hbar \Omega_{{\rm IR}}
\left( 
\hat c_{{\rm Vib}|{\bf q}}^\dag \hat a_{{\rm IR}|{\bf q}}
+
\hat a_{{\rm IR}|{\bf q}}^\dag \hat c_{{\rm Vib}|{\bf q}} 
\right),
\end{multline}
where the interaction constant with the coherent laser drive is expressed as $\Omega_{{\bf k}_i} = \Omega'\cos \varphi_{{\bf k}_i} $, introduced the collective Rabi frequency for IR cavity $\Omega_{\rm IR}=\sqrt{\sum_{j=1}^{N_{\rm exc}}\left( \Omega_{{\rm IR}j{\bf k}}^{(1)} \right)^2}$ which we assume to be $\bf k$-independent.
We set $\hbar\Omega_{\rm IR} = 16~{\rm meV}$, which is consistent with the experiments on strong vibrational coupling in Fabry--Perot cavities~\cite{simpkins2015spanning, simpkins2023control}. 
We also denoted $X_{s_{\rm U}|{\bf k}, {\rm Vis}|{\bf k}} = X_{s_{\rm U}|{\bf k}, {\rm VisR}|{\bf k}} + X_{s_{\rm U}|{\bf k}, {\rm VisL}|{\bf k}}$ and $X_{s_{\rm L}|{\bf k}, {\rm Vis}|{\bf k}} = X_{s_{\rm L}|{\bf k}, {\rm VisR}|{\bf k}} + X_{s_{\rm L}|{\bf k}, {\rm VisL}|{\bf k}}$ and set $\omega_\Omega = \omega_{s_{\rm U}|{\bf k}_{i}}$, such that the frequency of the external field matches the frequency of the lower polariton state with the wave vector ${\bf k}_i$.

Preserving only the resonance terms in Eq.~(\ref{H_weak}), corresponding to the phonon-assisted transitions from upper exciton-polaritons to lower exciton-polaritons, we obtain
\begin{multline} \label{H_weak}
\hat H 
=
\hat H_{\rm Dark}
+
\sum\limits_{\bf k} 
\hbar \omega_{s_{\rm U}|{\bf k}}
\hat s_{{\rm U}|{\bf k}}^\dag \hat s_{{\rm U}|{\bf k}}
+
\sum\limits_{\bf k} 
\hbar \omega_{s_{\rm L}|{\bf k}}
\hat s_{{\rm L}|{\bf k}}^\dag \hat s_{{\rm L}|{\bf k}}
+
\hbar 
\Omega_{{\bf k}_{i}}
\left(
\hat s_{{\rm U}|{\bf k}_{i}}^\dag e^{-i\omega_{s_{\rm U}|{\bf k}_{i}} t} + 
\hat s_{{\rm U}|{\bf k}_{i}} e^{i\omega_{s_{\rm U}|{\bf k}_{i}} t}
\right) 
\\
+
\sum_{{\bf k}'-{\bf k}={\bf q}}
\left(
\hbar 
g_{{\rm Vib}|{\bf kk'}}
\hat s_{{\rm U}|{\bf k'}}^\dag
\hat c_{{\rm Vib}|{\bf q}}
\hat s_{{\rm L}|{\bf k}}
+
h.c.
\right)
+
\sum_{{\bf k}'-{\bf k}={\bf q}}
\left(
\hbar 
g_{{\rm IR}|{\bf kk'}}
\hat s_{{\rm U}|{\bf k'}}^\dag
\hat a_{{\rm IR}|{\bf q}}
\hat s_{{\rm L}|{\bf k}}
+
h.c.
\right)
\\
+
\sum\limits_{\bf q} 
\hbar {\omega_{\rm Vib}}
\hat c_{{\rm Vib}|{\bf q}}^\dag \hat c_{{\rm Vib}|{\bf q}} 
+
\sum\limits_{\bf q} 
\hbar {\omega_{{\rm IR}|{\bf q}}}
\hat a_{{\rm IR}|{\bf q}}^\dag {{\hat a}_{{\rm IR}|{\bf q}}}
+
\sum\limits_{{\bf q}} 
\hbar \Omega_{{\rm IR}}
\left( 
\hat c_{{\rm Vib}|{\bf q}}^\dag \hat a_{{\rm IR}|{\bf q}}
+
\hat a_{{\rm IR}|{\bf q}}^\dag \hat c_{{\rm Vib}|{\bf q}} 
\right),
\end{multline}
where the optomechanical interaction constants of exciton-polaritons with bright vibrations and mid-IR cavity photons, $g_{{\rm Vib}|{\bf kk'}}$ and $g_{{\rm IR}|{\bf kk'}}$, are
\begin{equation} \label{g Vis}
g_{{\rm Vib}|{\bf kk'}} 
= 
\frac{\Lambda \Omega_{\rm Vis}}{\sqrt{N_{\rm exc}}}
\left(
X_{s_{\rm U}|{\bf k'}, {\rm Vis}|{\bf k'}}^*
X_{s_{\rm L}|{\bf k}, {\rm Exc}} 
-
X_{s_{\rm U}|{\bf k'}, {\rm Exc}}^* 
X_{s_{\rm L}|{\bf k}, {\rm Vis}|{\bf k}}
\right),
\end{equation}
\begin{equation} \label{g IR}
g_{{\rm IR}|{\bf kk'}} 
= 
-\frac{\Lambda \Omega_{\rm IR}}{\sqrt{N_{\rm exc}}}
X_{s_{\rm U}|{\bf k'}, {\rm Exc}}^*
X_{s_{\rm L}|{\bf k}, {\rm Exc}} .
\end{equation}
The term proportional to $g_{{\rm IR}|{\bf kk'}}$ is not trivial, as there is no direct coupling between the mid-IR cavity photons and either the visible cavity photons or the excitons in the initial Hamiltonian~(\ref{Full Hamiltonian simple form}).
Nevertheless, this term describes the coupling of mid-IR cavity photons with exciton-polariton states originating from the optomechanical interaction. 
This interaction is a direct consequence of the excitonic dressing of the vibrational states.
The term $\hat H_{\rm Dark}$ represents the dark states that act as a reservoir for both the exciton-polaritons and phonon-polaritons~\cite{shishkov2024room}.

To address the strong coupling between vibrational states and the mid-IR cavity photons, we introduce upper phonon-polaritons, $\hat v_{{\rm U}|{\bf q}}$, and lower phonon-polaritons, $\hat v_{{\rm L}|{\bf q}}$, as follows
\begin{equation}
\hat v_{{\rm U}|{\bf q}} =
\hat a_{{\rm IR}|{\bf q}} \sin \phi_{{\bf q}} +
\hat c_{{\rm Vib}|{\bf q}} \cos \phi_{{\bf q}},
\end{equation}
\begin{equation}
\hat v_{{\rm L}|{\bf q}} =
\hat a_{{\rm IR}|{\bf q}} \cos \phi_{{\bf q}} -
\hat c_{{\rm Vib}|{\bf q}} \sin \phi_{{\bf q}},
\end{equation}
where
\begin{equation} \label{TransformationAngleVP}
\phi_{{\bf q}} =
{\rm arctg}
\left[
\sqrt{
\frac
{(\omega_{\rm Vib} - \omega_{{\rm IR}|{\bf q}})^2+4\Omega_{\rm IR}^2}
{4\Omega_{\rm IR}^2}
}
-
\frac{\omega_{\rm Vib} - \omega_{{\rm IR}|{\bf q}}}{2 \Omega_{\rm IR}}
\right]
.
\end{equation}
The eigen-frequencies of the upper and lower phonon-polaritons are
\begin{equation}
\omega _{v_{\rm U}|{\bf q}} 
= 
\frac{\omega_{\rm Vib} + \omega_{{\rm IR}|{\bf q}} }{ 2}
+
\sqrt{
\frac
{
\left( 
\omega_{\rm Vib} - \omega _{{\rm IR}|{\bf q}} 
\right)^2 
}{
4
}
+
\Omega_{\rm IR}^2
}
,
\end{equation}
\begin{equation}
\omega_{v_{\rm L}|{\bf q}} 
= 
\frac{\omega_{\rm Vib} + \omega_{{\rm IR}|{\bf q}} }{ 2}
-
\sqrt{
\frac{
\left( 
\omega_{\rm Vib} - \omega _{{\rm IR}|{\bf q}} 
\right)^2 
}{
4
}
+
\Omega_{\rm IR}^2
}
.
\end{equation}
After transformation to upper and lower phonon-polaritons, the Hamiltonian~(\ref{H_weak}) can be reformulated as follows
\begin{multline} \label{H optomech}
\hat H 
=
\hat H_{\rm Dark}
+
\sum\limits_{\bf k} 
\hbar \omega_{s_{\rm U}|{\bf k}}
\hat s_{{\rm U}|{\bf k}}^\dag \hat s_{{\rm U}|{\bf k}}
+
\sum\limits_{\bf k} 
\hbar \omega_{s_{\rm L}|{\bf k}}
\hat s_{{\rm L}|{\bf k}}^\dag \hat s_{{\rm L}|{\bf k}}
+
\sum\limits_{\bf q} 
\hbar \omega_{v_{\rm U}|{\bf q}}
\hat v_{{\rm U}|{\bf q}}^\dag 
\hat v_{{\rm U}|{\bf q}}
+
\sum\limits_{\bf q} 
\hbar \omega_{v_{\rm L}|{\bf q}}
\hat v_{{\rm L}|{\bf q}}^\dag 
\hat v_{{\rm L}|{\bf q}}
\\
+
\hbar 
\Omega_{{\bf k}_i}
\left(
\hat s^\dag_{{\rm U}|{\bf k}_{i}} 
e^{-i\omega_{s_{\rm U}|{\bf k}_{i}} t} 
+  
\hat s_{{\rm U}|{\bf k}_{i}} 
e^{i\omega_{s_{\rm U}|{\bf k}_{i}} t} 
\right) 
+
\sum_{{\bf k}'-{\bf k}={\bf q}}
\left[
\hat s_{{\rm U}|{\bf k'}}^\dag
\left(
\hbar g_{{\rm U}|{\bf kk'}}
\hat v_{{\rm U}|{\bf q}}
+
\hbar g_{{\rm L}|{\bf kk'}}
\hat v_{{\rm L}|{\bf q}}
\right)
\hat s_{{\rm L}|{\bf k}} 
+
h.c.
\right]
,
\end{multline}
where the optomechanical interaction constants for phonon-polaritons are 
\begin{equation} \label{def gU}
g_{{\rm U}|{\bf kk'}} = 
g_{{\rm IR}|{\bf kk'}} \sin\phi_{{\bf k'}-{\bf k}} +
g_{{\rm Vib}|{\bf kk'}} \cos\phi_{{\bf k'}-{\bf k}},   
\end{equation}
\begin{equation} \label{def gL}
g_{{\rm L}|{\bf kk'}} = 
g_{{\rm IR}|{\bf kk'}} \cos\phi_{{\bf k'}-{\bf k}} -
g_{{\rm Vib}|{\bf kk'}} \sin\phi_{{\bf k'}-{\bf k}}.
\end{equation}
Thus, strong interaction between excitons and vibrations with cavity modes in the visible range and mid-IR range leads to the formation of new eigenstates, exciton- and phonon-polaritons~\cite{kirton2013nonequilibrium, wu2016polarons, tereshchenkov2024thermalization}.
The part of the Hamiltonian~(\ref{H optomech}) governing these states is analogous to those found in well-established multimode cavity optomechanics~\cite{shishkov2024room}, 
\begin{multline} \label{H_OM_appendix}
\hat H_{\rm OM}
=
\sum\limits_{\bf k} 
\hbar \omega_{s_{\rm L}|{\bf k}}
\hat s^\dag_{{\rm L}|{\bf k}}
\hat s_{{\rm L}|{\bf k}}
+
\sum\limits_{\bf k} 
\hbar \omega_{s_{\rm U}|{\bf k}}
\hat s^\dag_{{\rm U}|{\bf k}}
\hat s_{{\rm U}|{\bf k}}
+
\hbar 
\Omega_{{\bf k}_i}(t)
\left(
\hat s^\dag_{{\rm U}|{\bf k}_{i}} 
e^{-i\omega_{s_{\rm U}|{\bf k}_{i}} t} 
+  
h.c. 
\right) 
+
\sum\limits_{\bf q} 
\hbar \omega_{v_{\rm U}|{\bf q}}
\hat v_{{\rm U}|{\bf q}}^\dag 
\hat v_{{\rm U}|{\bf q}}
\\
+
\sum\limits_{\bf q} 
\hbar \omega_{v_{\rm L}|{\bf q}}
\hat v_{{\rm L}|{\bf q}}^\dag 
\hat v_{{\rm L}|{\bf q}}
+
\sum_{{\bf k}_i,{\bf k}_f}
\left[
\hbar 
g_{{\rm U}|{\bf k}_f{\bf k}_i}
\hat v_{{\rm U}|{\bf k}_i-{\bf k}_f}^\dag 
\hat s_{{\rm L}|{\bf k}_f}^\dag
\hat s_{{\rm U}|{\bf k}_i} 
+
h.c.
\right]
+
\sum_{{\bf k}_i,{\bf k}_f}
\left[
\hbar 
g_{{\rm L}|{\bf k}_f{\bf k}_i}
\hat v_{{\rm L}|{\bf k}_i-{\bf k}_f}^\dag 
\hat s_{{\rm L}|{\bf k}_f}^\dag
\hat s_{{\rm U}|{\bf k}_i} 
+
h.c.
\right],
\end{multline}

We discussed the dissipation rates of the visible cavity photons at the end of Appendix~\ref{appendix: coupled mode theory}.
We assume the dissipation rate of excitons is $\hbar\gamma_{\rm Exc}=0.01~{\rm meV}$, whereas the dephasing rate is $\hbar\Gamma_{\rm Exc}=50~{\rm meV}$, which is in agreement with the experimental data~\cite{kena2008strong, plumhof2014room, zasedatelev2019room, zasedatelev2021single, xia2023ladder}.
We assume that the dissipation rate of mid-IR cavity photons in $\hbar\gamma_{{\rm IR}|{\bf q}}=4~{\rm meV}$ and is independent of the wave vector $\bf q$, which agrees with the recent experiments~\cite{long2015coherent, simpkins2015spanning, simpkins2023control}.
We assume a quality factor ($Q$) for high-energy vibrational modes coupled to excitons is approximately $100$, which aligns well with experimental observations derived from spontaneous Raman and IR spectra~\cite{coles2011vibrationally, zasedatelev2019room, ahn2018vibrational, simpkins2023control}.
The ratio between the matter and light components in exciton-polaritons determines their dissipation rates. 
For higher, upper, and lower exciton-polaritons, the dissipation rates are
\begin{equation} \label{higher exciton-polaritons}
\gamma_{s_{\rm h}|{\bf k}} 
=
|X_{s_{\rm h}|{\bf k}, {\rm VisR}|{\bf k}}|^2
\gamma_{{\rm VisR}|{\bf k}}
+
|X_{s_{\rm h}|{\bf k}, {\rm VisL}|{\bf k}}|^2
\gamma_{{\rm VisL}|{\bf k}}
+
|X_{s_{\rm h}|{\bf k}, {\rm Exc}}|^2
\gamma_{\rm Exc} ,
\end{equation}
\begin{equation} \label{upper exciton-polaritons}
\gamma_{s_{\rm U}|{\bf k}} 
=
|X_{s_{\rm U}|{\bf k}, {\rm VisR}|{\bf k}}|^2
\gamma_{{\rm VisR}|{\bf k}}
+
|X_{s_{\rm U}|{\bf k}, {\rm VisL}|{\bf k}}|^2
\gamma_{{\rm VisL}|{\bf k}}
+
|X_{s_{\rm U}|{\bf k}, {\rm Exc}}|^2
\gamma_{\rm Exc} ,
\end{equation}
\begin{equation} \label{lower exciton-polaritons}
\gamma_{s_{\rm L}|{\bf k}} 
=
|X_{s_{\rm L}|{\bf k}, {\rm VisR}|{\bf k}}|^2
\gamma_{{\rm VisR}|{\bf k}}
+
|X_{s_{\rm L}|{\bf k}, {\rm VisL}|{\bf k}}|^2
\gamma_{{\rm VisL}|{\bf k}}
+
|X_{s_{\rm L}|{\bf k}, {\rm Exc}}|^2
\gamma_{\rm Exc} .
\end{equation}
Similarly, for upper and lower phonon-polaritons, the dissipation rate is given by 
\begin{equation}
\gamma_{v_{\rm U}|{\bf q}} = \gamma_{{\rm IR}|{\bf q}} \sin^2\phi_{\bf q} + \gamma_{\rm Vib} \cos^2\phi_{\bf q},    
\end{equation}
\begin{equation}
\gamma_{v_{\rm L}|{\bf q}} = \gamma_{{\rm IR}|{\bf q}} \cos^2\phi_{\bf q} + \gamma_{\rm Vib} \sin^2\phi_{\bf q}.
\end{equation}

We plot the parameters of the system in Fig.~\ref{fig:Fig1}, which includes eigen frequencies and dissipation rates for the upper and lower exciton-polaritons and phonon-polaritons, as well as the single-polariton optomechanical interaction energies between lower exciton-polaritons and phonon-polaritons.

\begin{figure}
\includegraphics[width=1\linewidth]{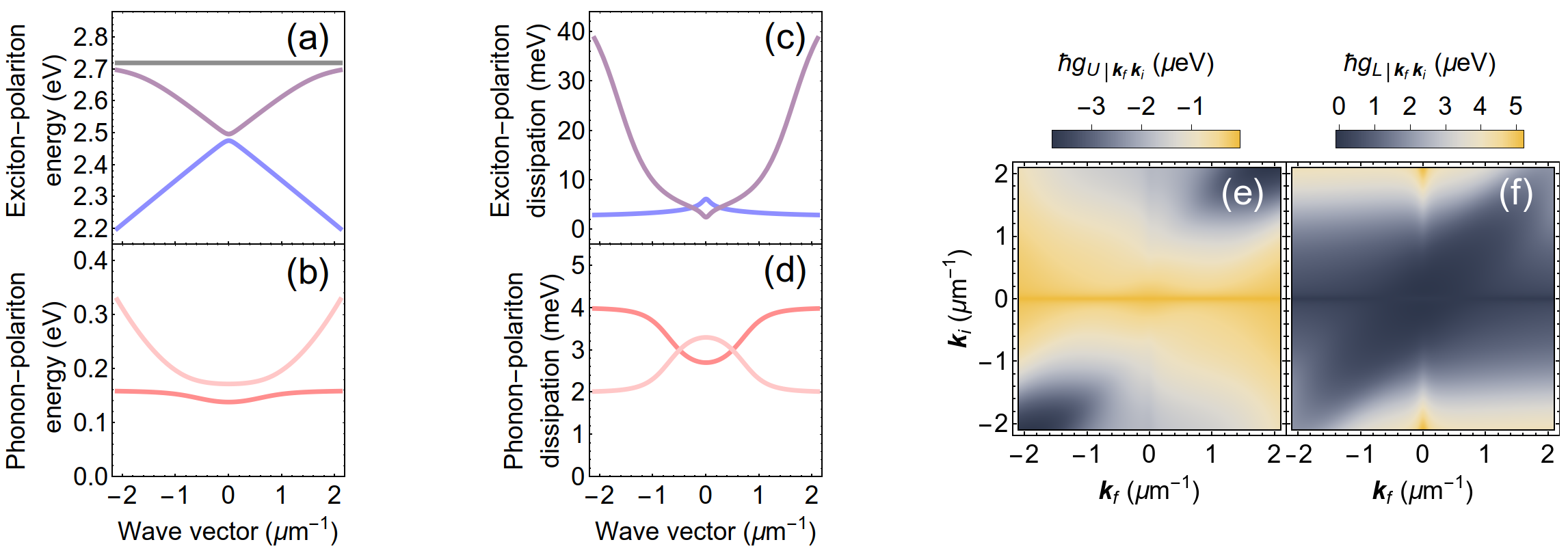}
\caption{
{\bf Parameters of the system.}
Dispersion (a) and dissipation (c) for upper (purple line) and lower (blue line) exciton-polarotons.
Dispersion (b) and dissipation (d) for upper (pink line) and lower (red line) phonon-polaritons.
The black line in (a) shows the resonant frequency of excitons. 
Optomechanical constant for upper phonon-polaritons (f) and lower phonon-polaritons (g).
We consider the system operating at room temperature, $k_BT = 25~{\rm meV}$.
The parameters of the system are specified in Sec.~\ref{appendix: coupled mode theory} and Sec.~\ref{appendix: Hamiltonian}.
}
    \label{fig:parameters}
\end{figure}

As we discussed in the main text, when the main channel of energy flow from directly driven upper exciton-polaritons is the dissipation, their dynamics become completely independent of the evolution of the lower exciton-polaritons and phonon-polaritons.
In this case, we can exclude upper exciton-polaritons adiabatically, replacing $\hat s_{{\rm U}|{\bf k}_i}(t)$ with $\sqrt{\langle \hat s_{{\rm U}|{\bf k}_i}^\dag(t) \hat s_{{\rm U}|{\bf k}_i}(t) \rangle}e^{-i\omega_{s_{\rm U}|{\bf k}_i}t} = \sqrt{n_{{\rm U}|{\bf k}_i}(t)}e^{-i\omega_{s_{\rm U}|{\bf k}_i}t}$ and disregarding the redundant constant phase of $\hat s_{{\rm U}|{\bf k}_i}(t)$.
Thus, we can exclude upper exciton-polaritons adiabatically and obtain the linearized optomechanical Hamiltonian
\begin{multline} \label{H_OM_lin_all_modes_1}
\hat H_{{\rm OM}|{\bf k}_i,{\bf k}_f}
\approx
\sum\limits_{{\bf k}_f} 
\hbar \omega_{s_{\rm L}|{\bf k}_f}
\hat s^\dag_{{\rm L}|{\bf k}_f}
\hat s_{{\rm L}|{\bf k}_f}
+
\sum\limits_{{\bf k}_f}
\hbar \omega_{v_{\rm U}|{\bf k}_i - {\bf k}_f}
\hat v_{{\rm U}|{\bf k}_i - {\bf k}_f}^\dag 
\hat v_{{\rm U}|{\bf k}_i - {\bf k}_f}
+
\sum\limits_{{\bf k}_f} 
\hbar \omega_{v_{\rm L}|{\bf k}_i - {\bf k}_f}
\hat v_{{\rm L}|{\bf k}_i - {\bf k}_f}^\dag 
\hat v_{{\rm L}|{\bf k}_i - {\bf k}_f}
\\
+
\sum_{{\bf k}_f}
\left[
\hbar 
g_{{\rm U}|{\bf k}_f{\bf k}_i}
\sqrt{n_{{\rm U}|{\bf k}_i}(t)}
\hat v_{{\rm U}|{\bf k}_i-{\bf k}_f}^\dag 
\hat s_{{\rm L}|{\bf k}_f}^\dag
e^{-i\omega_{s_{\rm U}|{\bf k}_i}t}
+
h.c.
\right]
+
\sum_{{\bf k}_f}
\left[
\hbar 
g_{{\rm L}|{\bf k}_f{\bf k}_i}
\sqrt{n_{{\rm U}|{\bf k}_i}(t)}
\hat v_{{\rm L}|{\bf k}_i-{\bf k}_f}^\dag 
\hat s_{{\rm L}|{\bf k}_f}^\dag
e^{-i\omega_{s_{\rm U}|{\bf k}_i}t}
+
h.c.
\right].
\end{multline}
We consider slow dynamics of lower exciton-polaritons, replacing $\hat s_{{\rm L}|{\bf k}_f} = \hat S_{{\rm L}|{\bf k}_f}e^{i\omega_{s_{\rm U}|{\bf k}_i}t}$.
As a result we obtain
\begin{multline}  \label{H_OM_lin_all_modes_2}
\hat H_{{\rm OM}|{\bf k}_i,{\bf k}_f}
\approx
\sum\limits_{{\bf k}_f} 
\hbar \left( \omega_{s_{\rm L}|{\bf k}_f} - \omega_{s_{\rm L}|{\bf k}_i} \right)
\hat S^\dag_{{\rm L}|{\bf k}_f}
\hat S_{{\rm L}|{\bf k}_f}
+
\sum\limits_{{\bf k}_f}
\hbar \omega_{v_{\rm U}|{\bf k}_i - {\bf k}_f}
\hat v_{{\rm U}|{\bf k}_i - {\bf k}_f}^\dag 
\hat v_{{\rm U}|{\bf k}_i - {\bf k}_f}
+
\sum\limits_{{\bf k}_f} 
\hbar \omega_{v_{\rm L}|{\bf k}_i - {\bf k}_f}
\hat v_{{\rm L}|{\bf k}_i - {\bf k}_f}^\dag 
\hat v_{{\rm L}|{\bf k}_i - {\bf k}_f}
\\
+
\sum_{{\bf k}_f}
\left[
\hbar 
g_{{\rm U}|{\bf k}_f{\bf k}_i}
\sqrt{n_{{\rm U}|{\bf k}_i}(t)}
\hat v_{{\rm U}|{\bf k}_i-{\bf k}_f}^\dag 
\hat S_{{\rm L}|{\bf k}_f}^\dag
+
h.c.
\right]
+
\sum_{{\bf k}_f}
\left[
\hbar 
g_{{\rm L}|{\bf k}_f{\bf k}_i}
\sqrt{n_{{\rm U}|{\bf k}_i}(t)}
\hat v_{{\rm L}|{\bf k}_i-{\bf k}_f}^\dag 
\hat S_{{\rm L}|{\bf k}_f}^\dag
+
h.c.
\right].
\end{multline}
For the sake of briefness in the {\it main text}, we use the notation $\hat s_{{\rm L}|{\bf k}_f}$ instead of $\hat S_{{\rm L}|{\bf k}_f}$ in Hamiltonian~(\ref{H_OM_lin_all_modes_2}).

\section{Covariance matrix of exciton-polaritons and phonon-polarions} \label{appendix: correlations}

In this Appendix, we use the Hamiltonian~(\ref{H_OM_lin_all_modes_2}) (or equivalently Hamiltonian~(\ref{H_OM_lin_all_modes_1})) to find the covariance matrix of exciton-polaritons and phonon-polarions, i.e. matrix $\langle \hat A(t) \hat B(t+\tau) \rangle$, where $\hat A$ and $\hat B$ can be one of the following operators $\hat s_{{\rm L}|{\bf k}_{f}}$, $\hat s_{{\rm L}|{\bf k}_{f}}^\dag$, $\hat v_{{\rm U}|{\bf k}_{i} - {\bf k}_{f}}$, $\hat v_{{\rm U}|{\bf k}_{i} - {\bf k}_{f}}^\dag$, $\hat v_{{\rm L}|{\bf k}_{i} - {\bf k}_{f}}$, or $\hat v_{{\rm L}|{\bf k}_{i} - {\bf k}_{f}}^\dag$.
We consider two limiting cases: CW excitations, when $n_{{\rm U}|{\bf k}_i}(t) = n_{{\rm U}|{\bf k}_i} = const$, and pulsed excitation with short pulse such that $n_{{\rm U}|{\bf k}_i}(t) = n_{{\rm U}|{\bf k}_i}(0) e^{-\gamma_{s_{\rm U}|{\bf k}_i}t}$, where $n_{{\rm U}|{\bf k}_i}(0)$ is the number of initially excited upper polaritons. 
Finally, we will discuss the applicability of the linearized approach.

\subsection{CW excitation}

To describe the quantum dynamics of the system in the CW excitation regime, we use Hamiltonian~(\ref{H_OM_lin_all_modes_1}) and obtain the following Heisenberg--Langevin equation for each wave vector ${\bf k}_f$ separately
\begin{equation} \label{eq for A}
\frac{d \hat {\bf A}(t)}{dt} = \mathcal{M} \hat {\bf A}(t) + \hat {\bf F}(t),
\end{equation}
where $\hat {\bf A}(t)$ is the vector of exciton-polariton and phonon-polariton operators
\begin{equation} \label{A def}
\hat {\bf A}(t) 
= 
\left( 
\hat s_{{\rm L}|{\bf k}_{f}}(t) e^{i\omega_{s_{\rm U}|{\bf k}_{i}} t}, 
\hat s^\dag_{{\rm L}|{\bf k}_{f}}(t) e^{-i\omega_{s_{\rm U}|{\bf k}_{i}} t},
\hat v_{{\rm U}|{\bf k}_i - {\bf k}_f}(t), 
\hat v^\dag_{{\rm U}|{\bf k}_i - {\bf k}_f}(t), 
\hat v_{{\rm L}|{\bf k}_i - {\bf k}_f}(t), 
\hat v^\dag_{{\rm L}|{\bf k}_i - {\bf k}_f}(t) 
\right)^{\rm T},
\end{equation}
and $\hat {\bf F}(t)$ is the vector of noise operators
\begin{equation} \label{F def}
\hat {\bf F}(t) 
=
\left( 
\hat f_{s_{\rm L}|{\bf k}_{f}}(t) e^{i\omega_{s_{\rm U}|{\bf k}_{i}} t}, 
\hat f^\dag_{s_{\rm L}|{\bf k}_{f}}(t) e^{-i\omega_{s_{\rm U}|{\bf k}_{i}} t}, 
\hat f_{v_{\rm U}|{\bf k}_i - {\bf k}_f}(t), 
\hat f^\dag_{v_{\rm U}|{\bf k}_i - {\bf k}_f}(t), 
\hat f_{v_{\rm L}|{\bf k}_i - {\bf k}_f}(t), 
\hat f^\dag_{v_{\rm L}|{\bf k}_i - {\bf k}_f}(t) 
\right)^{\rm T}.
\end{equation}
The noise operators in $\hat {\bf F}(t)$ have zero mean $\langle \hat {\bf F}(t) \rangle = 0$ and are delta-correlated, $\langle
\hat {\bf F}(t_1) \hat {\bf F}^{\rm T}(t_2)
\rangle
=
\mathcal{D}_F\delta(t_1-t_2)$~\cite{gardiner2004quantum, coffey2012langevin}
\begin{equation}
\mathcal{D}_F =
\begin{pmatrix}
\mathcal D_{s_{\rm L}|{\bf k}_{f}} & 0 & 0 \\
0 & \mathcal D_{v_{\rm U}|{\bf k}_i - {\bf k}_f} & 0 \\
0 & 0 & \mathcal D_{v_{\rm L}|{\bf k}_i - {\bf k}_f}  
\end{pmatrix},
\end{equation}
\begin{equation}
\mathcal D_{\alpha | {\bf q}} =
\begin{pmatrix}
0 & \gamma_{\alpha|{\bf q}} (1 + n^{\rm th}_{\alpha|{\bf q}}) \\
\gamma_{\alpha|{\bf q}} n^{\rm th}_{\alpha|{\bf q}} & 0  
\end{pmatrix},
\end{equation}
$n^{\rm th}_{\alpha|{\bf q}} = (e^{\hbar \omega_{\alpha|{\bf q}}/k_BT} - 1)^{-1}$, $k_B$ is the Boltzmann constant, and $T$ is the ambient temperature.

The matrix of the differential equation~(\ref{eq for A}) is
\begin{equation} \label{matrix of the linearized system}
\mathcal{M} = 
\begin{pmatrix}
\mathcal{L}_{s_{\rm L}|{\bf k}_{f}} + \mathcal{J} \omega_{s_{\rm U}|{\bf k}_{i}} & \mathcal{G}_{{\rm U}|{\bf k}_{f}{\bf k}_{i}} & \mathcal{G}_{{\rm L}|{\bf k}_{f}{\bf k}_{i}} \\
\mathcal{G}_{{\rm U}|{\bf k}_{f}{\bf k}_{i}} & \mathcal{L}_{v_{\rm U}|{\bf k}_i - {\bf k}_f} & 0 \\
\mathcal{G}_{{\rm L}|{\bf k}_{f}{\bf k}_{i}} & 0 & \mathcal{L}_{v_{\rm L}|{\bf k}_i - {\bf k}_f}  
\end{pmatrix},
\end{equation}
where
\begin{equation}
\mathcal{J} = 
\begin{pmatrix}
i & 0 \\
0 & -i
\end{pmatrix},
\end{equation}
\begin{equation}
\mathcal{L}_{\alpha|{\bf q}}  = 
\begin{pmatrix}
-i \omega_{\alpha|{\bf q}}-\frac{\gamma_{\alpha|{\bf q}}}{2} & 0 \\
0 & i\omega_{\alpha|{\bf q}} - \frac{\gamma_{\alpha|{\bf q}}}{2} 
\end{pmatrix},
\end{equation}
\begin{equation}
\mathcal{G}_{{\rm U/L}|{\bf k}_{f}{\bf k}_{i}}  = 
\begin{pmatrix}
0 & -i g_{{\rm U/L}|{\bf k}_{f}{\bf k}_{i}} \sqrt{n_{{\rm U}|{\bf k}_i} } \\
i g_{{\rm U/L}|{\bf k}_{f}{\bf k}_{i}}^* \sqrt{n_{{\rm U}|{\bf k}_i} } & 0
\end{pmatrix}.
\end{equation}

The solution of Eq.~(\ref{eq for A}) is $\hat {\bf A}(t) 
=
\int_{-\infty}^t 
e^{\mathcal{M}(t-t')}
\hat {\bf F}(t')
dt'$.
We use this solution to obtain all the paired correlations of the operators listed in Eq.~(\ref{A def}).
\begin{equation} \label{matrix of paired correlators}
\langle \hat {\bf A}(t_1) \hat {\bf A}^{\rm T}(t_2) \rangle
=
\int\limits_{-\infty}^{\min(t_1,t_2)} 
e^{\mathcal{M}(t_1-t')}
\mathcal{D}_F
e^{\mathcal{M}^{\rm T}(t_2-t')}
dt'
.
\end{equation}
It is evident that $\langle \hat s_{{\rm L}|{\bf k}_{f}}(t_1) \hat v_{{\rm U}|{\bf k}_i - {\bf k}_f}(t_2) \rangle e^{i\omega_{s_{\rm U}|{\bf k}_{i}} t_1}$ is the first row, third column of the matrix $\langle \hat {\bf A}(t_1) \hat {\bf A}^{\rm T}(t_2) \rangle$; $\langle \hat s_{{\rm L}|{\bf k}_{f}}(t_1) \hat v_{{\rm L}|{\bf k}_i - {\bf k}_f}(t_2) \rangle e^{i\omega_{s_{\rm U}|{\bf k}_{i}} t_1}$ is the first row, fifth column; $\langle \hat s^\dag_{{\rm L}|{\bf k}_{f}}(t_1) \hat s_{{\rm L}|{\bf k}_{f}}(t_2) \rangle e^{i\omega_{s_{\rm U}|{\bf k}_{i}} (t_2-t_1)}$ is the second row, first column; $\langle \hat v^\dag_{{\rm U}|{\bf k}_i - {\bf k}_f}(t_1) \hat v_{{\rm U}|{\bf k}_i - {\bf k}_f}(t_2) \rangle$ is the forth row, third column; and $\langle \hat v^\dag_{{\rm L}|{\bf k}_i - {\bf k}_f}(t_1) \hat v_{{\rm L}|{\bf k}_i - {\bf k}_f}(t_2) \rangle$ is the sixth row, fifth column of this matrix.

From Eq.~(\ref{matrix of paired correlators}), we obtain the linear equation for the equal time correlation matrix $\langle \hat {\bf A}(t) \hat {\bf A}^{\rm T}(t) \rangle$
\begin{equation}
\frac{d\langle \hat {\bf A}(t) \hat {\bf A}^{\rm T}(t) \rangle}{dt}
=
\mathcal{M} \langle \hat {\bf A}(t) \hat {\bf A}^{\rm T}(t) \rangle
+
\langle \hat {\bf A}(t) \hat {\bf A}^{\rm T}(t) \rangle \mathcal{M}^{\rm T}
+
\mathcal{D}_F,
\end{equation}
and the corresponding stationary solution for $\langle \hat {\bf A}(t) \hat {\bf A}^{\rm T}(t) \rangle$ at $t \to +\infty$, that we denote as $\langle \hat {\bf A} \hat {\bf A}^{\rm T} \rangle$
\begin{equation} \label{stationary eq. for AA}
\mathcal{M} \langle \hat {\bf A} \hat {\bf A}^{\rm T} \rangle
+
\langle \hat {\bf A} \hat {\bf A}^{\rm T} \rangle \mathcal{M}^{\rm T}
+
\mathcal{D}_F
=
0.
\end{equation}

The solution~(\ref{matrix of paired correlators}) also allow us to connect equal time correlation matrix $\langle \hat {\bf A}(t) \hat {\bf A}^{\rm T}(t) \rangle$ and two-time correlation matrix $\langle \hat {\bf A}(t) \hat {\bf A}^{\rm T}(t+\tau) \rangle$ as follows
\begin{equation}
\langle \hat {\bf A}(t) \hat {\bf A}^{\rm T}(t+\tau) \rangle
=
\langle \hat {\bf A}(t) \hat {\bf A}^{\rm T}(t) \rangle
e^{\mathcal{M}^{\rm T}\tau}.
\end{equation}
Below the threshold of optomechanical instability, we can use the approximate expression
\begin{equation}
\langle \hat {\bf A}(t) \hat {\bf A}^{\rm T}(t+\tau) \rangle
\approx
\langle \hat {\bf A}(t) \hat {\bf A}^{\rm T}(t) \rangle
e^{\mathcal{M}_{\rm free}^{\rm T}\tau},
\end{equation}
with
\begin{equation} \label{non-interacting matrix of the linearized system}
\mathcal{M}_{\rm free} = 
\begin{pmatrix}
\mathcal{L}_{s_{\rm L}|{\bf k}_{f}} + \mathcal{J} \omega_{s_{\rm U}|{\bf k}_{i}} & 0 & 0 \\
0 & \mathcal{L}_{v_{\rm U}|{\bf k}_i - {\bf k}_f} & 0 \\
0 & 0 & \mathcal{L}_{v_{\rm L}|{\bf k}_i - {\bf k}_f}  
\end{pmatrix},
\end{equation}

\subsection{Pulsed excitation}
To describe the quantum dynamics of the system in the pulsed excitation regime, we use Hamiltonian~(\ref{H_OM_lin_all_modes_2}) separately for each wave vector ${\bf k}_f$
\begin{multline}  \label{H_OM_lin_all_modes_3}
\hat H_{{\rm OM}|{\bf k}_f} (t)
\approx
\hbar \left( \omega_{s_{\rm L}|{\bf k}_f} - \omega_{s_{\rm L}|{\bf k}_i} \right)
\hat S^\dag_{{\rm L}|{\bf k}_f}
\hat S_{{\rm L}|{\bf k}_f}
+
\hbar \omega_{v_{\rm U}|{\bf k}_i - {\bf k}_f}
\hat v_{{\rm U}|{\bf k}_i - {\bf k}_f}^\dag 
\hat v_{{\rm U}|{\bf k}_i - {\bf k}_f}
+
\hbar \omega_{v_{\rm L}|{\bf k}_i - {\bf k}_f}
\hat v_{{\rm L}|{\bf k}_i - {\bf k}_f}^\dag 
\hat v_{{\rm L}|{\bf k}_i - {\bf k}_f}
\\
+
\left[
\hbar 
g_{{\rm U}|{\bf k}_f{\bf k}_i}
\sqrt{n_{{\rm U}|{\bf k}_i}(0)}
e^{-\gamma_{s_{\rm U}|{\bf k}_i}t/2}
\hat v_{{\rm U}|{\bf k}_i-{\bf k}_f}^\dag 
\hat S_{{\rm L}|{\bf k}_f}^\dag
+
h.c.
\right]
+
\left[
\hbar 
g_{{\rm L}|{\bf k}_f{\bf k}_i}
\sqrt{n_{{\rm U}|{\bf k}_i}(0)}
e^{-\gamma_{s_{\rm U}|{\bf k}_i}t/2}
\hat v_{{\rm L}|{\bf k}_i-{\bf k}_f}^\dag 
\hat S_{{\rm L}|{\bf k}_f}^\dag
+
h.c.
\right].
\end{multline}
and write the master equation for the density matrix, $\hat \rho$, of upper exciton-polarions with the wave vector ${\bf k}_f$, upper and lower phonon-polaritons with the wave vector ${\bf k}_i - {\bf k}_f$
\begin{multline} \label{master equation}
\frac{d \hat \rho(t)}{dt} 
=
-\frac{i}{\hbar}\left[ \hat H(t), \hat \rho(t) \right]
+
L_{\sqrt{\gamma_{s_{\rm L}|{\bf k}_f}}\hat S_{{\rm L}|{\bf k}_f}}
\left[ \hat\rho(t) \right]
\\
+
L_{\sqrt{\gamma_{v_{\rm U}|{\bf k}_i-{\bf k}_f}(1+n^{\rm th}_{v_{\rm U}|{\bf k}_i-{\bf k}_f})}\hat v_{{\rm U}|{\bf k}_i-{\bf k}_f}}
\left[ \hat\rho(t) \right]
+
L_{\sqrt{\gamma_{v_{\rm U}|{\bf k}_i-{\bf k}_f}n^{\rm th}_{v_{\rm U}|{\bf k}_i-{\bf k}_f}}\hat v_{{\rm U}|{\bf k}_i-{\bf k}_f}^\dag}
\left[ \hat\rho(t) \right]
\\
+
L_{\sqrt{\gamma_{v_{\rm L}|{\bf k}_i-{\bf k}_f}(1+n^{\rm th}_{v_{\rm L}|{\bf k}_i-{\bf k}_f})}\hat v_{{\rm L}|{\bf k}_i-{\bf k}_f}}
\left[ \hat\rho(t) \right]
+
L_{\sqrt{\gamma_{v_{\rm L}|{\bf k}_i-{\bf k}_f}n^{\rm th}_{v_{\rm L}|{\bf k}_i-{\bf k}_f}}\hat v_{{\rm L}|{\bf k}_i-{\bf k}_f}^\dag}
\left[ \hat\rho(t) \right],
\end{multline}
where $n^{\rm th}_{\alpha|{\bf q}} = (e^{\hbar \omega_{\alpha|{\bf q}}/k_BT} - 1)^{-1}$, $k_B$ is the Boltzmann constant, $T$ is the ambient temperature, and
\begin{equation}
L_{\hat A} 
\left[ \hat\rho(t) \right] 
= 
\hat A \hat\rho(t) \hat A^\dag 
-
\frac{1}{2} \hat A^\dag \hat A \hat\rho(t)
-
\frac{1}{2} \hat\rho(t) \hat A^\dag \hat A.
\end{equation}
We solve master equation~(\ref{master equation}) numerically and obtain covariance matrix for the operators $\hat s_{{\rm L}|{\bf k}_{f}}$, $\hat s_{{\rm L}|{\bf k}_{f}}^\dag$, $\hat v_{{\rm U}|{\bf k}_{i} - {\bf k}_{f}}$, $\hat v_{{\rm U}|{\bf k}_{i} - {\bf k}_{f}}^\dag$, $\hat v_{{\rm L}|{\bf k}_{i} - {\bf k}_{f}}$, and $\hat v_{{\rm L}|{\bf k}_{i} - {\bf k}_{f}}^\dag$, we also obtain the second-order cross-correlation function $g^{(2)}_{\rm Vis-IR}(t,0)$.

\subsection{Applicability of linearized approach}

In the case of CW pumping, the linearized theory is applicable below the threshold of optomechanical instability.
This puts limits on the number of the directly excited exciton-polaritons, $n_{s_{\rm U}|{\bf k}_{i}}$.
The optomechanical instability onsets when the real part of at least one of the eigenvectors of the matrix of the linearized system~(\ref{matrix of the linearized system}) crosses zero.
After the real part of one of the eigenvectors of matrix~(\ref{matrix of the linearized system}) become greater than zero, we cannot use the linearized Hamiltonian~(\ref{H_optomech_f}) and Eq.~(\ref{eq for A}) and should consider the dynamics of the full quantum system given by Hamiltonian~(\ref{H_strong}).
Fig.~(\ref{fig:eigenvalues}) shows the real part of the eigenvalues of the linearized matrix~(\ref{matrix of the linearized system}).
As we can see for the parameters corresponding to the red dot in Fig.~\ref{fig:correlations_CW}, the linearized theory applicable for $n_{s_{\rm U}|{\bf k}_{i}} < 4\cdot 10^6$.

In the case of pulsed pumping with short pulses, the linearized theory is applicable when the rate of stimulated transitions from the upper exciton-polariton state with wave vector ${\bf k}_i$ is lower than the dissipation rate $\gamma_{s_{\rm U}|{\bf k}_i}$.
Above the threshold of optomechanical instability, the rate of these stimulated transitions is $\max Re(\lambda)$, where $\lambda$ is the eigenfrequencies of the matrix~(\ref{matrix of the linearized system}) with $n_{s_{\rm U}|{\bf k}_{i}}$ replaced by $n_{s_{\rm U}|{\bf k}_{i}}(0)$.
Fig.~(\ref{fig:eigenvalues}) shows the real part of the eigenvalues of this linearized matrix with $x$-axis denoted as $n_{s_{\rm U}|{\bf k}_{i}}$ instead of $n_{s_{\rm U}|{\bf k}_{i}}(0)$.
As one can see, for the parameters corresponding to the red dot in Fig.~\ref{fig:correlations_CW} the linearized theory applicable for the initial occupations of the upper exciton-polariton state with the wave vector ${\bf k}_i$, $n_{s_{\rm U}|{\bf k}_{i}}(0) < 7\cdot 10^7$.

\begin{figure*}
\includegraphics[width=0.4\linewidth]{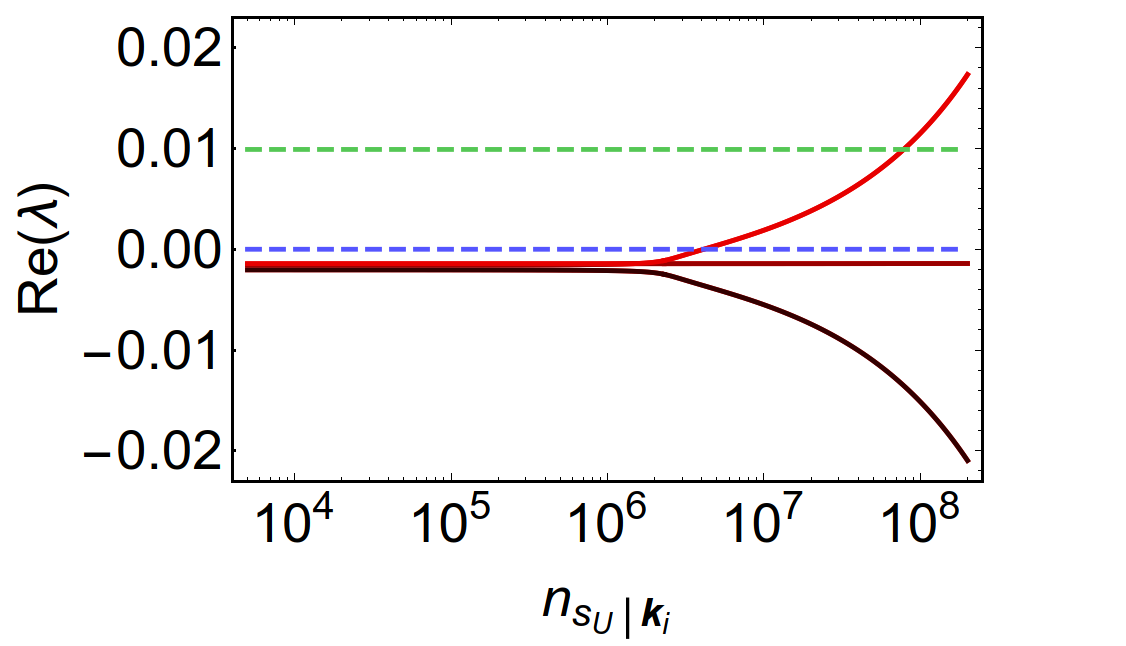}
\caption{
{\bf Real part of the eigenvalues of the matrix of the linearized system, $\mathcal{M}$.}
Real part of the eigenvalues of the linearized system matrix~(\ref{matrix of the linearized system}).
The parameters are specified in Fig.~\ref{fig:correlations_CW} the wave vectors ${\bf k}_i$ and ${\bf k}_f$ correspond to the red dot in Fig.~\ref{fig:correlations_CW}.
Blue dashed line marks ${\rm Re}(\lambda) = 0$ and green dashed line marks ${\rm Re}(\lambda) = \gamma_{s_{\rm U}|{\bf k}_i}$.
}
    \label{fig:eigenvalues}
\end{figure*}

\section{Two-partite entanglement between exciton-polaritons and phonon-polaritons, and between emitted visible and mid-IR light} \label{appendix: logneg}

In this Appendix, we consider the CW excitation regime.
As the results of the Appendix~\ref{appendix: correlations}, we obtained the correlation matrix~(\ref{matrix of paired correlators}).
From Eq.~(\ref{matrix of paired correlators}) we can easily reconstruct a steady state correlation matrix $\langle \hat{\bf a} \hat{\bf a}^{\rm T} \rangle$, where $\hat{\bf a}$ is 
\begin{equation} \label{a def}
\hat {\bf a}
= 
\left( 
\hat s_{{\rm L}|{\bf k}_{f}}, 
\hat s^\dag_{{\rm L}|{\bf k}_{f}},
\hat v_{{\rm U}|{\bf k}_i - {\bf k}_f}, 
\hat v^\dag_{{\rm U}|{\bf k}_i - {\bf k}_f}, 
\hat v_{{\rm L}|{\bf k}_i - {\bf k}_f}, 
\hat v^\dag_{{\rm L}|{\bf k}_i - {\bf k}_f} 
\right)^{\rm T}.
\end{equation}
The explicit form of the correlation matrix $\langle \hat{\bf a} \hat{\bf a}^{\rm T} \rangle$ is
\begin{multline} \label{aa matrix}
\langle \hat{\bf a} \hat{\bf a}^{\rm T} \rangle
= \\
\begin{pmatrix}
\langle \hat s_{{\rm L}|{\bf k}_f} \hat s_{{\rm L}|{\bf k}_f} \rangle & 
\langle \hat s_{{\rm L}|{\bf k}_f} \hat s_{{\rm L}|{\bf k}_f}^\dag \rangle & 
\langle \hat s_{{\rm L}|{\bf k}_f} \hat v_{{\rm U}|{\bf k}_i-{\bf k}_f} \rangle & 
\langle \hat s_{{\rm L}|{\bf k}_f} \hat v_{{\rm U}|{\bf k}_i-{\bf k}_f}^\dag \rangle & 
\langle \hat s_{{\rm L}|{\bf k}_f} \hat v_{{\rm L}|{\bf k}_i-{\bf k}_f} \rangle & 
\langle \hat s_{{\rm L}|{\bf k}_f} \hat v_{{\rm L}|{\bf k}_i-{\bf k}_f}^\dag \rangle
\\
\langle \hat s_{{\rm L}|{\bf k}_f}^\dag \hat s_{{\rm L}|{\bf k}_f} \rangle & 
\langle \hat s_{{\rm L}|{\bf k}_f}^\dag \hat s_{{\rm L}|{\bf k}_f}^\dag \rangle & 
\langle \hat s_{{\rm L}|{\bf k}_f}^\dag \hat v_{{\rm U}|{\bf k}_i-{\bf k}_f} \rangle & 
\langle \hat s_{{\rm L}|{\bf k}_f}^\dag \hat v_{{\rm U}|{\bf k}_i-{\bf k}_f}^\dag \rangle & 
\langle \hat s_{{\rm L}|{\bf k}_f}^\dag \hat v_{{\rm L}|{\bf k}_i-{\bf k}_f} \rangle & 
\langle \hat s_{{\rm L}|{\bf k}_f}^\dag \hat v_{{\rm L}|{\bf k}_i-{\bf k}_f}^\dag \rangle
\\
\langle \hat v_{{\rm U}|{\bf k}_i-{\bf k}_f} \hat s_{{\rm L}|{\bf k}_f} \rangle & 
\langle \hat v_{{\rm U}|{\bf k}_i-{\bf k}_f} \hat s_{{\rm L}|{\bf k}_f}^\dag \rangle & 
\langle \hat v_{{\rm U}|{\bf k}_i-{\bf k}_f} \hat v_{{\rm U}|{\bf k}_i-{\bf k}_f} \rangle & 
\langle \hat v_{{\rm U}|{\bf k}_i-{\bf k}_f} \hat v_{{\rm U}|{\bf k}_i-{\bf k}_f}^\dag \rangle & 
\langle \hat v_{{\rm U}|{\bf k}_i-{\bf k}_f} \hat v_{{\rm L}|{\bf k}_i-{\bf k}_f} \rangle & 
\langle \hat v_{{\rm U}|{\bf k}_i-{\bf k}_f} \hat v_{{\rm L}|{\bf k}_i-{\bf k}_f}^\dag \rangle
\\
\langle \hat v_{{\rm U}|{\bf k}_i-{\bf k}_f}^\dag  \hat s_{{\rm L}|{\bf k}_f} \rangle & 
\langle \hat v_{{\rm U}|{\bf k}_i-{\bf k}_f}^\dag  \hat s_{{\rm L}|{\bf k}_f}^\dag \rangle & 
\langle \hat v_{{\rm U}|{\bf k}_i-{\bf k}_f}^\dag \hat v_{{\rm U}|{\bf k}_i-{\bf k}_f} \rangle & 
\langle \hat v_{{\rm U}|{\bf k}_i-{\bf k}_f}^\dag \hat v_{{\rm U}|{\bf k}_i-{\bf k}_f}^\dag \rangle & 
\langle \hat v_{{\rm U}|{\bf k}_i-{\bf k}_f}^\dag \hat v_{{\rm L}|{\bf k}_i-{\bf k}_f} \rangle & 
\langle \hat v_{{\rm U}|{\bf k}_i-{\bf k}_f}^\dag \hat v_{{\rm L}|{\bf k}_i-{\bf k}_f}^\dag \rangle
\\
\langle \hat v_{{\rm L}|{\bf k}_i-{\bf k}_f} \hat s_{{\rm L}|{\bf k}_f} \rangle & 
\langle \hat v_{{\rm L}|{\bf k}_i-{\bf k}_f} \hat s_{{\rm L}|{\bf k}_f}^\dag \rangle & 
\langle \hat v_{{\rm L}|{\bf k}_i-{\bf k}_f} \hat v_{{\rm U}|{\bf k}_i-{\bf k}_f} \rangle & 
\langle \hat v_{{\rm L}|{\bf k}_i-{\bf k}_f} \hat v_{{\rm U}|{\bf k}_i-{\bf k}_f}^\dag \rangle & 
\langle \hat v_{{\rm L}|{\bf k}_i-{\bf k}_f} \hat v_{{\rm L}|{\bf k}_i-{\bf k}_f} \rangle & 
\langle \hat v_{{\rm L}|{\bf k}_i-{\bf k}_f} \hat v_{{\rm L}|{\bf k}_i-{\bf k}_f}^\dag \rangle
\\
\langle \hat v_{{\rm L}|{\bf k}_i-{\bf k}_f}^\dag  \hat s_{{\rm L}|{\bf k}_f} \rangle & 
\langle \hat v_{{\rm L}|{\bf k}_i-{\bf k}_f}^\dag  \hat s_{{\rm L}|{\bf k}_f}^\dag \rangle & 
\langle \hat v_{{\rm L}|{\bf k}_i-{\bf k}_f}^\dag \hat v_{{\rm U}|{\bf k}_i-{\bf k}_f} \rangle & 
\langle \hat v_{{\rm L}|{\bf k}_i-{\bf k}_f}^\dag \hat v_{{\rm U}|{\bf k}_i-{\bf k}_f}^\dag \rangle & 
\langle \hat v_{{\rm L}|{\bf k}_i-{\bf k}_f}^\dag \hat v_{{\rm L}|{\bf k}_i-{\bf k}_f} \rangle & 
\langle \hat v_{{\rm L}|{\bf k}_i-{\bf k}_f}^\dag \hat v_{{\rm L}|{\bf k}_i-{\bf k}_f}^\dag \rangle    
\end{pmatrix}.
\end{multline}
We use this matrix to analyze two-partite entanglement between lower exciton-polaritons and phonon-polaritons, as well as two-partite entanglement between emitted visible light and mid-IR light.

\subsection{Entanglement between lower exciton-polaritons and phonon-polaritons}
We analyze two-partite entanglement between exciton-polaritons and phonon-polaritons by calculating the logarithmic negativity between lower exciton-polaritons with the wave vector ${\bf k}_f$ and upper or lower phonon-polaritons with the wave vector ${\bf k}_i-{\bf k}_f$.
We transit from the correlation matrix $\langle \hat{\bf a} \hat{\bf a}^{\rm T} \rangle$ to the standard covariance matrix $\langle \hat {\bf x} \hat {\bf x}^{\rm T} \rangle$ with the transformation
\begin{equation} \label{aa to xx}
\langle \hat {\bf x} \hat {\bf x}^{\rm T} \rangle
=
U_{a \to x}
\langle \hat{\bf a} \hat{\bf a}^{\rm T} \rangle
U_{a \to x}^{\rm T},
\end{equation}
where 
\begin{equation}
U_{a \to x} 
= 
\frac{1}{\sqrt{2}}
\begin{pmatrix}
1 & 
1 & 
0 & 
0 & 
0 & 
0
\\
-i & 
i & 
0 & 
0 & 
0 & 
0
\\
0 & 
0 & 
1 & 
1 & 
0 & 
0
\\
0 & 
0 & 
-i & 
i & 
0 & 
0
\\
0 & 
0 & 
0 & 
0 & 
1 & 
1 
\\
0 & 
0 & 
0 & 
0 & 
-i & 
i
\end{pmatrix}.
\end{equation}
The explicit form of the vector $\hat {\bf x}$ is
\begin{multline} \label{x explicit}
\hat {\bf x}
= 
\left( 
\frac{ \hat s_{{\rm L}|{\bf k}_{f}} + \hat s^\dag_{{\rm L}|{\bf k}_{f}} }{\sqrt{2}}
, 
\frac{ \hat s_{{\rm L}|{\bf k}_{f}} - \hat s^\dag_{{\rm L}|{\bf k}_{f}} }{i\sqrt{2}}
,
\frac{ \hat v_{{\rm U}|{\bf k}_i - {\bf k}_f} + \hat v_{{\rm U}|{\bf k}_i - {\bf k}_f}^\dag }{\sqrt{2}}
, 
\frac{ \hat v_{{\rm U}|{\bf k}_i - {\bf k}_f} - \hat v_{{\rm U}|{\bf k}_i - {\bf k}_f}^\dag }{i\sqrt{2}}
, 
\right. \\ \left.
\frac{ \hat v_{{\rm L}|{\bf k}_i - {\bf k}_f} + \hat v_{{\rm L}|{\bf k}_i - {\bf k}_f}^\dag }{\sqrt{2}}
, 
\frac{ \hat v_{{\rm L}|{\bf k}_i - {\bf k}_f} - \hat v_{{\rm L}|{\bf k}_i - {\bf k}_f}^\dag }{i\sqrt{2}}
\right)^{\rm T},
\end{multline}
which, for the sake of briefness, we also write in the form
\begin{equation} \label{x explicit}
\hat {\bf x}
= 
\left( 
\hat x_{s_{\rm L}|{\bf k}_f}
, 
\hat p_{s_{\rm L}|{\bf k}_f}
,
\hat x_{v_{\rm U}|{\bf k}_i - {\bf k}_f}
, 
\hat p_{v_{\rm U}|{\bf k}_i - {\bf k}_f}
, 
\hat x_{v_{\rm L}|{\bf k}_i - {\bf k}_f}
, 
\hat p_{v_{\rm L}|{\bf k}_i - {\bf k}_f}
\right)^{\rm T}.
\end{equation}

To analyze two partite entanglement between lower exciton-polaritons and upper phonon-polaritons we consider the reduced covariance matrix 
\begin{equation}
\langle \hat {\bf x} \hat {\bf x}^{\rm T} \rangle_{s_{\rm L}v_{\rm U}}
=
\begin{pmatrix}
\langle \hat x_{s_{\rm L}|{\bf k}_f} \hat x_{s_{\rm L}|{\bf k}_f} \rangle & 
\langle \hat x_{s_{\rm L}|{\bf k}_f} \hat p_{s_{\rm L}|{\bf k}_f} \rangle & 
\langle \hat x_{s_{\rm L}|{\bf k}_f} \hat x_{v_{\rm U}|{\bf k}_i - {\bf k}_f} \rangle & 
\langle \hat x_{s_{\rm L}|{\bf k}_f} \hat p_{v_{\rm U}|{\bf k}_i - {\bf k}_f} \rangle 
\\
\langle \hat p_{s_{\rm L}|{\bf k}_f} \hat x_{s_{\rm L}|{\bf k}_f} \rangle & 
\langle \hat p_{s_{\rm L}|{\bf k}_f} \hat p_{s_{\rm L}|{\bf k}_f} \rangle & 
\langle \hat p_{s_{\rm L}|{\bf k}_f} \hat x_{v_{\rm U}|{\bf k}_i - {\bf k}_f} \rangle & 
\langle \hat p_{s_{\rm L}|{\bf k}_f} \hat p_{v_{\rm U}|{\bf k}_i - {\bf k}_f} \rangle
\\
\langle \hat x_{v_{\rm U}|{\bf k}_i - {\bf k}_f} \hat x_{s_{\rm L}|{\bf k}_f} \rangle & 
\langle \hat x_{v_{\rm U}|{\bf k}_i - {\bf k}_f} \hat p_{s_{\rm L}|{\bf k}_f} \rangle & 
\langle \hat x_{v_{\rm U}|{\bf k}_i - {\bf k}_f} \hat x_{v_{\rm U}|{\bf k}_i - {\bf k}_f} \rangle & 
\langle \hat x_{v_{\rm U}|{\bf k}_i - {\bf k}_f} \hat p_{v_{\rm U}|{\bf k}_i - {\bf k}_f} \rangle 
\\
\langle \hat p_{v_{\rm U}|{\bf k}_i - {\bf k}_f} \hat x_{s_{\rm L}|{\bf k}_f} \rangle & 
\langle \hat p_{v_{\rm U}|{\bf k}_i - {\bf k}_f} \hat p_{s_{\rm L}|{\bf k}_f} \rangle & 
\langle \hat p_{v_{\rm U}|{\bf k}_i - {\bf k}_f} \hat x_{v_{\rm U}|{\bf k}_i - {\bf k}_f} \rangle & 
\langle \hat p_{v_{\rm U}|{\bf k}_i - {\bf k}_f} \hat p_{v_{\rm U}|{\bf k}_i - {\bf k}_f} \rangle 
\end{pmatrix}.
\end{equation}
and denote the symmetric part of this matrix as $\mathcal{R}_{s_{\rm L}v_{\rm U}} = (\langle \hat {\bf x} \hat {\bf x}^{\rm T} \rangle_{s_{\rm L}v_{\rm U}} + \langle \hat {\bf x} \hat {\bf x}^{\rm T} \rangle_{s_{\rm L}v_{\rm U}}^{\rm T})/2$, writing it in a block form
\begin{equation}
\mathcal{R}_{s_{\rm L}v_{\rm U}} 
=
\begin{pmatrix}
\mathcal{C}_{s_{\rm L}s_{\rm L}} & \mathcal{C}_{s_{\rm L}v_{\rm U}} \\
\mathcal{C}_{s_{\rm L}v_{\rm U}} & \mathcal{C}_{v_{\rm U}v_{\rm U}}  
\end{pmatrix},
\end{equation}
where
\begin{equation}
\mathcal{C}_{s_{\rm L}s_{\rm L}}
=
\frac{1}{2}
\begin{pmatrix}
2\langle \hat x_{s_{\rm L}|{\bf k}_f} \hat x_{s_{\rm L}|{\bf k}_f} \rangle & 
\langle \hat x_{s_{\rm L}|{\bf k}_f} \hat p_{s_{\rm L}|{\bf k}_f} \rangle
+
\langle \hat p_{s_{\rm L}|{\bf k}_f} \hat x_{s_{\rm L}|{\bf k}_f} \rangle
\\
\langle \hat x_{s_{\rm L}|{\bf k}_f} \hat p_{s_{\rm L}|{\bf k}_f} \rangle
+
\langle \hat p_{s_{\rm L}|{\bf k}_f} \hat x_{s_{\rm L}|{\bf k}_f} \rangle
& 
2\langle \hat p_{s_{\rm L}|{\bf k}_f} \hat p_{s_{\rm L}|{\bf k}_f} \rangle 
\end{pmatrix}
\end{equation}
\begin{equation}
\mathcal{C}_{v_{\rm U}v_{\rm U}}
=
\frac{1}{2}
\begin{pmatrix}
2\langle \hat x_{v_{\rm U}|{\bf k}_i - {\bf k}_f} \hat x_{v_{\rm U}|{\bf k}_i - {\bf k}_f} \rangle & 
\langle \hat x_{v_{\rm U}|{\bf k}_i - {\bf k}_f} \hat p_{v_{\rm U}|{\bf k}_i - {\bf k}_f} \rangle
+
\langle \hat p_{v_{\rm U}|{\bf k}_i - {\bf k}_f} \hat x_{v_{\rm U}|{\bf k}_i - {\bf k}_f} \rangle
\\
\langle \hat p_{v_{\rm U}|{\bf k}_i - {\bf k}_f} \hat x_{v_{\rm U}|{\bf k}_i - {\bf k}_f} \rangle
+
\langle \hat x_{v_{\rm U}|{\bf k}_i - {\bf k}_f} \hat p_{v_{\rm U}|{\bf k}_i - {\bf k}_f} \rangle
& 
2\langle \hat p_{v_{\rm U}|{\bf k}_i - {\bf k}_f} \hat p_{v_{\rm U}|{\bf k}_i - {\bf k}_f} \rangle
\end{pmatrix},
\end{equation}
\begin{equation}
\mathcal{C}_{s_{\rm L}v_{\rm U}}
=
\frac{1}{2}
\begin{pmatrix}
\langle \hat x_{s_{\rm L}|{\bf k}_f} \hat x_{v_{\rm U}|{\bf k}_i - {\bf k}_f} \rangle
+
\langle \hat x_{v_{\rm U}|{\bf k}_i - {\bf k}_f} \hat x_{s_{\rm L}|{\bf k}_f} \rangle
& 
\langle \hat x_{s_{\rm L}|{\bf k}_f} \hat p_{v_{\rm U}|{\bf k}_i - {\bf k}_f} \rangle
+
\langle \hat p_{v_{\rm U}|{\bf k}_i - {\bf k}_f} \hat x_{s_{\rm L}|{\bf k}_f} \rangle
\\
\langle \hat p_{s_{\rm L}|{\bf k}_f} \hat x_{v_{\rm U}|{\bf k}_i - {\bf k}_f} \rangle
+
\langle \hat x_{v_{\rm U}|{\bf k}_i - {\bf k}_f} \hat p_{s_{\rm L}|{\bf k}_f} \rangle
& 
\langle \hat p_{s_{\rm L}|{\bf k}_f} \hat p_{v_{\rm U}|{\bf k}_i - {\bf k}_f} \rangle
+
\langle \hat p_{v_{\rm U}|{\bf k}_i - {\bf k}_f} \hat p_{s_{\rm L}|{\bf k}_f} \rangle
\end{pmatrix},
\end{equation}
The logarithmic negativity for lower exciton-polaritons and phonon-polaritons, $E_N$, is
\begin{equation}
E_{N}
=
\max \left( 0, -\ln2|\xi| \right),
\end{equation}
where $\xi$ is the smallest solution of the equation~\cite{vidal2002computable}
\begin{equation}
\xi^4 
+ 
\xi^2 
\left( 
\det \mathcal{C}_{s_{\rm L}s_{\rm L}} 
+
\det \mathcal{C}_{v_{\rm U}v_{\rm U}} 
-
2 \det \mathcal{C}_{s_{\rm L}v_{\rm U}} 
\right) 
+ 
\det \mathcal{R}_{s_{\rm L}v_{\rm U}} 
=
0
\end{equation}

The analysis of the entanglement between lower exciton-polaritons and lower phonon-polaritons can be considered in a similar manner.

\subsection{Entanglement between visible light and mid-IR light} 
To analyze two-partite entanglement between emitted visible and mid-IR light, we apply the transformation matrix
\begin{equation}
U_\text{Vis-IR} 
= 
\begin{pmatrix}
1 & 
0 & 
0 & 
0 & 
0 & 
0
\\
0 & 
1 & 
0 & 
0 & 
0 & 
0
\\
0 & 
0 & 
\sin \phi_{{\bf k}_i-{\bf k}_f} & 
0 & 
\cos \phi_{{\bf k}_i-{\bf k}_f} & 
0
\\
0 & 
0 & 
0 & 
\sin \phi_{{\bf k}_i-{\bf k}_f} & 
0 & 
\cos \phi_{{\bf k}_i-{\bf k}_f}
\end{pmatrix}.
\end{equation}
which reduces phonon-polariton part if the matrix $\langle \hat{\bf a} \hat{\bf a}^{\rm T} \rangle$ to the correlations between operator $\hat s_{{\rm L}|{\bf k}_f}$ representing the emitted visible light and operator $\hat a_{{\rm IR}|{\bf k}_i-{\bf k}_f}$ representing all the light emitted by the mid-IR cavity
\begin{equation}
\langle \hat{\bf a} \hat{\bf a}^{\rm T} \rangle_\text{Vis-IR}
=
\begin{pmatrix}
\langle \hat s_{{\rm L}|{\bf k}_f} \hat s_{{\rm L}|{\bf k}_f} \rangle & 
\langle \hat s_{{\rm L}|{\bf k}_f} \hat s_{{\rm L}|{\bf k}_f}^\dag \rangle & 
\langle \hat s_{{\rm L}|{\bf k}_f} \hat a_{{\rm IR}|{\bf k}_i-{\bf k}_f} \rangle & 
\langle \hat s_{{\rm L}|{\bf k}_f} \hat a_{{\rm IR}|{\bf k}_i-{\bf k}_f}^\dag \rangle 
\\
\langle \hat s_{{\rm L}|{\bf k}_f}^\dag \hat s_{{\rm L}|{\bf k}_f} \rangle & 
\langle \hat s_{{\rm L}|{\bf k}_f}^\dag \hat s_{{\rm L}|{\bf k}_f}^\dag \rangle & 
\langle \hat s_{{\rm L}|{\bf k}_f}^\dag \hat a_{{\rm IR}|{\bf k}_i-{\bf k}_f} \rangle & 
\langle \hat s_{{\rm L}|{\bf k}_f}^\dag \hat a_{{\rm IR}|{\bf k}_i-{\bf k}_f}^\dag \rangle 
\\
\langle \hat a_{{\rm IR}|{\bf k}_i-{\bf k}_f} \hat s_{{\rm L}|{\bf k}_f} \rangle & 
\langle \hat a_{{\rm IR}|{\bf k}_i-{\bf k}_f} \hat s_{{\rm L}|{\bf k}_f}^\dag \rangle & 
\langle \hat a_{{\rm IR}|{\bf k}_i-{\bf k}_f} \hat a_{{\rm IR}|{\bf k}_i-{\bf k}_f} \rangle & 
\langle \hat a_{{\rm IR}|{\bf k}_i-{\bf k}_f} \hat a_{{\rm IR}|{\bf k}_i-{\bf k}_f}^\dag \rangle 
\\
\langle \hat a_{{\rm IR}|{\bf k}_i-{\bf k}_f}^\dag \hat s_{{\rm L}|{\bf k}_f} \rangle & 
\langle \hat a_{{\rm IR}|{\bf k}_i-{\bf k}_f}^\dag \hat s_{{\rm L}|{\bf k}_f}^\dag \rangle & 
\langle \hat a_{{\rm IR}|{\bf k}_i-{\bf k}_f}^\dag \hat a_{{\rm IR}|{\bf k}_i-{\bf k}_f} \rangle & 
\langle \hat a_{{\rm IR}|{\bf k}_i-{\bf k}_f}^\dag \hat a_{{\rm IR}|{\bf k}_i-{\bf k}_f}^\dag \rangle 
\end{pmatrix}.
\end{equation}
To calculate logarithmic negativity, we adjust correlation matrix $\langle \hat{\bf a} \hat{\bf a}^{\rm T} \rangle_\text{Vis-IR}$ to include background noises in visible spectral range, $N^{(\rm bg)}_{\rm Vis}$, and mid-IR spectral range, $N^{(\rm bg)}_{\rm IR}$, and then we transit from the  to the standard covariance matrix $\langle \hat {\bf x} \hat {\bf x}^{\rm T} \rangle_\text{Vis-IR}$
\begin{equation} \label{aa to xx}
\langle \hat {\bf x} \hat {\bf x}^{\rm T} \rangle_\text{Vis-IR}
=
V_{a \to x}
\left[
\langle \hat{\bf a} \hat{\bf a}^{\rm T} \rangle_\text{Vis-IR}
+
\begin{pmatrix}
0 & 
N^{(\rm bg)}_{\rm Vis} & 
0 & 
0 
\\
N^{(\rm bg)}_{\rm Vis} & 
0 & 
0 & 
0
\\
0 & 
0 & 
0 & 
N^{(\rm bg)}_{\rm IR} 
\\
0 & 
0 & 
N^{(\rm bg)}_{\rm IR} & 
0 
\end{pmatrix}
\right]
V_{a \to x}^{\rm T},
\end{equation}
with the transformation matrix
\begin{equation}
V_{a \to x} 
= 
\frac{1}{\sqrt{2}}
\begin{pmatrix}
1 & 
1 & 
0 & 
0 & 
\\
-i & 
i & 
0 & 
0 & 
\\
0 & 
0 & 
1 & 
1 & 
\\
0 & 
0 & 
-i & 
i 
\end{pmatrix},
\end{equation}
and denote the symmetric part of this matrix as $\mathcal{R}_\text{Vis-IR} = (\langle \hat {\bf x} \hat {\bf x}^{\rm T} \rangle_\text{Vis-IR} + \langle \hat {\bf x} \hat {\bf x}^{\rm T} \rangle_\text{Vis-IR}^{\rm T})/2$, writing it in a block form
\begin{equation}
\mathcal{R}_{s_{\rm L}v_{\rm U}} 
=
\begin{pmatrix}
\mathcal{C}_\text{Vis-Vis} & \mathcal{C}_\text{Vis-IR} \\
\mathcal{C}_\text{Vis-IR} & \mathcal{C}_\text{IR-IR}  
\end{pmatrix},
\end{equation}
where 
\begin{equation}
\mathcal{C}_\text{Vis-Vis}
=
\frac{1}{2}
\begin{pmatrix}
2
\left(
\langle \hat x_{s_{\rm L}|{\bf k}_f} \hat x_{s_{\rm L}|{\bf k}_f} \rangle 
+ 
N^{(\rm bg)}_{\rm Vis}
\right) & 
\langle \hat x_{s_{\rm L}|{\bf k}_f} \hat p_{s_{\rm L}|{\bf k}_f} \rangle
+
\langle \hat p_{s_{\rm L}|{\bf k}_f} \hat x_{s_{\rm L}|{\bf k}_f} \rangle
\\
\langle \hat x_{s_{\rm L}|{\bf k}_f} \hat p_{s_{\rm L}|{\bf k}_f} \rangle
+
\langle \hat p_{s_{\rm L}|{\bf k}_f} \hat x_{s_{\rm L}|{\bf k}_f} \rangle
& 
2
\left(
\langle \hat p_{s_{\rm L}|{\bf k}_f} \hat p_{s_{\rm L}|{\bf k}_f} \rangle 
+ 
N^{(\rm bg)}_{\rm Vis}
\right)
\end{pmatrix},
\end{equation}
\begin{equation}
\mathcal{C}_\text{IR-IR}
=
\frac{1}{2}
\begin{pmatrix}
2
\left(
\langle \hat x_{{\rm IR}|{\bf k}_i-{\bf k}_f} \hat x_{{\rm IR}|{\bf k}_i-{\bf k}_f} \rangle 
+ 
N^{(\rm bg)}_{\rm IR}
\right)
& 
\langle \hat x_{{\rm IR}|{\bf k}_i-{\bf k}_f} \hat p_{{\rm IR}|{\bf k}_i-{\bf k}_f} \rangle
+
\langle \hat p_{{\rm IR}|{\bf k}_i-{\bf k}_f} \hat x_{{\rm IR}|{\bf k}_i-{\bf k}_f} \rangle
\\
\langle \hat p_{{\rm IR}|{\bf k}_i-{\bf k}_f} \hat x_{{\rm IR}|{\bf k}_i-{\bf k}_f} \rangle
+
\langle \hat x_{{\rm IR}|{\bf k}_i-{\bf k}_f} \hat p_{{\rm IR}|{\bf k}_i-{\bf k}_f} \rangle
& 
2
\left(
\langle \hat p_{{\rm IR}|{\bf k}_i-{\bf k}_f} \hat p_{{\rm IR}|{\bf k}_i-{\bf k}_f} \rangle
+ 
N^{(\rm bg)}_{\rm IR}
\right)
\end{pmatrix},
\end{equation}
\begin{equation}
\mathcal{C}_\text{Vis-IR}
=
\frac{1}{2}
\begin{pmatrix}
\langle \hat x_{s_{\rm L}|{\bf k}_f} \hat x_{{\rm IR}|{\bf k}_i-{\bf k}_f} \rangle
+
\langle \hat x_{{\rm IR}|{\bf k}_i-{\bf k}_f} \hat x_{s_{\rm L}|{\bf k}_f} \rangle
& 
\langle \hat x_{s_{\rm L}|{\bf k}_f} \hat p_{{\rm IR}|{\bf k}_i-{\bf k}_f} \rangle
+
\langle \hat p_{{\rm IR}|{\bf k}_i-{\bf k}_f} \hat x_{s_{\rm L}|{\bf k}_f} \rangle
\\
\langle \hat p_{s_{\rm L}|{\bf k}_f} \hat x_{{\rm IR}|{\bf k}_i-{\bf k}_f} \rangle
+
\langle \hat x_{{\rm IR}|{\bf k}_i-{\bf k}_f} \hat p_{s_{\rm L}|{\bf k}_f} \rangle
& 
\langle \hat p_{s_{\rm L}|{\bf k}_f} \hat p_{{\rm IR}|{\bf k}_i-{\bf k}_f} \rangle
+
\langle \hat p_{{\rm IR}|{\bf k}_i-{\bf k}_f} \hat p_{s_{\rm L}|{\bf k}_f} \rangle
\end{pmatrix},
\end{equation}
and we denoted $\hat x_{{\rm IR}|{\bf k}_i-{\bf k}_f} = (\hat a_{{\rm IR}|{\bf k}_i-{\bf k}_f} + \hat a_{{\rm IR}|{\bf k}_i-{\bf k}_f}^\dag)/\sqrt{2}$ and  $\hat p_{{\rm IR}|{\bf k}_i-{\bf k}_f} = (\hat a_{{\rm IR}|{\bf k}_i-{\bf k}_f} - \hat a_{{\rm IR}|{\bf k}_i-{\bf k}_f}^\dag)/(i\sqrt{2})$.

The logarithmic negativity for emitted visible and mid-IR light, $E_N$, is
\begin{equation}
E_{N}
=
\max \left( 0, -\ln2|\xi| \right),
\end{equation}
where $\xi$ is a solution of the equation with the smallest absolute value~\cite{vidal2002computable}
\begin{equation}
\xi^4 
+ 
\xi^2 
\left( 
\det \mathcal{C}_\text{Vis-Vis} 
+
\det \mathcal{C}_\text{IR-IR} 
-
2 \det \mathcal{C}_\text{Vis-IR} 
\right) 
+ 
\det \mathcal{R}_\text{Vis-IR} 
=
0.
\end{equation}

\section{Isserlis's theorem for the correlation functions} \label{appendix: g2}

\subsection{Second-order cross-correlation function}

The second-order cross-correlation function between visible light with the wave vector ${\bf k}_{f}$ and the mid-IR light with the wave vector ${\bf k}_{i} - {\bf k}_{f}$ is
\begin{equation} \label{g2 def}
g^{(2)}_\text{Vis-IR}(\tau) 
=
\frac
{\left\langle 
\hat s^\dag_{{\rm L}|{\bf k}_{f}}(t) 
\hat a^\dag_{{\rm IR}|{\bf k}_{i} - {\bf k}_{f}}(t+\tau)
\hat a_{{\rm IR}|{\bf k}_{i} - {\bf k}_{f}}(t+\tau)
\hat s_{{\rm L}|{\bf k}_{f}}(t)
\right\rangle}
{\left\langle 
\hat s^\dag_{{\rm L}|{\bf k}_{f}}(t) 
\hat s_{{\rm L}|{\bf k}_{f}}(t)
\right\rangle \left\langle  
\hat a^\dag_{{\rm IR}|{\bf k}_{i} - {\bf k}_{f}}(t+\tau)
\hat a_{{\rm IR}|{\bf k}_{i} - {\bf k}_{f}}(t+\tau)
\right\rangle}.
\end{equation}
Below the optomechanical instability threshold, we can use the linearized version of Hamiltonian~(\ref{H optomech}) over the coherent drive~\cite{aspelmeyer2014cavity}.
In this case, the operators $\hat s_{{\rm L}|{\bf k}_f}$, $\hat v_{{\rm U}|{\bf k}_{i} - {\bf k}_{f}}$, and $\hat v_{{\rm L}|{\bf k}_{i} - {\bf k}_{f}}$ are proportional to the noise operators.
In the Born--Markov approximation, these noise operators obey Isserlis's theorem~\cite{coffey2012langevin}.
Therefore, Isserlis's theorem also applies to the operators $\hat s_{{\rm L}|{\bf k}_f}$, $\hat v_{{\rm U}|{\bf k}_{i} - {\bf k}_{f}}$, and $\hat v_{{\rm L}|{\bf k}_{i} - {\bf k}_{f}}$.
We apply Isserlis's theorem to the numerator of Eq.~(\ref{g2 def}) and obtain
\begin{multline}
{\left\langle 
\hat s^\dag_{{\rm L}|{\bf k}_{f}}(t) 
\hat a^\dag_{{\rm IR}|{\bf k}_{i} - {\bf k}_{f}}(t+\tau)
\hat a_{{\rm IR}|{\bf k}_{i} - {\bf k}_{f}}(t+\tau)
\hat s_{{\rm L}|{\bf k}_{f}}(t)
\right\rangle}
=
\left\langle 
\hat s^\dag_{{\rm L}|{\bf k}_{f}}(t) 
\hat s_{{\rm L}|{\bf k}_{f}}(t)
\right\rangle \left\langle  
\hat a^\dag_{{\rm IR}|{\bf k}_{i} - {\bf k}_{f}}(t+\tau)
\hat a_{{\rm IR}|{\bf k}_{i} - {\bf k}_{f}}(t+\tau)
\right\rangle
\\
+
\left|\left\langle 
\hat a_{{\rm IR}|{\bf k}_{i} - {\bf k}_{f}}(t+\tau)
\hat s_{{\rm L}|{\bf k}_{f}}(t)
\right\rangle\right|^2
+
\left|\left\langle 
\hat a^\dag_{{\rm IR}|{\bf k}_{i} - {\bf k}_{f}}(t+\tau)
\hat s_{{\rm L}|{\bf k}_{f}}(t)
\right\rangle\right|^2.
\end{multline}
The solution of the Heisenberg--Langevin equation provided in Appendix~\ref{appendix: correlations} implies $\left|\left\langle \hat a^\dag_{{\rm IR}|{\bf k}_{i} - {\bf k}_{f}}(t+\tau) \hat s_{{\rm L}|{\bf k}_{f}}(t) \right\rangle\right|^2 = 0$.
Thus, we obtain
\begin{equation}
g^{(2)}_\text{Vis-IR}(\tau) = 1 +
\frac
{
|\left\langle 
\hat a_{{\rm IR}|{\bf k}_{i} - {\bf k}_{f}}(t+\tau)
\hat s_{{\rm L}|{\bf k}_{f}}(t)
\right\rangle|^2
}
{\left\langle 
\hat s^\dag_{{\rm L}|{\bf k}_{f}}(t) 
\hat s_{{\rm L}|{\bf k}_{f}}(t)
\right\rangle \left\langle  
\hat a^\dag_{{\rm IR}|{\bf k}_{i} - {\bf k}_{f}}(t+\tau)
\hat a_{{\rm IR}|{\bf k}_{i} - {\bf k}_{f}}(t+\tau)
\right\rangle}.
\end{equation}

\subsection{Second-order autocorrelation function of the visible and mid-IR light}

The second-order autocorrelation functions of visible light with the wave vector ${\bf k}$ and the mid-IR light with the wave vector ${\bf q}$ are
\begin{equation} \label{g2 Vis def}
g^{(2)}_\text{Vis}(\tau) 
=
\frac
{\left\langle 
\hat s^\dag_{{\rm L}|{\bf k}}(t) 
\hat s^\dag_{{\rm L}|{\bf k}}(t+\tau)
\hat s^\dag_{{\rm L}|{\bf k}}(t+\tau)
\hat s_{{\rm L}|{\bf k}}(t)
\right\rangle}
{\left\langle 
\hat s^\dag_{{\rm L}|{\bf k}}(t) 
\hat s_{{\rm L}|{\bf k}}(t)
\right\rangle \left\langle  
\hat s^\dag_{{\rm L}|{\bf k}}(t+\tau)
\hat s_{{\rm L}|{\bf k}}(t+\tau)
\right\rangle},
\end{equation}

\begin{equation} \label{g2 IR def}
g^{(2)}_\text{IR}(\tau) 
=
\frac
{\left\langle 
\hat a^\dag_{{\rm IR}|{\bf q}}(t) 
\hat a^\dag_{{\rm IR}|{\bf q}}(t+\tau)
\hat a_{{\rm IR}|{\bf q}}(t+\tau)
\hat a_{{\rm IR}|{\bf q}}(t)
\right\rangle}
{\left\langle 
\hat a^\dag_{{\rm IR}|{\bf q}}(t) 
\hat a_{{\rm IR}|{\bf q}}(t)
\right\rangle \left\langle  
\hat a^\dag_{{\rm IR}|{\bf q}}(t+\tau)
\hat a_{{\rm IR}|{\bf q}}(t+\tau)
\right\rangle}.
\end{equation}
We apply Isserlis's theorem~\cite{coffey2012langevin} to the numerator of Eq.~(\ref{g2 Vis def})~and~(\ref{g2 IR def}) and obtain
\begin{equation} 
g^{(2)}_\text{Vis}(\tau) 
=
1
+
\frac
{
\left|
\left\langle 
\hat s^\dag_{{\rm L}|{\bf k}}(t+\tau)
\hat s_{{\rm L}|{\bf k}}(t)
\right\rangle
\right|^2
}
{\left\langle 
\hat s^\dag_{{\rm L}|{\bf k}}(t) 
\hat s_{{\rm L}|{\bf k}}(t)
\right\rangle \left\langle  
\hat s^\dag_{{\rm L}|{\bf k}}(t+\tau)
\hat s_{{\rm L}|{\bf k}}(t+\tau)
\right\rangle},
\end{equation}

\begin{equation} \label{g2 IR def}
g^{(2)}_\text{IR}(\tau) 
=
1
+
\frac
{
\left|\left\langle 
\hat a^\dag_{{\rm IR}|{\bf q}}(t+\tau)
\hat a_{{\rm IR}|{\bf q}}(t)
\right\rangle
\right|^2
}
{\left\langle 
\hat a^\dag_{{\rm IR}|{\bf q}}(t) 
\hat a_{{\rm IR}|{\bf q}}(t)
\right\rangle \left\langle  
\hat a^\dag_{{\rm IR}|{\bf q}}(t+\tau)
\hat a_{{\rm IR}|{\bf q}}(t+\tau)
\right\rangle}.
\end{equation}
At $\tau = 0$, we obtain $g^{(2)}_\text{Vis}(0) = 2$ and $g^{(2)}_\text{IR}(0) = 2$.

\subsection{Second-order autocorrelation function of heralded mid-IR light}
The second-order cross-correlation function of a heralded single-photon source of mid-IR light is~\cite{panyukov2022heralded}
\begin{equation}
g^{(2)}_\text{IR(her)}(0) =
\frac{g^{(3)}_{\rm Vis-IR}(0)}{\left[g^{(2)}_{\rm Vis-IR}(0)\right]^2},
\end{equation}
where
\begin{equation} \label{g3 def}
g^{(3)}_\text{Vis-IR}(0) =
\frac
{
\left\langle 
\hat s^\dag_{{\rm L}|{\bf k}_{f}}(t) 
\hat a^\dag_{{\rm IR}|{\bf k}_{i} - {\bf k}_{f}}(t)
\hat a^\dag_{{\rm IR}|{\bf k}_{i} - {\bf k}_{f}}(t)
\hat a_{{\rm IR}|{\bf k}_{i} - {\bf k}_{f}}(t)
\hat a_{{\rm IR}|{\bf k}_{i} - {\bf k}_{f}}(t)
\hat s_{{\rm L}|{\bf k}_{f}}(t)
\right\rangle
}
{
\left\langle 
\hat s^\dag_{{\rm L}|{\bf k}_{f}}(t) \hat s_{{\rm L}|{\bf k}_{f}}(t) 
\right\rangle
\left\langle 
\hat a^\dag_{{\rm IR}|{\bf k}_{i} - {\bf k}_{f}}(t)
\hat a_{{\rm IR}|{\bf k}_{i} - {\bf k}_{f}}(t)
\right\rangle^2
}.
\end{equation}
We apply Isserlis's theorem~\cite{coffey2012langevin} to the numerator of Eq.~(\ref{g3 def}) and obtain
\begin{multline}
\left\langle 
\hat s^\dag_{{\rm L}|{\bf k}_{f}}(t) 
\hat a^\dag_{{\rm IR}|{\bf k}_{i} - {\bf k}_{f}}(t)
\hat a^\dag_{{\rm IR}|{\bf k}_{i} - {\bf k}_{f}}(t)
\hat a_{{\rm IR}|{\bf k}_{i} - {\bf k}_{f}}(t)
\hat a_{{\rm IR}|{\bf k}_{i} - {\bf k}_{f}}(t)
\hat s_{{\rm L}|{\bf k}_{f}}(t)
\right\rangle
=
\\
2
\left\langle 
\hat s^\dag_{{\rm L}|{\bf k}_{f}}(t) 
\hat s_{{\rm L}|{\bf k}_{f}}(t)
\right\rangle 
\left\langle  
\hat a^\dag_{{\rm IR}|{\bf k}_{i} - {\bf k}_{f}}(t)
\hat a_{{\rm IR}|{\bf k}_{i} - {\bf k}_{f}}(t)
\right\rangle^2
+
4
\left|\left\langle 
\hat a_{{\rm IR}|{\bf k}_{i} - {\bf k}_{f}}(t)
\hat s_{{\rm L}|{\bf k}_{f}}(t)
\right\rangle\right|^2
\left\langle  
\hat a^\dag_{{\rm IR}|{\bf k}_{i} - {\bf k}_{f}}(t)
\hat a_{{\rm IR}|{\bf k}_{i} - {\bf k}_{f}}(t)
\right\rangle.
\end{multline}
As a result, we obtain
\begin{equation}
g^{(2)}_\text{IR(her)}(0) = 
\frac{4}{g^{(2)}_\text{Vis-IR}(0)}
-
\frac{2}{[g^{(2)}_\text{Vis-IR}(0)]^2}.
\end{equation}

\section{Exciton-polariton and phonon-polariton entanglement witness in the visible spectral range.} \label{appendix: witness}

\begin{figure}
\includegraphics[width=0.38\linewidth]{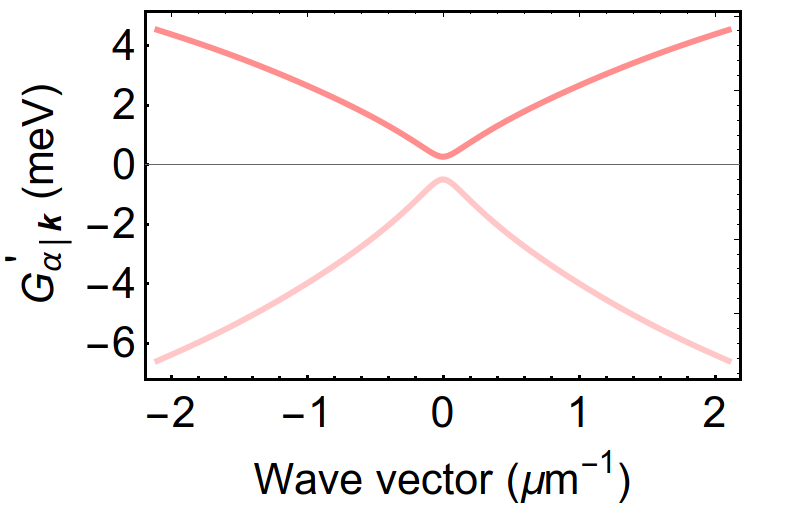}
\caption{
{\bf Collective optomechanical interaction energy for read process.}
Interaction constants $G_{{\rm L}|{\bf k}}'$~(red line) and $G_{{\rm U}|{\bf k}}'$~(pink line) defined by Eq.~(\ref{G_L}) and Eq.~(\ref{G_U}), respectively for the parameters specified in Appendix~\ref{appendix: Hamiltonian} and $n_{s_{\rm L}|{\bf k}}^{\rm drive} = 10^5$.
}
    \label{fig:G_prime}
\end{figure}

To transfer the entanglement between exciton-polaritons and phonon-polaritons to the entanglement between exciton-polaritons with different wave vectors, we use {\it write} and {\it read} light (Fig.~\ref{fig:WitnessJoined}(a)).
The {\it write light} and corresponding {\it write process} is captured by the Hamiltonian~(\ref{H_weak}).
To include the {\it read light} into our model, we add the term
\begin{equation} \label{read light}
\hat H_{\rm read} 
=
\hbar \Omega_{\rm read}
\left(
\hat c_{{\rm Exc}|{\bf k}}^\dag
e^{-i \omega_{\rm read} t}
+
\hat c_{{\rm Exc}|{\bf k}}
e^{i \omega_{\rm read} t}
\right)
\end{equation}
to the Hamiltonian~(\ref{H_weak}).
This term reflects our suggestion that the {\it read light} predominantly interacts with the excitons hosted by media and creates the polarisation oscillating with the frequency $\omega_{\rm read} = 2\omega_{s_{\rm L}|{\bf -k}} - \omega_{s_{\rm U}|{\bf -k}}$.

Adding the Hamiltonian~(\ref{read light}) to the Hamiltonian~(\ref{H_weak}) and repeating all the steps that previously led us to the Hamiltonian~(\ref{H_OM_appendix}), we obtain
\begin{equation} \label{Hamiltonian: witness}
\hat H_{\rm witness} = \hat H_{\rm write} + \hat H_{\rm read},    
\end{equation}
where $\hat H_{\rm write}$ describes the {\it write process} and is determined by Eq.~(\ref{H_OM_lin_all_modes_2})
\begin{equation} \label{H_write}
\hat H_{\rm write}
=
\hbar 
\left(
\omega_{s_{\rm L}|{\bf k}} - \omega_{s_{\rm U}|{\bf k}}
\right)
\hat s^\dag_{{\rm L}|{\bf k}}
\hat s_{{\rm L}|{\bf k}}
+
\sum_{\alpha={\rm L,U}}
\left[
\hbar \omega_{v_{\rm \alpha}|{\bf 0}}
\hat v_{{\rm \alpha}|{\bf 0}}^\dag 
\hat v_{{\rm \alpha}|{\bf 0}}
+
\left(
\hbar 
G_{{\rm \alpha}|{\bf k}{\bf k}}
\hat v_{{\rm \alpha}|{\bf 0}}^\dag 
\hat s_{{\rm L}|{\bf k}}^\dag
+
h.c.
\right)
\right],
\end{equation}
and $\hat H_{\rm read}$ describes the {\it read process}.
The full Hamiltonian (before linearization) of the {\it read process}, $\hat H_{\rm read}^{\rm (full)}$, is
\begin{multline} \label{H read full}
\hat H_{\rm read}^{\rm (full)}
=
\sum_{\alpha={\rm L,U,h}}
\left[
\hbar \omega_{{\rm s_\alpha}|{\bf -k}} \hat n_{{\rm s_\alpha}|{\bf -k}} 
+
\hbar 
\Omega_{\rm read}
\left(
X_{{\rm s_\alpha}|{\bf -k},{\rm Exc}}^* 
\hat s^\dag_{{\alpha}|{\bf -k}} 
e^{-i\omega_{\rm read} t} 
+
\hat s_{{\rm L}|{-\bf k}}^\dag
\sum_{\beta={\rm L,U}} 
\hbar g_{\beta\alpha|{-\bf k}}' 
\hat v_{\beta|{-\bf k}} \hat s_{\alpha|{-\bf k}}
+
h.c.
\right)
\right],
\end{multline}
where
\begin{equation}
g_{{\rm LL}|{\bf k}}' = 
\frac{\Lambda \Omega_{\rm Vis}}{\sqrt{N_{\rm exc}}}
\sin\phi_{\bf 0} 
\left(
X^*_{s_{\rm L}|{\bf k},{\rm Exc}} X_{s_{\rm L}|{\bf k},{\rm Vis}|{\bf k}}
- X^*_{s_{\rm L}|{\bf k},{\rm Vis}|{\bf k}} X_{s_{\rm L}|{\bf k},{\rm Exc}}
\right)
-
\frac{\Lambda \Omega_{\rm IR}}{\sqrt{N_{\rm exc}}}
\cos\phi_{\bf 0} X^*_{s_{\rm L}|{\bf k},{\rm Exc}} X_{s_{\rm L}|{\bf k},{\rm Exc}},
\end{equation}
\begin{equation}
g_{{\rm LU}|{\bf k}}' = 
\frac{\Lambda \Omega_{\rm Vis}}{\sqrt{N_{\rm exc}}}
\sin\phi_{\bf 0} 
\left(
 X^*_{s_{\rm L}|{\bf k},{\rm Exc}} X_{s_{\rm U}|{\bf k},{\rm Vis}|{\bf k}}
- X^*_{s_{\rm L}|{\bf k},{\rm Vis}|{\bf k}} X_{s_{\rm U}|{\bf k},{\rm Exc}}
\right)
-
\frac{\Lambda \Omega_{\rm IR}}{\sqrt{N_{\rm exc}}}
\cos\phi_{\bf 0} X^*_{s_{\rm L}|{\bf k},{\rm Exc}} X_{s_{\rm U}|{\bf k},{\rm Exc}},
\end{equation}
\begin{equation}
g_{{\rm Lh}|{\bf k}}' = 
\frac{\Lambda \Omega_{\rm Vis}}{\sqrt{N_{\rm exc}}}
\sin\phi_{\bf 0} 
\left(
X^*_{s_{\rm L}|{\bf k},{\rm Exc}} X_{s_{\rm h}|{\bf k},{\rm Vis}|{\bf k}}
- 
X^*_{s_{\rm L}|{\bf k},{\rm Vis}|{\bf k}} X_{s_{\rm h}|{\bf k},{\rm Exc}}
\right)
-
\frac{\Lambda \Omega_{\rm IR}}{\sqrt{N_{\rm exc}}}
\cos\phi_{\bf 0} 
X^*_{s_{\rm L}|{\bf k},{\rm Exc}} X_{s_{\rm h}|{\bf k},{\rm Exc}},
\end{equation}
\begin{equation}
g_{{\rm UL}|{\bf k}}' = 
-\frac{\Lambda \Omega_{\rm Vis}}{\sqrt{N_{\rm exc}}}
\cos\phi_{\bf 0} 
\left(
 X^*_{s_{\rm L}|{\bf k},{\rm Exc}} X_{s_{\rm L}|{\bf k},{\rm Vis}|{\bf k}}
- X^*_{s_{\rm L}|{\bf k},{\rm Vis}|{\bf k}} X_{s_{\rm L}|{\bf k},{\rm Exc}}
\right)
-
\frac{\Lambda \Omega_{\rm IR}}{\sqrt{N_{\rm exc}}}
\sin\phi_{\bf 0} X^*_{s_{\rm L}|{\bf k},{\rm Exc}} X_{s_{\rm L}|{\bf k},{\rm Exc}},
\end{equation}
\begin{equation}
g_{{\rm UU}|{\bf k}}' = 
-\frac{\Lambda \Omega_{\rm Vis}}{\sqrt{N_{\rm exc}}}
\cos\phi_{\bf 0} 
\left(
X^*_{s_{\rm L}|{\bf k},{\rm Exc}} X_{s_{\rm U}|{\bf k},{\rm Vis}|{\bf k}}
-
X^*_{s_{\rm L}|{\bf k},{\rm Vis}|{\bf k}} X_{s_{\rm U}|{\bf k},{\rm Exc}}
\right)
-
\frac{\Lambda \Omega_{\rm IR}}{\sqrt{N_{\rm exc}}}
\sin\phi_{\bf 0} X^*_{s_{\rm L}|{\bf k},{\rm Exc}} X_{s_{\rm U}|{\bf k},{\rm Exc}},
\end{equation}
\begin{equation}
g_{{\rm Uh}|{\bf k}}' = 
-\frac{\Lambda \Omega_{\rm Vis}}{\sqrt{N_{\rm exc}}}
\cos\phi_{\bf 0} 
\left(
X^*_{s_{\rm L}|{\bf k},{\rm Exc}} X_{s_{\rm h}|{\bf k},{\rm Vis}|{\bf k}}
-
X^*_{s_{\rm L}|{\bf k},{\rm Vis}|{\bf k}} X_{s_{\rm h}|{\bf k},{\rm Exc}}
\right)
-
\frac{\Lambda \Omega_{\rm IR}}{\sqrt{N_{\rm exc}}}
\sin\phi_{\bf 0} 
X^*_{s_{\rm L}|{\bf k},{\rm Exc}} X_{s_{\rm h}|{\bf k},{\rm Exc}}.
\end{equation}
The linearization of (\ref{H read full}) leads to the Hamiltonian 
%describing the {\it write} and {\it read} processes is $\hat H_{\rm witness} = \hat H_{\rm write} + \hat H_{\rm read}$ with $\hat H_{\rm write} = \hat H_{{\rm OM}|{\bf k},{\bf k}}$  and 
\begin{equation} \label{H read linear}
\hat H_{\rm read} 
=
\hbar (\omega_{{\rm s_L}|{\bf -k}} - \omega_{\rm read}) 
\hat n_{{\rm s_L}|{\bf -k}}
+
\left(
\hbar G'_{{\rm L}|{\bf -k}} \hat s_{{\rm L}|{\bf -k}}^\dag \hat v_{{\rm L}|{\bf 0}}
+
\hbar G'_{{\rm U}|{\bf -k}} \hat s_{{\rm L}|{\bf -k}}^\dag \hat v_{{\rm U}|{\bf 0}}
+
h.c.
\right),
\end{equation}
where $G'_{L|{\bf k}}$ and $G'_{U|{\bf k}}$ are the linearized optomechanical constants of the read process proportional to the square root of the intensity of the read light
%(Appendix~\ref{appendix: witness}).
%We illustrate the processes governed by these Hamiltonians in Fig.~\ref{fig:WitnessJoined}(a).
%Eq.~(\ref{H read linear}) with optomechanical interaction constants
\begin{equation} \label{G_L}
G_{{\rm L}|{\bf k}}' 
= 
g_{{\rm LL}|{\bf k}}' \sqrt{n_{s_{\rm L}|{\bf k}}^{\rm drive}} 
+
g_{{\rm LU}|{\bf k}}' \sqrt{n_{s_{\rm U}|{\bf k}}^{\rm drive}} 
+
g_{{\rm Lh}|{\bf k}}' \sqrt{n_{s_{\rm h}|{\bf k}}^{\rm drive}} ,
\end{equation}
\begin{equation} \label{G_U}
G_{{\rm U}|{\bf k}}' 
= 
g_{{\rm UL}|{\bf k}}' \sqrt{n_{s_{\rm L}|{\bf k}}^{\rm drive}} 
+
g_{{\rm UU}|{\bf k}}' \sqrt{n_{s_{\rm U}|{\bf k}}^{\rm drive}} 
+
g_{{\rm Uh}|{\bf k}}' \sqrt{n_{s_{\rm h}|{\bf k}}^{\rm drive}} 
\end{equation}
and
\begin{equation}
n_{s_{\rm L}|{\bf k}}^{\rm drive} 
= 
|X_{s_{\rm L}|{\bf k},{\rm Exc}}|^2
\left( \frac{\Omega_{\rm read}}{\omega_{\rm L2} - \omega_{\rm read}} \right)^2,
\end{equation}
\begin{equation}
n_{s_{\rm U}|{\bf k}}^{\rm drive}
= 
|X_{s_{\rm U}|{\bf k},{\rm Exc}}|^2
\left( \frac{\Omega_{\rm read}}{\omega_{\rm U2} - \omega_{\rm read}} \right)^2,
\end{equation}
\begin{equation}
n_{s_{\rm h}|{\bf k}}^{\rm drive}
= 
|X_{s_{\rm h}|{\bf k},{\rm Exc}}|^2
\left( \frac{\Omega_{\rm read}}{\omega_{\rm h2} - \omega_{\rm read}} \right)^2.
\end{equation}
We plot $G_{{\rm L}|{\bf k}}'$ and $G_{{\rm U}|{\bf k}}'$ as the functions of $\bf k$ in Fig.~\ref{fig:G_prime}.

The Hamiltonian~(\ref{Hamiltonian: witness}) allows us to obtain the following Heisenberg--Langevin equation, determining the quantum dynamics of the system
\begin{equation} \label{eq for A witness}
\frac{d \hat {\bf A}(t)}{dt} = \mathcal{M} \hat {\bf A}(t) + \hat {\bf F}(t),
\end{equation}
where $\hat {\bf A}(t)$ is the vector of exciton-polariton and phonon-polariton operators
\begin{equation} \label{A def witness}
\hat {\bf A}(t) 
= 
\left( 
\hat s_{{\rm L}|{\bf k}}(t) e^{i\omega_{\rm write} t}, 
\hat s^\dag_{{\rm L}|{\bf k}}(t) e^{-i\omega_{\rm write} t},
\hat s_{{\rm L}|{\bf -k}}(t) e^{i\omega_{\rm read} t}, 
\hat s^\dag_{{\rm L}|{\bf -k}}(t) e^{-i\omega_{\rm read} t},
\hat v_{{\rm U}|{\bf 0}}(t), 
\hat v^\dag_{{\rm U}|{\bf 0}}(t), 
\hat v_{{\rm L}|{\bf 0}}(t), 
\hat v^\dag_{{\rm L}|{\bf 0}}(t) 
\right)^{\rm T},
\end{equation}
and $\hat {\bf F}(t)$ is the vector of noise operators
\begin{equation} \label{F def}
\hat {\bf F}(t) 
=
\left( 
\hat f_{s_{\rm L}|{\bf k}}(t) e^{i\omega_{\rm write} t}, 
\hat f^\dag_{s_{\rm L}|{\bf k}}(t) e^{-i\omega_{\rm write} t}, 
\hat f_{s_{\rm L}|{\bf -k}}(t) e^{i\omega_{\rm read} t}, 
\hat f^\dag_{s_{\rm L}|{\bf -k}}(t) e^{-i\omega_{\rm read} t}, 
\hat f_{v_{\rm U}|{\bf 0}}(t), 
\hat f^\dag_{v_{\rm U}|{\bf 0}}(t), 
\hat f_{v_{\rm L}|{\bf 0}}(t), 
\hat f^\dag_{v_{\rm L}|{\bf 0}}(t) 
\right)^{\rm T}.
\end{equation}
The noise operators in $\hat {\bf F}(t)$ have zero mean $\langle \hat {\bf F}(t) \rangle = 0$ and are delta-correlated, $\langle
\hat {\bf F}(t_1) \hat {\bf F}^{\rm T}(t_2)
\rangle
=
\mathcal{D}_F\delta(t_1-t_2)$~\cite{gardiner2004quantum, coffey2012langevin}
\begin{equation}
\mathcal{D}_F =
\begin{pmatrix}
\mathcal D_{s_{\rm L}|{\bf k}} & 0 & 0 & 0 \\
0 & \mathcal D_{s_{\rm L}|{\bf -k}} & 0 & 0 \\
0 & 0 & \mathcal D_{v_{\rm U}|{\bf 0}} & 0 \\ 
0 & 0 & 0 & \mathcal D_{v_{\rm L}|{\bf 0}} \\ 
\end{pmatrix},
\end{equation}
\begin{equation}
\mathcal D_{\alpha} =
\begin{pmatrix}
0 & \gamma_{\alpha} (1 + n^{\rm th}_{\alpha}) \\
\gamma_{\alpha} n^{\rm th}_{\alpha} & 0  
\end{pmatrix},
\end{equation}
$n^{\rm th}_{\alpha} = (e^{\hbar \omega_{\alpha}/k_BT} - 1)^{-1}$, $k_B$ is the Boltzmann constant, and $T$ is the ambient temperature.

The matrix of the differential equation~(\ref{eq for A witness}) is
\begin{equation} \label{matrix of the linearized system witness}
\mathcal{M} = 
\begin{pmatrix}
\mathcal{L}_{s_{\rm L}|{\bf k}} + \mathcal{J} \omega_{\rm write} & 0 & \mathcal{G}_{{\rm U}|{\bf k}} & \mathcal{G}_{{\rm L}|{\bf k}} 
\\
0 & \mathcal{L}_{s_{\rm L}|{\bf -k}} + \mathcal{J} \omega_{\rm read} & \mathcal{G}'_{{\rm U}|{\bf -k}} & \mathcal{G}'_{{\rm L}|{\bf -k}} 
\\
\mathcal{G}_{{\rm U}|{\bf k}} & \mathcal{G}'_{{\rm U}|{\bf -k}} & \mathcal{L}_{v_{\rm U}|{\bf 0}} & 0 
\\
\mathcal{G}_{{\rm L}|{\bf k}} & \mathcal{G}'_{{\rm L}|{\bf -k}} & 0 & \mathcal{L}_{v_{\rm L}|{\bf 0}}
\end{pmatrix},
\end{equation}
where
\begin{equation}
\mathcal{J} = 
\begin{pmatrix}
i & 0 \\
0 & -i
\end{pmatrix},
\end{equation}
\begin{equation}
\mathcal{L}_{\alpha}  = 
\begin{pmatrix}
-i \omega_{\alpha}-\frac{\gamma_{\alpha}}{2} & 0 \\
0 & i\omega_{\alpha} - \frac{\gamma_{\alpha}}{2} 
\end{pmatrix},
\end{equation}
\begin{equation}
\mathcal{G}_{{\rm U/L}|{\bf k}}  = 
\begin{pmatrix}
0 & -i G_{{\rm U/L}|{\bf k}{\bf k}} \\
i G_{{\rm U/L}|{\bf k}{\bf k}}^* & 0
\end{pmatrix},
\end{equation}
and
\begin{equation}
\mathcal{G}_{{\rm U/L}|{\bf -k}}'  = 
\begin{pmatrix}
-i G'_{{\rm U/L}|{\bf -k}} & 0 \\
0 & i (G'_{{\rm U/L}|{\bf -k}})^{*}
\end{pmatrix}.
\end{equation}

We use the Eq.~(\ref{stationary eq. for AA}) to obtain the stationary correlation matrix $\langle \hat {\bf A} \hat {\bf A}^{\rm T} \rangle = \langle \hat {\bf A}(t) \hat {\bf A}^{\rm T}(t) \rangle_{t\to+\infty}$.
This stationary correlation matrix allows us to find the second-order cross-correlation function between lower exciton-polaritons with the wave vectors $\bf k$ and $-{\bf k}$~(\ref{Vis-Vis correlations}),
the logarithmic negativity $E_N$ between these exciton-polaritons, and the Wigner function~\cite{weedbrook2012gaussian} of symmetric and antisymmetric exciton-polariton modes, corresponding to the annihilation operators $(\hat s_{{\rm L}|{\bf k}} + \hat s_{{\rm L}|{\bf -k}})/\sqrt2$ and $(\hat s_{{\rm L}|{\bf k}} - \hat s_{{\rm L}|{\bf -k}})/\sqrt2$, respectively.

\end{widetext}

%\nocite{apsrev41Control}
%\bibliographystyle{apsrev4-1}
%\bibliographystyle{apsrev4-2}
\bibliography{refs}% Produces the bibliography via BibTeX.

\end{document}